\documentclass{vldb}
\usepackage{graphicx,times}
\usepackage{balance}  % for  \balance command ON LAST PAGE  (only there!)
\usepackage{amsmath, amssymb}
\usepackage{bm,url,paralist,algorithm,algorithmic}
\usepackage{epstopdf}
\input{Definitions}

\newcommand{\subparagraph}{}
\usepackage[small,compact]{titlesec}
\usepackage[small, bf]{caption}
\DeclareCaptionType{copyrightbox}
\newcommand{\hw}{\hat{\wvec}}
\newcommand{\hww}{\hat{W}}
\newcommand{\hh}{\hat{\hvec}}
\newcommand{\thh}{\tilde{\hvec}}
\newcommand{\thhh}{\tilde{H}}
\newcommand{\tw}{\tilde{\wvec}}
\newcommand{\tww}{\tilde{W}}
\newcommand{\bh}{\overline{\hvec}^*}
\newcommand{\bhh}{\overline{H}^*}
\newcommand{\bw}{\overline{\wvec}^*}

\newcommand{\hhh}{\hat{H}}

\begin{document}

\title{A Learning Framework for Self-Tuning Histograms}

\numberofauthors{4} 

\author{
% 1st. author
\alignauthor
Raajay Viswanathan\\
       \affaddr{Microsoft Research India}\\
       \affaddr{Bangalore, India}\\
       \email{t-rviswa@microsoft.com}
% 2nd. author
\alignauthor Prateek Jain\\
       \affaddr{Microsoft Research India}\\
       \affaddr{Bangalore, India}\\
       \email{prajain@microsoft.com}
% 3rd. author
\alignauthor Srivatsan Laxman \\
       \affaddr{Microsoft Research India}\\
       \affaddr{Bangalore, India}\\
       \email{slaxman@microsoft.com}
%\and  % use '\and' if you need 'another row' of author names
% 4th. author
\alignauthor Arvind Arasu\\
       \affaddr{Microsoft Research}\\
       \affaddr{Redmond, WA, USA}\\
       \email{arvinda@microsoft.com}
}

% \author{\hspace*{-40pt}Raajay Viswanathan\\\hspace*{-40pt}Microsoft Research India\\\hspace*{-40pt}Bangalore, India\\\hspace*{-40pt}t-rviswa@microsoft.com \and \hspace*{-20pt}Prateek Jain\\\hspace*{-20pt}Microsoft Research India\\\hspace*{-20pt}Bangalore, India\\\hspace*{-20pt}prajain@microsoft.com \and \hspace*{-20pt}Srivatsan Laxman \\\hspace*{-20pt}Microsoft Research India\\\hspace*{-20pt}Bangalore, India\\\hspace*{-20pt}slaxman@microsoft.com\and \hspace*{-20pt}Arvind Arasu\\\hspace*{-20pt}Microsoft Research\\\hspace*{-20pt}Redmond, USA\\\hspace*{-20pt}arvinda@microsoft.com}
\date{}

\maketitle

\begin{abstract}
  
% Arvind: Changes: general edits for conciseness. Removed implication that SpHist is order-of-magnitude better
% than ISOMER in efficiency. Also, we can be better even for 1-d data, so removed high-dim dataset
% qualification.
In this paper, we consider the problem of estimating self-tuning histograms using query workloads. 
To this end, we propose a general learning theoretic formulation. Specifically, we use query feedback from a workload as {\em training data} to estimate a histogram with a small memory footprint that minimizes the
  expected error on future queries. Our formulation provides a framework in which different approaches can be studied and developed. %Our formulation is flexible in the sense that it allows to design and  compare different methods (possibly specialized for different settings). 
We first study the simple
  class of equi-width histograms and present a learning algorithm, {\em EquiHist}, that is
  competitive in many settings. We also provide formal guarantees for equi-width histograms that highlight scenarios in which equi-width histograms can be expected to succeed or fail.
% Ours is the first work to provide formal error guarantees for equi-width histograms. 
We then go beyond equi-width histograms
  and present a novel learning algorithm, {\em SpHist}, for estimating general histograms. Here we use 
  Haar wavelets to reduce the problem of learning histograms to that of learning a sparse vector.  Both 
  algorithms have
  multiple advantages over existing methods: 1) simple and scalable extensions to multi-dimensional data, 2)
  scalability with number of histogram buckets and size of query feedback, 3) natural extensions to incorporate new
  feedback and handle database updates.  We demonstrate these advantages over the current state-of-the-art,
  ISOMER, through detailed experiments on real and synthetic data. In particular, we show that SpHist obtains up to $50\%$ less error than ISOMER on real-world multi-dimensional datasets. %For example, for a three-dimensional projection of Census data, our SpHist method consistently outperforms ISOMER, obtaining 
\end{abstract}

%%% Local Variables: 
%%% mode: latex
%%% TeX-master: "histestimation-main"
%%% End: 

\section{Introduction}
\label{sec:intro}
Histograms are a central component of modern databases. They are used
to summarize data and estimate cardinalities of sub-expressions during
query optimization. Typically, histograms are constructed solely from
data and are not \emph{workload-aware}. This approach has known
limitations~\cite{BrunoC02, SrivastavaHMKT06}: First, a workload may
access data non-uniformly %more often than others 
 due to which a
workload-oblivious histogram might waste resources (e.g., space) on
infrequently accessed parts of the data. Second, constructing a
histogram from data can be expensive, requiring a scan or a sample of
data; further, maintaining a histogram in the presence of updates is
nontrivial. The standard approach is to rebuild histograms from
scratch after a certain number of updates, resulting in possibly
inaccurate histograms between builds~\cite{Ioannidis03}.

%\vspace{1ex} \noindent {\bf Self-Tuning Histograms:} 
To address these
limitations, prior work has proposed \emph{self-tuning histograms}
\cite{AboulnagaC99, BrunoC02, SrivastavaHMKT06}. Briefly, the idea is
to collect \emph{query feedback} information during query execution
and to use this information to build and refine histograms. Query
feedback is typically cardinalities of filter expressions over tables
(e.g., $|\sigma_{5 \leq A \leq 10 \wedge 2 \leq B \leq 3} (R)| = 10$);
following prior work~\cite{SrivastavaHMKT06} we call such
cardinalities together with the corresponding query expressions as \emph{query feedback records} (or
\emph{QFRs}). Such feedback can be collected with minimal overhead
during query execution by maintaining counters at operators in the
query plan~\cite{AboulnagaC99, MarklLR03}. The overall idea is that since
query feedback captures data characteristics relevant to a workload,
histograms built using query feedback would be more accurate for similar
workloads. Also, histograms can be refined as new feedback information
is available and hence one can track changes in data characteristics arising
from updates.

Over the years, several self-tuning histograms have been introduced, such as STGrid \cite{AboulnagaC99}, STHoles \cite{BrunoC02}, ISOMER \cite{SrivastavaHMKT06}. Each of these methods uses an interesting way of selecting  histogram bucket boundaries as well as fixing histogram bucket heights. However, most of the existing methods lack theoretical analyses/guarantees and do not scale well with high dimensions or large number of QFRs. Of particular interest is ISOMER, \cite{SrivastavaHMKT06}, the current state-of-the-art in self-tuning histograms. ISOMER uses query feedback to compute a ``consistent'' and unbiased histogram based on the {\em maximum-entropy (maxent)} principle. However, the obtained histogram might have $\Theta(N)$ buckets, given $N$ QFRs. To get a histogram with $k\ll N$ buckets, ISOMER heuristically eliminates up to $(N - k)$ feedback records. This step discards valuable feedback information and can have an adverse impact on quality, as empirically demonstrated in Section~\ref{sec:exp}. Furthermore, it hinders the method's scalability to high-dimensions or large number of QFRs. Another limitation of this approach is that it is not robust to database updates. Updates can produce \emph{inconsistent} query feedback for which the maxent distribution is undefined. Again, ISOMER heuristically discards potentially useful QFRs to get a consistent subset.

In this paper, we propose and study a simple learning-theoretic formalization of
self-tuning histograms. Informally, we model the QFRs as \emph{training examples} drawn from some unknown
distribution and the goal is to learn a $k$-bucket histogram that
minimizes the expected cardinality estimation error on future queries. %We consider both single- and multi-dimensional histograms. 
 Our formalization is based on standard learning principles and confers several advantages: %We note that this is a direct formalization of histogram learning unlike ISOMER which relies on an ``intermediate'' maximum entropy step, and this formalization confers several advantages: 
(1) Our learning algorithms leverage all available
feedback information (unlike ISOMER) and in many scenarios this
additional information translates to dramatic (order-of-magnitude)
reductions in cardinality estimation errors (see
Section~\ref{sec:exp}). (2) Our framework lends itself to efficient algorithms that are scalable to multiple dimensions and large number of QFRs (3) Our formalization is database-update friendly: it is inherently robust to inconsistent
QFRs and can easily incorporate natural strategies
such as using higher weights for recent QFRs compared to older
ones. 

We next list our main algorithmic contributions:

\vspace{1ex} \noindent \emph{1. Equi-width histograms}: We begin by
studying $1$-dimensional equi-width histograms in our learning framework and provide an efficient algorithm (EquiHist) for the same.  When the number of buckets is reasonably large (relative to how spiky the true distribution is),
this approach performs well in-practice. We also present
a theoretical analysis that shows that the error incurred by the learned
equi-width histogram is arbitrarily close to that of the
best overall histogram under
reasonable assumptions. This result is of independent theoretical
interest; we know of no prior result that analyzes cases when equi-width histograms can be expected to succeed/fail.

\vspace{1ex} \noindent \emph{2. Sparse-vector Recovery based method:} One of the main contributions of this paper is a novel reduction from the general histogram learning problem to a sparse-vector recovery problem. Informally, using Haar wavelets, we represent a histogram as a sparse vector. We then cast our learning problem as that of learning a sparse-vector. To this end, we provide an efficient algorithm (SpHist) by adapting techniques from \emph{compressed sensing} \cite{TroppG07}.
%\vspace{1ex} \noindent \emph{2. Sparse-vector Recovery based method:} One of the
%main contributions of this paper is a novel reduction from the general histogram learning problem to a sparse-vector recovery problem. We then provide eff%icient learning algorithm (SpHist) for learning general histograms based on adaptation of techniques from \emph{sparse vector recovery} \cite{Candes08, TroppG07}. Informally, we show that a
%histogram with a small number of buckets can be viewed as a sparse
%vector under \emph{Haar wavelet transform}~\cite{Mallat97} and this
%sparse vector can be ``recovered'' using query feedback via recent
%techniques in compressed sensing.

\vspace{1ex} \noindent \emph{3. Multi-dimensional histograms:} Our
equi-width and sparse vector recovery algorithms admit straightforward
 generalizations to multi-dimensional settings. We also show that the error bounds for equi-width histograms extend to multiple dimensions. Also, SpHist admits a powerful class of histograms characterized by sparsity under Haar wavelet transformations. Using results in \cite{VitterW99}, we show that not only does this class of histograms have small memory footprint, but it can also estimate cardinality for high dimensional range queries as efficiently as existing self-tuning histograms. 
%Interestingly, the
%multi-dimensional histograms learned by the sparse recovery approach
%does not have a simple \emph{structural characterization} (e.g.,
%STHoles in \cite{BrunoCG01} and \cite{SrivastavaHMKT06}); it is a new
%class of histograms characterized by sparsity under Haar wavelet
%transformations.

\vspace{1ex} \noindent \emph{4. Dynamic QFRs and Database Updates:} We present online variants
of our algorithms that maintain a histogram in the presence of new
QFRs. We also present extensions to incorporate database updates that ensure that our learned
histograms remain accurate.% in presence of data updates.

Finally, we include extensive empirical evaluation of our proposed
techniques over real and standard synthetic datasets, including comparisons with
prior work such as ISOMER. Our empirical results demonstrate significant improvement over ISOMER in terms of accuracy in query cardinality estimation for several scenarios and databases. %Furthermore, we demonstrate that our methods are scalable: a) they are able to handle truly multi-dimensional queries, 2) scales well with range of attribute values, number of buckets, as well as number of QFRs. 

{\bf Outline}: We present notations and preliminaries in Section~\ref{sec:prelim}.  We then present our equi-width as well sparse-recovery based approaches in Section~\ref{sec:method}. We provide empirical evaluation of our methods in Section~\ref{sec:exp} . In Section~\ref{sec:related}, we present some of the related works to our work and contrast them against our methods, and finally conclude with Section~\ref{sec:conclusion}. 
%%% Local Variables: 
%%% mode: latex
%%% TeX-master: "histestimation-main"
%%% End: 

\section{Notation and Preliminaries}
\label{sec:prelim}
In this section we introduce notation and review concepts from
learning used in the paper.

For 1-dimensional histograms,  $R$ and $A$ denote the relation and the column, respectively, over which a histogram is defined. We assume throughout that the domain of column $A$ is $[1,\ldots, r]$; our algorithms can be generalized to handle categorical and other numeric domains. Also, let $M$ be the number of records in $R$, i.e., $M=|R|$. A histogram over $A$ consists of $k$
buckets $B_1, \ldots, B_k$. Each bucket $B_j$ is associated with an
interval $[\ell_j, u_j]$ and a count $n_j$ representing (an estimate of)
 the number of values in $R(A)$ that belong to interval $[\ell_j,
u_j]$. The intervals $[\ell_j, u_j] (1\leq j \leq k)$ are
non-overlapping and partition the domain $[1, r]$. We say that the
\emph{width} of bucket $B_j$ is $(u_j - \ell_j + 1)$. A histogram
represents an \emph{approximate} distribution of values in $R(A)$; the
estimated frequency of value $i \in [\ell_j, u_j]$ is $\frac{n_j} {u_j
  - \ell_j + 1}$. We use interval $[\ell, u]$ to represent the range
query $\sigma_{A \in [\ell, u]} (R)$.

% Arvind: Defining k-piecewise constant right here. This helps us
% compress the Haar wavelet discussion
For conciseness, we use a vector notation to represent queries and
histograms.  We denote vectors by lower-case bold letters
(e.g. $\wvec$) and matrices by upper-case letters (e.g., $A$). The
term $w_i$ denotes the $i$-th component of $\wvec$ and $\avec^T\bvec$
(or $\avec \cdot \bvec$) $ = \sum_i a_i b_i$ denotes inner product
between vectors $\avec$ and $\bvec$. We represent a histogram as a
vector $\hvec\in \mathbb{R}^r$ specifying its estimated distribution,
i.e., $h_i = \frac{n_j} {u_j - \ell_j + 1}$. By definition $\hvec$ is
\emph{constant} in each of $k$ bucket intervals and we refer to such
vectors as $k$-\emph{piecewise constant}. We represent a range query
$q = [\ell, u]$ in unary form as $\qvec \in \mathbb{R}^r$ where
$q_i=1, \forall i\in [\ell, u]$ and $q_i=0$, otherwise.  Hence, the
estimated cardinality of $q$ using $\hvec$, denoted $\hat{s}_\qvec$, is
given by $\hat{s}_\qvec=\qvec^T\hvec. $

% Arvind: added notation for multi-dim histogram for consistency. This
% should also help us save space in method section.
When discussing multi-dimensional histograms, we use $A_1, \ldots,$
$A_d$ to denote the $d$ columns over which a histogram is
defined. For ease of exposition, we assume all column domains are $[1, r]$. We first
present notation for $d = 2$: A histogram is a (estimated) value
distribution over every possible assignment of values to columns $A_1$
and $A_2$, and can be represented as a matrix $H \in \mathbb{R}^{r
  \times r}$. A $k$-bucket histogram has $k$ non-overlapping
rectangles with uniform estimated frequency within each rectangle; we
also consider other kinds of ``sparse'' histograms that we define in
Section \ref{sec:multi}. A query $Q$ is of the form $\sigma_{A_1 \in
  [\ell_1, u_1] \wedge A_2 \in [\ell_2, u_2]} (R)$ and can be
represented in unary form as a matrix $Q\in \mathbb{R}^{r \times r}$. The
estimated cardinality of query $Q$ using histogram $H$ is given by
their inner product $\hat{s}_Q=\langle Q, H \rangle = Tr(Q^T H)$. For $d > 2$,
histograms and queries are $d$-dimensional \emph{tensors} ($\in
\mathbb{R}^{r \times \cdots \times r}$); estimated cardinality of query $Q\in \mathbb{R}^{r \times \cdots \times r}$ for histogram $H\in \mathbb{R}^{r \times \cdots \times r}$ is given by tensor inner product $\hat{s}_Q=\langle Q, H \rangle$. 

\vspace{1ex} \noindent {\bf $L_p$-norm}: $\|\xvec\|_p$ denotes $L_p$ norm of $\xvec\in \mathbb{R}^r$ and is given by $\|\xvec\|_p=(\sum_{i=1}^r|x_i|^p)^{1/p}$.

\vspace{1ex}\noindent {\bf Lipschitz Functions}: A function $f:\mathbb{R}^r\rightarrow
\mathbb{R}$ is $L$-Lipschitz continuous if:
$\forall \xvec,\yvec,\ |f(\xvec)-f(\yvec)|\leq \|\xvec-\yvec\|_2\cdot L.$

\vspace{1ex}\noindent {\bf Convex Functions}: A function $f:\mathbb{R}^r\rightarrow \mathbb{R}$ is convex if:\vspace*{-3pt}
$$\forall~ 0\leq \lambda \leq 1, \xvec,\yvec\in \mathbb{R}^r,\ f(\lambda \xvec+ (1-\lambda)\yvec)\leq \lambda f(\xvec)+(1-\lambda) f(\yvec).\vspace*{-3pt}$$
Furthermore, a $f:\mathbb{R}^r\rightarrow \mathbb{R}$ is $\alpha$-{\em strongly} convex ($\alpha > 0$) w.r.t. 
$L_2$ norm if, $\forall~ 0\leq \lambda \leq 1, \xvec,\yvec\in \mathbb{R}^r$: \vspace*{-3pt}
$$f(\lambda \xvec+ (1-\lambda)\yvec)\leq \lambda f(\xvec)+(1-\lambda) f(\yvec)-\alpha\frac{\lambda(1-\lambda)}{2}\|\xvec-\yvec\|_2^2.\vspace*{-3pt}$$
Let $H\in \mathbb{R}^{r\times r}$ be the Hessian of $f$, then $f$ is $\alpha$-strongly convex iff smallest eigenvalue of $H$ is greater than $\alpha$. 

\vspace{1ex}\noindent {\bf Empirical-risk minimization}: In many learning applications, the input is a set of training examples $\Xcal=\{\xvec_i,1\leq i\leq n\}$ and their labels/predictions $\Yvec=\{y_i, 1\leq i\leq n\}$. The goal is to minimize expected 
error on unseen test points after training on a small training set $\Xcal$. Empirical-risk minimization (ERM) is a canonical algorithm to provably achieve this goal for a setting given below. %and several results show that ERM leads to provably small expected error on unseen test samples for this setting.

Let each training sample $\xvec_i$ be sampled i.i.d. from a fixed distribution $\Dcal$, i.e. $\xvec_i\sim \Dcal$. Let $\ell(\wvec;\xvec):\mathbb{R}^{r}\times \mathbb{R}^{r}\rightarrow \mathbb{R}$ be the loss function that provides loss incurred by model parameters $\wvec$ for a given point $\xvec$. Then, the goal is to minimize expected loss, i.e.,\vspace*{-3pt}
\begin{equation}
  \label{eq:l_prob}
  \min_{\wvec\in \Wcal} F(\wvec)=\mathbb{E}_{\Dcal}\left[\ell(\wvec;\xvec)\right]. \vspace*{-3pt}
\end{equation}
Let $\wvec^*$ be the optimal solution to \eqref{eq:l_prob}. 

Typically, the distribution $\Dcal$ is unknown. To address this, the \emph{empirical risk minimization} (ERM)approach uses the empirical distribution derived from the training data in lieu of $\Dcal$. Formally, ERM solves for:\vspace*{-3pt}
% To address this, ERM obtains $\wvec$ by minimizing loss function over a finite training sample only i.e., \vspace*{-3pt}
\begin{equation}
  \label{eq:erm}
  \hat{\wvec}=\min_{\wvec\in \Wcal} \hat{F}(\wvec)=\frac{1}{n}\sum_{i=1}^n \ell(\wvec;\xvec_i).\vspace*{-3pt}
\end{equation}
If loss function $\ell$ satisfies certain properties then we can prove bounds relating the quality (objective function value $F(\cdot)$) of $\hat{\wvec}$ and $\wvec^*$. In particular, \cite{Shalev-ShwartzSSS09} provided a bound on $F(\hat{\wvec})$ for the case of strongly-convex loss functions: %, several results exist for ``goodness'' of $\hat{\wvec}$ for the learning problem \eqref{eq:l_prob}. 
%In this work, we only need to consider the case when $\ell$ is a strongly-convex function. Recently, \cite{Shalev-ShwartzSSS09} provided a bound on $F(\hat{\wvec})$ for the case of strongly-convex loss functions. Specifically, 
\begin{theorem}[Stochastic Convex Optimization \cite{Shalev-ShwartzSSS09}]
Let $\ell(\wvec;\xvec)=f(\wvec^T\xvec;\xvec)+h(\wvec)$, where $h:\mathbb{R}^r\rightarrow \mathbb{R}$ is a $\alpha$-strongly convex regularization function. Let $f(u;\xvec)$ be a convex $L_f$-Lipschitz continuous function in $u$ and let $\|\xvec\|_2\leq R$. Let $\wvec^*$ be the optimal solution of Problem~\eqref{eq:l_prob} and $\hat{\wvec}$ be the optimal solution of Problem~\eqref{eq:erm}. Then, for any distribution over $\xvec$ and any $\delta>0$, with probability at least $1-\delta$ over a sample $\Xcal$ of size $n$:\vspace*{-3pt}
\begin{equation}
  \label{eq:sco}
  F(\hat{\wvec})-F(\wvec^*)\leq O\left(\frac{R^2L_f^2\log(1/\delta)}{\alpha n}\right). \vspace*{-3pt}
\end{equation}
\label{thm:sco}
\end{theorem}
Hence, the above theorem shows that solving ERM (i.e., Problem~\eqref{eq:erm}) serves as a good ``proxy'' for solving the original problem \eqref{eq:l_prob} and also the additional expected error can be decreased linearly by increasing the number of training samples, i.e., $n$. 

% Arvind: A lot of wordsmithing to compress this discussion.
\vspace{1ex} \noindent {\bf Haar-wavelets}: Wavelets serve as a
popular tool for compressing a regular signal (a vector in finite
dimensions for our purposes) \cite{Mallat97}. In particular, Haar
wavelets can compress a piece-wise constant vector effectively and can
therefore be used for a parsimonious representation of $k$-bucket
histograms.

Haar wavelet performs a linear orthogonal transformation of a vector
to obtain \emph{wavelet coefficients}.  In particular, given a vector
$\xvec\in \mathbb{R}^r$, we obtain a vector $\alphavec\in
\mathbb{R}^r$ of wavelet coefficient using:
\begin{equation}
  \label{eq:wavelet}
  \alphavec=\Psi \xvec,
\end{equation}
where $\Psi\in \mathbb{R}^{r\times r}$ is the \emph{wavelet transform
  matrix} given by:
\begin{equation}
  \label{eq:wavelet_matrix}
  \Psi=\begin{cases}
    \frac{1}{\sqrt{r}} & \text{ if } i = 1, 0 < j \leq r, \\
    +\sqrt{\frac{2^{\ceil{\log_2 i}}}{r}} &\text{ if }1<i \leq r,\frac{r}{2^{\ceil{\log_2 i}}}(i-2^{\ceil{\log_2 i}})< j \\&\qquad \qquad\leq \frac{r}{2^{\ceil{\log_2 i}}}(i+\frac{1}{2}-2^{\ceil{\log_2 i}}),\\
-\sqrt{\frac{2^{\ceil{\log_2 i}}}{r}} &\text{ if }1<i \leq r,\frac{r}{2^{\ceil{\log_2 i}}}(i+\frac{1}{2}-2^{\ceil{\log_2 i}})< j \\&\qquad \qquad\leq \frac{r}{2^{\ceil{\log_2 i}}}(i+1-2^{\ceil{\log_2 i}}),\\
0&\text{ otherwise. }
  \end{cases}
\end{equation}
We can show that if $\xvec$ is $k$-piecewise constant in
Equation~\ref{eq:wavelet}, $\alphavec$ has at most $k\log r$ non-zero
coefficients.

For signals in higher dimensions, wavelet transform can be obtained by
first \emph{vectorizing the signal} and then applying the
transformation of Equation~\ref{eq:wavelet} (see \cite{Mallat97} for more details). Several existing studies show that high-dimensional real-life data mostly resides in a small number of clusters and hence most of the wavelet coefficients are nearly zero \cite{ChakrabartiGRS01}, i.e., vector of coefficients is sparse. 
%Arvind: this statement does not seem very interesting since
%$k$-piecewise constant hyper rectangles are less general than STHoles
%& ISOMER.
%For a $d$-dimensional
%signal that is represented by $k$-piece-wise constant hyper-rectangles
%the wavelet coefficients can have at most $kr^{d-1}log(r)$ non-zero
%coefficients.  
% for a detailed specification of
%high-dimensional Haar wavelet transform.
%%% Local Variables: 
%%% mode: latex
%%% TeX-master: "histestimation-main"
%%% End: 

\section{Method}
\label{sec:method}
In this section, we present our learning theoretic framework and algorithms. We first review the architectural
context for self-tuning histogram learning.

\vspace{1ex}\noindent {\bf Architecture:} We assume the architectural
context of prior work in self-tuning
histograms~\cite{SrivastavaHMKT06}. In particular, we assume operators
in query plans are instrumented to maintain counters and produce query
feedback records (QFRs) at the end of query execution. Recall that a
QFR is a filter sub-expression and its cardinality. (In the following,
we abuse notation and refer to such sub-expressions as \emph{queries}
although they are actually parts of a query.)  These QFRs are
available as input to our learning system either continuously or in a
batched fashion (e.g., during periods of low system load). Although we
present our learning framework assuming QFRs are the only input, we
can extend our framework to incorporate workload-independent data
characteristics, e.g., an initial histogram constructed from
data. %We begin by assuming that input QFRs and
%database are fixed (unchanging); we discuss dynamically generated QFRs and database updates in
%Section~\ref{sec:online}.

First, in Section~\ref{sec:probform}, we present our learning theoretic framework for self-tuning histograms. 
%Arvind These two issues are common to all histograms, may not be worth emphasizing
%Our formulation clearly highlights the two main problems in estimating histogram: a) fixing bucket
%boundaries, b) fixing bucket heights. 
In Section~\ref{sec:equi} we study equi-width histograms in our framework and present a learning algorithm
for the same (EquiHist). We also present formal error analysis for histograms learned by EquiHist. Equi-width histograms are
known to be unsuitable in many settings, such as sparse high-dimensional datasets. To handle this, in Section~\ref{sec:sparse}, we present an algorithm (SpHist) for learning general histograms that relies on a reduction from histogram learning to sparse vector recovery. For
presentational simplicity, Sections~\ref{sec:probform}-\ref{sec:sparse} assume static QFRs, no database
updates, and 1-dimensional histograms. We extend our algorithms to multidimensional histograms in
Section~\ref{sec:multi} and dynamic data and QFRs in Section~\ref{sec:online}.

% Arvind: may not be worth discussing details such as OMP in this roadmap paragraph.
%and provide a simple algorithm to
%solve the problem. We analyze the approach of equi-width histograms in our framework and bound expected error
%incurred on unseen queries. Naturally, the equi-width approach suffers in high-dimensions as high-dimensional
%data typically lies small and dense pockets. To handle this problem, in Subsection~\ref{sec:sparse} we provide
%a novel reduction of the histogram estimation problem to a sparse-vector recovery problem. We adapt the
%popular Orthogonal Matching Pursuit (OMP) algorithm for our problem and provide a simple greedy
%algorithm. Finally, we extend our methods to higher dimensions (Subsection~\ref{sec:multi}), streaming queries
%scenario as well as online database update scenario (Subsection~\ref{sec:online}).

\subsection{Problem Formulation}
\label{sec:probform}
%Consider an integer-valued attribute (in a database) whose range is of size $r$. For simplicity, assume that
%the attribute takes values in the interval $[1,r]$. Now, given a query workload, the goal is to estimate a
%histogram $\hvec\in\mathbb{R}^r$ ($i^\mathrm{th}$ component $h_i$ denotes the number of records where the
%attribute takes value $i$) with at most $k$ buckets that would incur small error on unseen test
%queries. Recall that, in this paper we focus on self-tuning histograms only, hence we assume that the
%underlying database cannot be accessed, instead only query workload can be accessed. 

We now formalize the histogram estimation problem. Our formulation is
based on standard learning assumptions where we assume a training
query workload of QFRs which is sampled from a fixed distribution and the goal
is to estimate a $k$-bucket histogram that incurs small expected error for
unseen queries from the same distribution. We consider histograms over column $A$
of relation $R$. Recall from Section~\ref{sec:prelim} that domain of
$A$ is $[1,r]$ and a $k$-bucket histogram $\hvec \in \mathbb{R}^r$ is
a $k$-piecewise constant vector.

Let $\Dcal$ be a fixed (unknown) distribution of range queries over
$R(A)$. Let $\Qcal=\{(\qvec_1, s_{\qvec_1}), \dots,
(\qvec_N,s_{\qvec_N})\}$ be a query workload used for training where
each $\qvec_i\sim \Dcal, \forall 1\leq i\leq N$ and $s_\qvec$ is the
cardinality of query $\qvec$ when evaluated over $R$.

Let $f(\hat{s}_\qvec;s_\qvec):\mathbb{R}\times \mathbb{R}\rightarrow \mathbb{R}$ be a
\emph{loss function} that measures the error between the estimated
cardinality, $\hat{s}_\qvec$, and actual cardinality, $s_\qvec$, of
query $\qvec$. Since the estimated cardinality of $\qvec$ using
histogram $\hvec$ is $\hat{s}_\qvec=\qvec^T\hvec$, the error incurred
by $\hvec$ on $\qvec$ is $f(\qvec^T\hvec;s_\qvec)$.  Example loss
functions include $L_1$ loss ($f(\qvec^T\hvec;s_\qvec) =
|\qvec^T\hvec-s_\qvec|$) and $L_2$ loss ($f(\qvec^T\hvec;s_\qvec) =
(\qvec^T\hvec-s_\qvec)^2$).

Our goal is to learn a $k$-bucket histogram $\hvec\in \mathbb{R}^r$
that minimizes the expected error $F(\hvec) = \mathbb{E}_{\qvec\sim
  \Dcal}[f(\qvec^T\hvec;s_{\qvec})]$ incurred by $\hvec$ on test
queries sampled from $\Dcal$. 
Formally, we define our histogram estimation problem as:
\begin{align}
  \label{eq:prob_main}
  \min_{\hvec}\ \ F_\Dcal(\hvec),\quad 
  \text{s.t.}\ \  \hvec \in \Ccal,
\end{align}
and let $\hvec^*$ be the optimal solution to the above problem, i.e., 
\begin{equation}
\hvec^*=\argmin_{\hvec\in \Ccal} F_\Dcal(\hvec),
\label{eq:h*}
\end{equation}
where $\Ccal$ represents the following set of histograms:
\begin{align}
\Ccal = \{\hvec\::\: & \hvec\in\mathbb{R}^r \text{ is a histogram over
  range } [1,r] \text{with} \nonumber\\
\                & \text{at most } k \text{ buckets and minimum bucket-width } \Delta\}
\label{eq:c_1}
\end{align}
Note that $\Ccal$ only contains histograms whose each bucket is of width at least $\Delta$. Parameter $\Delta$ can be arbitrary; we introduce it for the purpose of analysis only. While our analysis do not make any assumption on $\Delta$, naturally, bounds would be better if $\Delta$ of the optimal histogram is large, i.e., the optimal histogram is relatively flat. %In practice, if some attribute values are significantly more frequent than others, then those attribute values can be treated separately and the true distribution over the remaining attributes should have a histogram with large $\Delta$, leading to better bounds for our method. 
We next study equi-width histograms in our framework and provide an
efficient algorithm for the same.
% Arvind: we seem to have stated this point several times already; One
% option is to uncomment the following and remove the same point made
% in the beginning of Sec 3

%However, for small number of buckets or high-dimensions, the error incurred by Equi-width histograms might be large. We then introduce a novel method based on sparse-vector recovery techniques (see Subsection~\ref{sec:sparse}), that can efficiently learn histograms in high-dimensional as well as small bucket-number ($k$) setting. %However, currently we do not have theoretical guarantees for this method and leave its analysis as topic for future research. 
% minimizing the expected error directly is not possible. So, in the next section we describe a method that first tries to minimize expected error over a finite sample and then provide guarantee for our original problem. 
\subsection{Equi-width Approach}
\label{sec:equi}
In this section, we study equi-width histograms for solving Problem~\eqref{eq:prob_main} and also provide approximation guarantees for the obtained method. 

Observe that set $\Ccal$ (Equation \ref{eq:c_1}) is a non-convex set, hence we cannot apply standard convex optimization techniques to obtain the optimal solution to (\ref{eq:prob_main}). To handle this, we relax the problem by fixing the bucket boundaries to be equi-spaced. That is, we first consider the class of histograms with $b$ equal-width buckets:
\begin{align}
\Ccal' = \{\hvec\::\: & \hvec\in\mathbb{R}^r \text{ is a histogram over integer range } [1,r]\nonumber\\
\                & \text{with } b \text{ equal-width buckets}\}.
\label{eq:c1_1}
\end{align}
For ease of exposition, we assume $r$ is divisible by $b$. 
%In this section, we describe an approach that can provably estimate a histogram that satisfies the three criteria specified above. 
%Now, recall that  $\Ccal=\{\hvec: \hvec \text{ has at most k bins and minimum}$  $\ \text{ bin size is } \Delta\}$. As mentioned above, it is not clear how to minimize the expected error directly. Instead, we use standard learning approach where we minimize empirical estimate of the expected error and using standard result from learning theory, we then show that the obtained solution by minimizing empirical error is a good solution for the original problem as well. 
%Formally, let $$\hat{F}(\hvec)=\frac{1}{N}\sum_{i=1}^N |s_{\qvec_i}-\qvec_i^T\hvec|,$$ where each $\qvec_i \sim \Dcal$. Hence, now the goal is to minimize $\hat{F}(\hvec)$ over $\Ccal$. Unfortunately, $\Ccal$ is a non-convex set and hence we cannot use standard optimization tools to guarantee optimality. Instead, we form a new set $\Ccal'$ which contains $b$ equally binned histograms, i.e, $\Ccal'=\{\hvec: \hvec \text{ has b equi-spaced bins}\}$. Now, note that since bin boundaries are fixed in $\Ccal'$, hence only bin heights are the variables. 
Note that, for any $\hvec\in \Ccal'$, we can find $\wvec\in\mathbb{R}^b$ such that $\hvec = B \wvec$, where $B\in \mathbb{R}^{r\times b}$ and 
\begin{equation}
  \label{eq:B}
  B_{ij} = \begin{cases} 1& \text{ if } \frac{r}{b}\cdot (j-1)< i \leq \frac{r}{b}\cdot j, \\0 & \text{otherwise}. \end{cases}
\end{equation}
For illustration (with $\frac{r}{b}=2$), 
\begin{equation}
  \label{eq:B_eg}
  \hvec=\left[\begin{array}{c}w_1\\w_1\\w_2\\w_2\\\vdots\end{array}\right]=\left[\begin{array}{ccc}1&0&\cdots\\1&0&\cdots\\0&1&\cdots\\0&1&\cdots\\\vdots&\vdots&\vdots\end{array}\right]\cdot \left[\begin{array}{c}w_1\\w_2\\\vdots\end{array}\right]. 
\end{equation}
Therefore, searching for a histogram $\hvec$ over $\Ccal'$ is equivalent to searching for a $\wvec\in\mathbb{R}^b$. Furthermore, optimal $\hvec^*$ to \eqref{eq:prob_main} should satisfy: $\|\hvec^*\|_1=\sum_i h_i=M$, i.e., total number of database records. Also, $\hvec^*$ should have minimum bucket width $\Delta$, hence $\|\hvec^*\|_\infty\leq \frac{M}{\Delta}$. Using these observations, we can constraint $\wvec$ to belong to a convex set $\Kcal$:
\begin{equation}
\Kcal=\{\wvec\in \mathbb{R}^b|\ \|\wvec\|_1=\frac{Mb}{r} \text{ and } \|\wvec\|_\infty \leq \frac{M}{\Delta}\}.
\label{eq:kcal}
\end{equation}
%Let $\Kcal \subseteq \mathbb{R}^b$ be a convex set, to which $\wvec$ must belong. 
%Since we can find a compact convex set $\Kcal\subset\mathbb{R}^b$ to which $\wvec$ must belong,
%we can now rewrite $\Ccal'$ as follows:
%\begin{align}
%\Ccal' = \{\hvec\::\: & \hvec\in\mathbb{R}^r \text{ is a histogram over  range } [1,r]\nonumber\\
%\                & \text{such that } \hvec=B\wvec, \ \wvec\in\Kcal, \ \Kcal\subset \mathbb{R}^b \nonumber\\
%\                & \text{where } \Kcal \text{ is compact convex set} \}.
%\end{align}
%\begin{equation}
%\label{eq:c}
%\Ccal'=\{\hvec\::\:\hvec=B\wvec, \wvec\in \Kcal\}.
%\end{equation}
Thus, $\Ccal'$ can be redefined as:
\begin{equation}
\label{eq:c2_1}
\Ccal'=\{\hvec\::\:\hvec=B\wvec, \wvec\in \Kcal\},
\end{equation}
and searching for a histogram $\hvec$ over $\Ccal'$ induces a corresponding search for $\wvec$ over $\Kcal$, which is a {\em convex} set.

Using the above observations, we obtain the following relaxed problem:
\begin{equation}
  \label{eq:prob1d_rel}
  \min_{\wvec\in \Kcal} F_\Dcal(B\wvec). 
\end{equation}
To avoid overfitting, we add entropy regularization to the above objective function. This leads to the following relaxed problem:
%We choose the $L_1$-loss function $f(\qvec^T B\wvec;s_\qvec)=|s_\qvec-\qvec^T B\wvec|$ and add entropy regularization to avoid overfitting, which gives us the following regularized objective:
\begin{equation}
\min_{\wvec\in \Kcal} G_\Dcal(\wvec) = E_\Dcal\left[f(\qvec^T B\wvec; s_\qvec)\right] - \lambda H(\frac{r}{Mb}\wvec),
\label{eq:g}
\end{equation}
%Also, $$\argmin_{\hvec\in \Ccal'} \hat{F}(h)= \argmin_{\wvec\in \Kcal} \frac{1}{N}\sum_{i=1}^N|s_{\qvec_i}-\qvec_i^TB\wvec|.$$
%By abuse of notation, $$\hat{F}(\wvec)=\frac{1}{N}\sum_{i=1}^N|s_{\qvec_i}-\qvec_i^TB\wvec|.$$
%As, the above problem is a convex program we can solve it using standard convex programming tools. However, if the number of queries is small then the above problem might be unstable and lead to over-fitting. Hence, we add an entropy regularization to the above objective function, i.e., 
%\begin{equation}
%  \label{eq:g}
%  \hat{G}(\wvec)=\frac{1}{N}\sum_{i=1}^N|s_{\qvec_i}-\qvec_i^TB\wvec|-\lambda H(\frac{R}{Mb}\wvec),%\sum_{j=1}^{b} \frac{w_iR}{Tb} \log \frac{w_iR}{Tb}, 
%\end{equation}
where $M$ is the size of the relation $R$, i.e., $M=|R|$, $\lambda>0$ is a constant specified later and $H(\frac{r}{Mb}\wvec)=-\sum_{j=1}^b \frac{w_ir}{Mb}$ $\times\log \frac{w_ir}{Mb}$ is the entropy of $\frac{r}{Mb}\wvec$. Note that we normalized each $w_i$ by multiplying by $r/(Mb)$ to make it a probability distribution.

The distribution $\Dcal$ is unknown except through the example queries in $\Qcal$, and so we cannot directly solve \eqref{eq:g}. As mentioned in Section~\ref{sec:prelim}, we instead optimize an empirical estimate ($\hat{G}(\cdot)$) of the objective $G_\Dcal(\cdot)$, which finally leads us to the following  problem: %We consider the empirical version $\hat{G}(\cdot)$ of the objective $G_\Dcal(\cdot)$ which finally leads us to the following optimization problem:
\begin{equation}
  \label{eq:g_new}
  \min_{\wvec\in \Kcal} \hat{G}(\wvec)=\frac{1}{N}\sum_{i=1}^Nf(\qvec_i^TB\wvec; s_{\qvec_i})-\lambda H(\frac{r}{Mb}\wvec),
\end{equation}
where $\Kcal$ is given by \eqref{eq:kcal}. Let $\hat{\wvec}$ be the optimal solution to \eqref{eq:g_new}, i.e., $$\hat{\wvec}=\argmin_{\wvec\in \Kcal} \hat{G}(\wvec),\quad \hh = B\hw.$$ 
Now note that the above relaxed problem is a convex program and can be solved optimally and efficiently using standard convex optimization methods. However, the obtained solution need not be optimal for our original problem \eqref{eq:prob_main}.

Interestingly, in the following theorem, we show that optimal equi-width histogram $\hh\in\Ccal'$ is a provably approximate solution to the original problem \eqref{eq:prob_main}. In particular, the theorem shows that by training with a finite number of queries ($N$) and selecting number of buckets $b$ to be a multiplicative factor larger than the required number of buckets $k$, the objective function (in Problem~\ref{eq:prob_main}) at $\hh$ is at most $\epsilon$ larger than the optimal value. 
\begin{theorem}
\label{thm:1d}
Let $f:\mathbb{R}\times \mathbb{R}\rightarrow \mathbb{R}$ be a convex $L_f$-Lipschitz continuous loss function. Let $\hat{\wvec}$ be the optimal solution to \eqref{eq:g_new}, $\hat{\hvec}=B\hat{\wvec}$ and let each query $\qvec_i\sim \Dcal$.  Let $\hvec^*$ be the optimal solution to \eqref{eq:prob_main} and has minimum bucket width $\Delta$. %Let $Kcal\subseteq\mathbb{R}^b$ be as defined in the intersection of the $L_1$ ball of radius $\frac{Mb}{r}$ and $L_\infty$ ball of radius $\frac{M}{\Delta}$. 
Let $|Q|=\max_{\qvec\sim \Dcal}\|\qvec\|_1$, i.e., the largest range of any query and let $C_1, C_2>0$ be universal constants.
Then, if the number of training queries ($N$) satisfies: $$N\geq C_1\left(\frac{|Q|}{\Delta}\right)^3 \frac{L_f^4\log \frac{1}{\delta}}{\epsilon^4},$$ and if the number of buckets ($b$) in $\hh$ satisfies: $b\geq C_2k\frac{|Q|L_f^2}{\Delta\epsilon^2},$
we have
$$F_\Dcal(\hat{\hvec})=\mathbb{E}_{\Dcal}[f(\qvec^T\hh; s_\qvec)]\leq F(\hvec^*)+M\epsilon.$$
\end{theorem}
See Appendix~\ref{sec:proof1d} of our full-version \cite{ViswanathanJLA11} for a detailed proof.

Note that the above theorem shows that the histogram $\hh$ that we learn satisfy both the required properties:
%\begin{compactitem}

\noindent {\bf \textbullet \ } Number of buckets ($b$) in $\hh$ is given by $b=C_2k\frac{|Q|L_f^2}{\Delta\epsilon^2}$. Hence, number of buckets in $\hh$ are larger than $k$ by a small approximation factor. In fact if queries are generally ``short'', i.e., $|Q|$ is smaller than $\Delta$, then the approximation factor is a constant dependent only on the accuracy parameter $\epsilon$. 

\noindent {\bf \textbullet \ } Relative Expected Error incurred by $\hh$ is only $\epsilon$, while sampling $N=C_1\left(\frac{|Q|}{\Delta}\right)^3 \frac{L_f^4\log \frac{1}{\delta}}{\epsilon^4}$ queries. Note that our bound on $N$ is {\em independent} of $r$, hence the number of queries is not dependent on the range of the space, but only on the ``complexity'' of histograms and queries considered, i.e., on $\Delta$ and $|Q|$. Our bound confirms the intuition that if $\Delta$ is smaller, that is, the optimal histograms have more buckets and is more ``spiky'', then the number of queries needed is also very large. However, if $\Delta$ is a constant factor of range $r$, then the number of queries required is a {\em constant}. %\end{compactitem}

Now, the above bounds depend critically upon the loss function $f$ through its Lipschitz constant $L_f$. In the following corollary, we provide bounds for loss function $f(\qvec^T\hvec;s_\qvec)=|\qvec^T\hvec-s_\qvec|$. 
\begin{corollary}
  Let $f(\qvec^T\hvec;s_\qvec)=|\qvec^T\hvec-s_\qvec|$, then under the assumptions of Theorem~\ref{thm:1d} and by select $N,b$ to be:
$$N\geq C_1\left(\frac{|Q|}{\Delta}\right)^3 \frac{\log \frac{1}{\delta}}{\epsilon^4},\ \ , b\geq C_2k\frac{|Q|}{\Delta\epsilon^2}.$$
Then, 
$$F_\Dcal(\hat{\hvec})=\mathbb{E}_{\Dcal}[|\qvec^T\hh-s_\qvec|]\leq F(\hvec^*)+M\epsilon.$$
\end{corollary}
Note that $f(\qvec^T\hvec;s_\qvec)=|\qvec^T\hvec-s_\qvec|$ is $1$-Lipschitz convex function. The above corollary now follows directly from Theorem~\ref{thm:1d}. \\[2pt]
{\bf EquiHist Method}: While selecting the loss function to be $L_1$ loss ($f(\qvec^T\hvec;s_\qvec)=|\qvec^T\hvec-s_\qvec|$) provides tight bounds, in practice optimization with $L_1$ loss is expensive as it is not a smooth differentiable function. Instead, for our implementation, we use $L_2$ loss and select regularization parameter $\lambda=0$. Hence, the empirical risk minimization problem that our Equi-width Histogram method (EquiHist) solves is given by:
\begin{equation}
  \label{eq:eh_1d}
  \hat{G}(\wvec)=\min_{\wvec\in\mathbb{R}^b}\frac{1}{N}\sum_{i=1}^N (\qvec_i^TB\wvec-s_{\qvec_i})^2. 
\end{equation}
Using techniques similar to Theorem~\ref{thm:1d}, we can easily obtain approximation guarantees for the optima of the above problem. Also, the above optimization problem is the well-known Least-squares problem and its solution can be obtained in closed form. Algorithm~\ref{alg:eh_1d} provides a pseudo-code of our method (EquiHist). Here, $\Qcal$ is a matrix whose each row contains the training query $\qvec_i$ and $s$ is the column vector containing corresponding query cardinalities. 
\begin{algorithm}[tb]
  \caption{EquiHist: Equi-width histogram based method for Histogram Estimation (1-dimensional case)}
  \begin{algorithmic}[1]
    \STATE {\bfseries Input}: Training Queries: $\Qcal\in \mathbb{R}^{r\times N}$ where $i$-th column $\Qcal_i\in \mathbb{R}^r$ is the $i$-th query $\qvec_i$. $\svec=[s_{\qvec_1};s_{\qvec_2};\dots;s_{\qvec_N}]\in \mathbb{R}^N$ is the column vector of training query cardinalities
  \STATE {\bfseries Parameters}: $k$: number of histogram buckets
  \STATE $B\in \mathbb{R}^{r\times k}$ is as given \eqref{eq:B} (with $b=k$).
  \STATE $\wvec\leftarrow (B^T\Qcal\Qcal^TB)^{-1}B^T\Qcal \svec$ (solution to \eqref{eq:eh_1d})
  \STATE $\hvec=B\wvec$
  \STATE {\bfseries Output}: $\hvec$
  \end{algorithmic}
\label{alg:eh_1d}
\end{algorithm}
\subsection{Sparse-vector Recovery based Approach}
\label{sec:sparse}
In the previous subsection, we provided an approximation algorithm for Problem~\ref{eq:prob_main}, by fixing bucket boundaries to be equi-width. 
%Recall that our goal is to minimize $F(\hvec)$ such that $\hvec$ incurs low expected error in cardinality estimation and has small number of bins. To achieve that goal, in previous subsection we minimize $\hat{F}(\hvec)$. However, as the set of histograms ($\Ccal$) with unknown bin boundaries and bin heights is a non-convex set, we cannot directly use standard convex programming methods to achieve this goal. In the previous subsection, we handled this problem by fixing bin boundaries using equi-spaced bins. 
 However, when the number of buckets required is extremely small then selecting large equi-width buckets might incur heavy error in practice. Furthermore, in high-dimensions the histograms can be very ``spiky'', hence minimum bucket width $\Delta$ might be small, leading to poor accuracies both theoretically as well in practice. 

To alleviate the above mentioned problem, we formulate a sparse-vector recovery based method that is able to use recently developed methods from sparse vector recovery domain. For this purpose, we use the $L_2$ loss for our objective function:
\begin{equation}
\label{eq:F_2}
  \hat{F}(\hvec)=\frac{1}{N}\sum_{i=1}^N(s_{\qvec_i}-\qvec_i^T\hvec)^2. 
\end{equation}
Now, we use wavelet basis to transform $\hvec$ into its wavelet coefficients. Let $\Psi$ be the Haar wavelet basis, and $\alphavec = \Psi \hvec$ be the wavelet transform of $\hvec$. Since $\Psi$ is orthonormal, we can rewrite cardinality estimation using $\hvec$ as:
%Let h be any histogram and recall that \psi is the haar wavelet transform matrix. Let \alpha = \psi h be the wavelet transform of h. Since \psi is orthonormal, we can rewrite cardinality estimation using h as:as $\Psi$ is a orthnormal basis:
\begin{equation}
\qvec^T\hvec=\qvec^T\Psi^T\Psi\hvec=\qvec^T\Psi^T\alphavec,
\label{eq:qh}
\end{equation}
where $\alphavec$ is the vector of Haar wavelet coefficients of
$\hvec$.  Furthermore, using standard results in wavelet transforms
\cite{Mallat97}, if $\hvec$ is $k$-piecewise constant then the wavelet
transform has at most $k\log r$ non-zero coefficients. As $k$ is
significantly smaller than $R$, hence wavelet transform of $\hvec$
should be sparse and we can use sparse-vector recovery techniques from
compressed sensing community to recover these wavelet coefficients.

%Unfortunately, random range queries do not satisfy necessary conditions for sparse-vector recovery and hence formal guarantees for this approach do not follow directly from existing proof techniques. We leave proof of our approach as future work. %however for a restricted case of uniformly random point queries, we show that our approach leads to provably optimal histogram estimation (see Subsubsection~\ref{sec:category}).  

We now describe our sparse-vector recovery based approach to estimate histograms. Below, we formally specify our sparse-wavelet coefficient recovery problem:
\begin{equation}
  \label{eq:sp1}
  \alphavec^*=\argmin_{\text{supp}(\alphavec)\leq k}\frac{1}{N}\sum_{i=1}^N(s_{\qvec_i}-\qvec_i^T\Psi^T\alphavec)^2,
\end{equation}
where $\text{supp}(\alphavec)$ is the number of non-zeros in $\alphavec$.%, $\Psi_{\beta}$ is the haar wavelet basis corresponding to top-$\beta$ basis elements; haar wavelet transform has a binary tree structure, and the coefficients that are near root are denoted as ``top'' coefficients. 

Note that the above problem is in general NP-hard. However, several recent work in the area of compressed sensing \cite{CandesT05, Candes08} show that under certain settings $\alphavec^*$ can be obtained up to an approximation factor. Unfortunately, random range queries do not satisfy necessary conditions for sparse-vector recovery and hence formal guarantees for this approach do not follow directly from existing proof techniques. We leave proof of our approach as future work. %however for a restricted case of uniformly random point queries, we show that our approach leads to provably optimal histogram estimation (see Subsubsection~\ref{sec:category}).  

Instead, we use sparse-recovery algorithms as heuristics for our problem. In particular, we use one of the most popular sparse-recovery algorithm, Orthogonal Matching Pursuit (OMP) \cite{TroppG07}. OMP is a greedy technique that starts with an empty set of coefficients (i.e. $\text{supp}(\alphavec)=0$). Now, at each step OMP adds a coefficient to the support set which leads to largest decrease in the objective function value. After greedily selecting $k$ coefficients, we  obtain $\alphavec$ and its support set with at most $k$ coefficients. 

Let, $\hat{\alphavec}$ be computed using OMP method, then we obtain our estimated histogram $\hat{\hvec}$ using:
$$\hat{\hvec}=\Psi^T \hat{\alphavec}.$$
Note that, if $\hat{\alphavec}$ has $k$ non-zeros then $\hat{\hvec}$ will have at most $3k$ non-zeros \cite{MatiasVW98}, hence our estimated histogram has small number of buckets. To further decrease the number of buckets to $k$, we use the dynamic programming based method by \cite{JagadishKMPSS98} that produces small number of buckets if heights (or probability density value) for each attribute value is provided. %Empirically, we observe that wavelets tend to overfit for some of the ``lower'' coefficients, and hence might give ``spiky'' histograms. To avoid this problem, we use $L_1$ loss based dynamic programming method of \cite{JagadishKMPSS98}. 
Also, the method of \cite{JagadishKMPSS98} runs in time quadratic in the number of attribute values, i.e., range $r$. However, since our frequency distribution (histogram $\hat{\hvec}$) has only $3k$ buckets, we can modify the Dynamic Programming based algorithm of  \cite{JagadishKMPSS98} so that it obtains the optimal solution in time $O(k^2)$. Algorithm~\ref{alg:sp} provides a pseudo-code of our algorithm. $\Qcal$ denotes training queries matrix and $A_\Scal\in \mathbb{R}^{N\times |\Scal|} $ represents a sub-matrix of $A$ formed by $A$'s columns indexed by $\Scal$. 

%Furthermore, since $\alphavec^*$ is the vector of wavelet coefficients, hence there is further more structure to $\alphavec^*$. Recently, \cite{BaraniukCDH10} showed that using slight modifications to sparse-recovery techniques, this structure can be further exploited. In our work, we use an algorithm similar to the one of \cite{BaraniukCDH10} but with simple modifications to make the method stable for our problem. We provide detailed algorithm in Algorithm~\ref{alg:sp}. Our algorithm is inspired by \cite{GargK09} and is just a projected gradient descent method where projection onto set of sparse wavelet coefficients is done using a technique developed by \cite{BaraniukCDH10}. 

\begin{algorithm}[tb]
  \caption{SpHist: Sparse-recovery based Histogram Estimation (1-dimensional case)}
  \begin{algorithmic}[1]
  \STATE {\bfseries Input}: Training Queries: $\Qcal\in \mathbb{R}^{r\times N}$ where $i$-th column $\Qcal_i\in \mathbb{R}^r$ is the $i$-th query $\qvec_i$. $\svec=[s_{q_1};s_{q_2};\dots;s_{q_N}]\in \mathbb{R}^N$ is the column vector of training query cardinalities
  \STATE {\bfseries Parameters}: $k$: number of histogram buckets%, $k'$: number of wavelet coefficients estimated, $\beta$: number of wavelet coefficients to be considered
  \STATE Set support set $\Scal=\phi$, residual $\zvec_0=\svec$. 
  \STATE Set $A=\Qcal^T\Psi^T$ (note that $A\in \mathbb{R}^{N\times r}$)
  \STATE $t=1$
  \REPEAT
  \STATE Find index $I_t=\argmax_{j=1,...,r}\zvec_{t-1}^TA_j$
  \STATE $\Scal=\Scal\cup \{I_t\}$
  \STATE $\alphavec^t=\bm{0}$ ($\alphavec^t\in \mathbb{R}^r$)
  \STATE Least Squares Solution: $\alphavec^t_\Scal=\argmin_{\alphavec_\Scal\in \mathbb{R}^{|\Scal|}}\|A_\Scal\alphavec_\Scal-\svec\|_2$. $//$ $A_\Scal$ is the column submatrix of $A$ whose columns are listed in $\Scal$. $\alphavec^t_\Scal$ is the sub-vector of $\alphavec^t$ with components listed in set $\Scal$.
  \STATE Update residual: $\zvec_t=\svec-A\alphavec^t$.
  \STATE $t=t+1$
  \UNTIL($t\leq k$)
  \STATE $\hat{\alphavec}=\alphavec^{k}$
  \STATE Form $\hat{\hvec}=\Psi^T\hat{\alphavec}$
  \STATE Apply modified version of DP Method of \cite{JagadishKMPSS98} to $\hat{\hvec}$ to obtain $\hvec$ with $k$ buckets
  \STATE {\bfseries Output}: Histogram $\hvec$ with $k$ buckets
  \end{algorithmic}
  \label{alg:sp}
\end{algorithm}
\subsection{Multi-dimensional Histograms}
\label{sec:multi}
In the previous two subsections, we discussed our two approaches for $1$-dimensional case. In this section, we extend both the approaches for multi-dimensional case as well. In next subsection we discuss EquiHist generalization to multiple dimensions and in Subsection~\ref{sec:sp_multi}, we generalize our sparse recovery based approach. 
%Below, we In Subsection~\ref{sec:sp_multi}, we generalize our sparse recovery based approach to multiple dimensions. 
\subsubsection{Equi-width Approach}
\label{sec:eq_multi}
We first provide an extension of the equi-width approach to the 2-d case and then briefly discuss extensions to general multi-dimensional case. 

Recall that, given a set of range queries $\Qcal=\{Q_1, \dots, Q_N\}$ and their cardinality $\svec=\{s_{Q_1}, s_{Q_2}, \dots, s_{Q_n}\}$ where $Q_i\in \mathbb{R}^{r\times r}$ and $Q_i\sim \Dcal$, the goal is to learn histogram $H\in \mathbb{R}^{r\times r}$ such that $H$ has at most $k$ buckets. For 2-D case, we consider a bucket to be a rectangle only. Note that $H$ can have arbitrary rectangular buckets, hence the class of $H$ considered is more general than STGrid. But, our class of $H$ is restricted compared to STHoles, which has an extra ``universal'' bucket.  Now, as for the $1$-d case, the goal is to minimize expected error in cardinality estimation, i.e., 
\begin{equation}
\min_{H\in \Ccal} E_{\Dcal}\left[f(\langle Q,H \rangle; s_Q)\right],
\label{eq:prob_2d}
\end{equation}
where $\langle Q, H \rangle=Tr(Q^TH)$ denotes the inner product between $Q$ and $H$, and $\Ccal$ is given by:
\begin{multline}
  \label{eq:C_2}
  \Ccal=\{H\in \mathbb{R}^{r\times r}: \text{ H has } k \text{ rectangular buckets } \\\text{ and minimum bucket size } \Delta\times \Delta\}
\end{multline}
%\begin{figure}[ht]
%  \centering
%  \includegraphics[width=\columnwidth]{2d_bins}
%  \caption{Division of Bins in 2-d case. There are $b_1\times b_1=b$ rectangular buckets for 2 attributes $[1, r]\times [1,r]$}
%  \label{fig:2d_bins}
%\end{figure}
Similar to $1$-d case, we restrict histograms to set $\Ccal'$ that consists of $b=b_1\times b_1$ equi-width buckets. Now it is easy to verify that for any $H\in \Ccal'$, we can find a matrix $W\in \mathbb{R}^{b_1\times b_1}$ s.t., 
\begin{equation}
  \label{eq:2d_wh}
  H=BWB^T,
\end{equation}
where $B\in \mathbb{R}^{r\times b_1}$ is as defined in \eqref{eq:B}. 

Hence, $\Ccal'$ can be defined as:
\begin{align}
\Ccal' = \{H: H=BWB^T, W\in \Kcal\},
\end{align}
where $\Kcal$ is a convex set defined analogously to \eqref{eq:kcal}. 
%Now the class of histograms $H$ with $b$ equally size buckets is given by the following set $\Ccal'$:
%Now, as in 1-d case, we divide the entire range into $b$ bins of equal size (see Figure~\ref{fig:2d_bins}), i.e., we restrict the class of histograms that we consider to the set $\Ccal'$: 
%\begin{align}
%\Ccal' = \{H\::\: & H\in\mathbb{R}^{r\times r} \text{ is a histogram over integer range } [1,r]\times [1,r]\nonumber\\
%\                & \text{with } b_1\times b_1=b \text{ equally-spaced buckets}\}.\nonumber
%\end{align}
%Now, it is easy to 

Selecting entropy regularization and using empirical estimate for optimization, we reduce \eqref{eq:prob_2d} to the following problem:
%Again selecting entropy regularization, our objective function becomes:
%\begin{equation}
%\label{eq:g_2d}
%G_\Dcal(W)=E_\Dcal\left[f(\langle Q,BWB^T\rangle;s_Q)\right]-\lambda H(\frac{r\cdot r}{Mb}W),  
%\end{equation}
%where $M=|R|$, i.e. relation cardinality and $\lambda>0$ is a constant. Similarly, we use empirical estimate $G_\Dcal$ for optimization. That is, 
\begin{equation}
  \label{eq:ge_2d}
  \min_{W\in \Kcal}\hat{G}(W)=\frac{1}{N}\sum_{i=1}^N f(\langle Q,BWB^T\rangle;s_Q)-\lambda H\left(\frac{r\cdot r}{Mb}W\right),
\end{equation}
where $M=|R|$, i.e. relation cardinality and $\lambda>0$ is a constant. 

Let $\hat{W}$ be the optimal solution to the above problem and let $\hat{H}=B\hat{W}B^T$. Now, similar to 1-D case, we bound the expected error incurred by $\hat{H}$ when compared to the optimal histogram $H^*\in \Ccal$ to Problem~\ref{eq:prob_2d}.% with $k$ bins and each bin of size at least $\Delta\times \Delta$. 
\begin{theorem}
Let $f:\mathbb{R}\times \mathbb{R}\rightarrow \mathbb{R}$ be a convex $L_f$-Lipschitz continuous loss function. Let $\hat{W}=\argmin_{W\in \Kcal} \hat{G}(W)$ and $\hat{H}=B\hat{W}B^T$ and each $Q_i\sim \Dcal$. Let $\Kcal\subset\mathbb{R}^{b_1\times b_1}$ be a convex set; if each $W\in \Kcal$ is treated as a $b_1\times b_1=b$-dimensional vector, then $\Kcal$ is selected to be the intersection of the $L_1$ ball of radius $\frac{Mb}{r}$ and $L_\infty$ ball of radius $\frac{M}{\Delta}$. If we are given that $$N\geq \left(\frac{|Q|}{\Delta}\right)^3 \frac{L_f^4\log \frac{1}{\delta}}{\epsilon^4},\ \ \  b\geq k^2\frac{|Q|^2L_f^2}{\epsilon^4}.$$
then we have
$$F(\hat{H})=\mathbb{E}_{\Dcal}[f(\langle Q,\hat{H}\rangle;s_Q)]\leq F(H^*)+M\epsilon.$$
\label{thm:2d}
\end{theorem}
See Appendix~\ref{sec:proof2d} of our full-version \cite{ViswanathanJLA11} for a detailed proof. 

Similar to $1$-dimensional case, we can obtain tighter bounds for Problem~\eqref{eq:prob_2d} using $L_1$ loss functions, but for implementation ease we select $L_2$ loss. %That is, our EquiHist implementation solves the following $2$-dimensional least squares problem:
%\begin{equation}
%  \label{eq:eh_2d}
%  \min_{\wvec\in \mathbb{R}^{b_1\times b_1}} \frac{1}{N} \sum_{i=1}^N (s_Q-\langle Q, BWB^T\rangle)^2. 
%\end{equation}
For ease of exposition, we stated our problem formulation and analysis with equal range $r$ for both the attributes and equal number of buckets $b_1$ along each dimension. However, our method can be easily generalized to different range sizes and bucket sizes along each dimension.

Also, note that for extension from $1$-dimensional case to $2$ dimensional case, we just rewrote the query cardinality estimation as a linear function of our restricted set of parameters $W$, i.e., $$\hat{s}_Q=\langle Q, B W B^T\rangle.$$ Similarly, for $d$-dimensions, $$\hat{s}_Q=\langle Q, W\times_1 B \times_2 B \dots \times_d B \rangle,$$
where ``$\times_i$'' is $i$-th mode tensor product and $\langle A,B\rangle$ represents tensor inner product of tensors $A$ and $B$ in $d$-dimensions. Hence, for $d$-dimensions, the corresponding least squares problem for our EquiHist method would be:
\begin{equation}
  \label{eq:eh_multi}
  \min_{W\in \mathbb{R}^{b_1\times b_1\times b_1\dots \times b_1}} \frac{1}{N} \sum_{i=1}^N (s_{Q_i}-\langle Q_i, W\times_1 B \times_2 B \dots \times_d B\rangle)^2. 
\end{equation}
See Algorithm~\ref{alg:eh_multi} for pseudo-code of our general $d$-dimensional EquiHist method. 
\begin{algorithm}[tb]
  \caption{EquiHist: Equi-width histogram based method for Histogram Estimation ($d$-dimensional case)}
  \begin{algorithmic}[1]
    \STATE {\bfseries Input}: Training Queries: $Q_i\in \mathbb{R}^{r\times r\times ... \times r}, 1\leq i\leq N$, $s_{Q_i}$: response cardinality for query $Q_i$
  \STATE {\bfseries Parameters}: $k$: number of histogram buckets
  \STATE $W\leftarrow $ solution to \eqref{eq:eh_multi} (a $d$-dimensional tensor)
  \STATE $H=W\times_1 B\times_2 B \dots \times_d B$, where $B$ is as given in \eqref{eq:B}
  \STATE {\bfseries Output}: $H$ ($d$-dimensional histogram with $k$ buckets)
  \end{algorithmic}
\label{alg:eh_multi}
\end{algorithm}
\subsubsection{Sparse-recovery Approach}
\label{sec:sp_multi}
In Subsection~\ref{sec:sparse}, we introduced a technique for estimating $1$-dimensional histograms using sparse-vector recovery techniques. In this section, we briefly discuss extension of our approach to multiple dimensions. Recall that, we use wavelet transform of a histogram to convert it into a sparse-vector, i.e., $\alphavec=\Psi \hvec$. Similarly, for any general $d$-dimensional histogram $H$, $H$ can be vectorized and then multi-dimensional wavelet transform can again be viewed as a linear orthogonal transform. That is, let $\hvec^d\in \mathbb{R}^{r^d}$ be an appropriately vectorized version of histogram $H\in \mathbb{R}^{r\times r\dots \times r}$. Then, wavelet coefficients $\alphavec^d\in \mathbb{R}^{r^d}$ can be obtained by applying an orthogonal transform $\Psi^d\in \mathbb{R}^{r^d\times r^d}$, i.e., $$\alphavec^d=\Psi^d\hvec.$$
We omit details for forming $\Psi^d$ and refer interested readers to \cite{Mallat97}. 

Now, as in $1$-dimensional case, we can show that if there are at most $k$-cuboidal buckets in the histogram $H$, then the number of non-zero wavelet coefficients is at most $O(kr^{d-1}\log r)$. In fact, in practice the number of non-zero coefficients turn out to be even smaller. This can be explained by the fact that in practice most of the data is clustered in small pockets and hence the number of non-zero coefficients at lowest levels is significantly smaller than theoretical bounds. 

Hence, similar to $1$-dimensional case, our histogram learning problem is reduced to:
\begin{equation}
  \label{eq:prob_sp_multi}
  \argmin_{\text{supp}(\alphavec^d)\leq k} \frac{1}{N}\sum_{i=1}^N \left(s_{Q_i}-(\qvec_i^d)^T(\Psi^d)^T\alphavec^d\right)^2,
\end{equation}
where $\qvec_i\in \mathbb{R}^{r^d}$ is the vectorized tensor $Q_i$ in $d$-dimensions. Now, similar to $1$-dimensional case, sparse wavelet coefficients $\alphavec^d$ are estimated using Orthogonal Matching Pursuit algorithm and then the histogram $H$ is obtained after inverse wavelet transform of $\alphavec^d$ and re-arranging coefficients appropriately (See Algorithm~\ref{alg:sp}). 

Recall that our sparse-recovery method represents histograms by their corresponding wavelet coefficients $\alphavec^d$. Since, $\alphavec^d$ has only $k$ non-zero coefficients, the memory footprint of this representation is small. But for computing cardinality of an unseen test query, the time requirement might be large, especially for large dimensions. However, \cite{VitterW99} showed that for range queries, cardinality estimation from $k$ non-zero wavelet coefficients can be performed in $O(kd)$ time using error tree data structure, but with $O(kd)$ space overhead. %however, the memory footprint increases to $O(kd)$ which is reasonable as typically $k$ is small and dimensionality $d$ is less than 10. 
%For the case of $1$-d, we avoid this problem altogether as we can run an efficient DP based algorithm to obtain $k$-bucket histogram itself. However, in higher-dimensions, DP algorithm might be expensive.  
\subsection{Dynamic QFRs and Database Updates}
\label{sec:online}

In previous sections, we assumed a static set of input QFRs and a static database. We now present extensions
to our algorithms that relax these assumptions. 

Dynamic QFRs and updates introduce several engineering challenges: (1) Do we keep histograms continuously
up-to-date as new QFRs are available or update them in a batch fashion periodically or when system load is
low? (2) How and at what level of detail is information about updates conveyed to the learning system? A
comprehensive study of such engineering considerations is beyond the scope of this paper. However, we believe
that the extensions we present below can form a conceptual basis for implementing many engineering design
choices addressing the questions above.

Our extensions are based on two ideas: making the learning algorithms \emph{online} and modifying the
empirical query distribution by biasing it towards recent QFRs. 

\vspace{1ex}\noindent{\bf Online learning:} Online learning algorithms \cite{Rakhlin09}, at every \emph{time
  step} $t$, maintain a current histogram $\hvec_t$. In response to a new QFR $\langle \qvec_t, s_{\qvec t}
\rangle$, they suitably modify $\hvec_t$ to produce $\hvec_{t+1}$. Recall that in EquiHist algorithm, a
histogram $\hvec_t$ is parametrized by $\wvec_t \in \mathbb{R}^b$ such that $\hvec_{t}=B\wvec_{t}$. To update
$\wvec_{t}$ to $\wvec_{t+1}$ in response to a new QFR $\langle \qvec_t, s_{\qvec t} \rangle$, we use a
well-known strategy called \emph{Follow the Regularized Leader (FTRL)} \cite{Rakhlin09}. Formally, the update
step is given by:
\begin{equation}
  \label{eq:eh_online}
  \wvec_{t+1}=\min_{\wvec\in\mathbb{R}^b}\left[\frac{1}{t}\sum_{i=1}^t (s_{\qvec_i}-\qvec_i^TB\wvec)+\lambda \|\wvec\|_2^2\right]
\end{equation}
where $\lambda \geq 0$ is an appropriately selected constant and note that $\hvec_{t+1} = B \wvec_{t+1}$. From
Equation~\ref{eq:eh_online}, it might seem that we are just ``relearning'' a histogram from scratch at every
time step. However, we can show that $\wvec_{t+1}$ can be computed from $\wvec_t$ using $O(k^2)$ time
(\emph{independent} of $t$) by maintaining appropriate data structures. We can also prove formal guarantees on
the error incurred by this approach using techniques in \cite{Rakhlin09}. We omit these details due to space
considerations.

%Note that although the above problem involves a least squares problem over $t$ points, however by maintaining appropriate data-structures, we can compute $\wvec_{t+1}$ using only $O(k^2)$ operations, which is {\em independent} of $t$. Furthermore, we can provide theroetical bound on extra loss incurred by our online algorithm when compared to the batch algorithm (algorithm with all the queries and cardinalities available beforehand). Our bound follows from standard regret analysis of FTRL method \cite{Rakhlin09} and we ommit details due to lack of space. 

Similar to EquiHist, SpHist also solves a least squares problem once it greedily selects a small set $\Scal$ of
non-zero wavelet coefficients.  To make SpHist online, we propose modifying $\Scal$ only infrequently (say
every night using all QFRs accumulated that day). In between these modification, $\Scal$ remains unchanged
and we can update the current histogram using techniques similar to ones we presented above for EquiHist.

Since new QFRs capture changes to workload and data characteristics, the histogram maintained by online
learning algorithms can adapt changes to workload and data characteristics. The online learning algorithms,
however, weigh older QFRs, which might contain outdated information, and newer QFRs equally. For faster
adaptation to changes, it might be useful to assign a higher weight to recent QFRs as we discuss how to do
this next.

\vspace{1ex}\noindent {\bf Biasing for recency:} Recall that our learning formulation involves a query
distribution $\Dcal$ and that our algorithms approximate $\Dcal$ using an empirical distribution $\hat{\Dcal}$
that assigns an equal probability that each training sample. To bias for recency, we simply use an alternate
empirical distribution that assigns a higher probability for recent training QFRs compared to older QFRs. The
modifications to our algorithms to incorporate this change are straightforward and we omit the details. 

% Finally, in practice, not only queries are streaming but also the database might be updated frequently. We can easily modify our techniques to handle database updates as well. For example, if the database is updated at $\tau$-th step
% , then the update for $\wvec_{t+1}$, $t+1\geq \tau$ is given by:
% \begin{multline}
%   \label{eq:eh_db}
%   \wvec_{t+1}=\min_{\wvec\in\mathbb{R}^k}\frac{1}{t}\left(\sum_{i=1}^\tau \eta(s_{\qvec_i}-\qvec_i^TB\wvec)^2+\sum_{i=\tau+1}^t (s_{\qvec_i}-\qvec_i^TB\wvec)^2\right)\\+\lambda \|\wvec\|_2^2,
% \end{multline}
% where $0<\eta<1$ is a scaling factor that decreases importance of older training queries (before database updates). Note that the above updates can be easily modified to handle multiple database updates. Also, similar updates can be used for updating wavelet coefficients in the sparse-recovery based approach. 
%\subsection{Extension to Online Adverserial Query Workload}
%%% Local Variables: 
%%% mode: latex
%%% TeX-master: "histestimation-main"
%%% End: 

\section{Experiments}
\label{sec:exp}

In this section, we empirically evaluate our algorithms (EquiHist and
Sphist) and present comparison against ISOMER \cite{SrivastavaHMKT06},
current state-of-the-art in self-tuning histograms. In particular, we
compare our algorithms and ISOMER on quality of learned histograms and
on various performance parameters including scalability with number of
histogram buckets, training data size, dimensionality of histograms,
and size of attribute domain. We use both real and synthetic data for
our evaluation.

Our experiments involve using one of the algorithms above to learn a
histogram from an input training set of QFRs. We use a separate
\emph{test set} of QFRs to measure the quality of learned
histograms. In particular, we measure quality using percentage {\em
  Average Relative Error} achieved over the test QFRs:
\begin{equation}
  \label{eq:re}
  \text{Avg. Rel. Error}=\frac{1}{N_\text{test}}\sum_{i=1}^{N_\text{test}}\frac{|s_{\qvec_i}-\widehat{s}_{\qvec_i}|}{\max\{100,s_{\qvec_i}\}}\times 100, 
\end{equation}
where $s_{\qvec_i}$ and $\widehat{s}_{\qvec_i}$ denote respectively
the actual and estimated cardinalities of test query $\qvec_i$ and
$N_\text{test}$ denotes the number of test queries. The same measure
is used in ISOMER \cite{SrivastavaHMKT06}.

%where $s_{\qvec_i}$ is the actual cardinality of query response to query $\qvec_i$, $\widehat{s}_{\qvec_i}$ is the estimated query response cardinality by a given method, $N_\text{test}$ is the number of test queries, and finally $\qvec_i$ is the $i$-th test query in the dataset. %This was the measure used by \cite{SrivastavaHMKT06} in their experiments as well. 
%Note that the $\max\{100,s_{\qvec_i}\}$ in the denominator of (\ref{eq:re}) helps to penalize larger errors ($>100$) more severely than smaller ones (This is reasonable since all our databases have ~100K records or more).

In Section~\ref{sec:dbandqueryworkloads}, we discuss details of data
and query workloads used in our experiments; we also present
implementation details of algorithms in this section. We present
results for 1-dimensional histograms in Section~\ref{sec:1D-results}
and multi-dimensional histograms, in
Section~\ref{sec:multiD-results}. Finally, in
Section~\ref{sec:results_streaming} we report results relating to
online learning for dynamic QFRs.

%provide details of the databases we consider, query workloads used and implementation details. Results for learning 1-D histograms are described in Section~\ref{sec:1D-results}, for multi-dimensional histograms are reported in Section~\ref{sec:multiD-results}. Finally, in Section~\ref{sec:results_streaming} we report results on experiments concerning streaming queries setting. 

\subsection{Data, Workload, and Implementation}
\label{sec:dbandqueryworkloads}
For real-world data, we used the Census dataset from the UCI Machine
Learning Repository \cite{FrankA10} also used in STHoles
\cite{BrunoC02}. For synthetic data, we used the data generator used
by STHoles \cite{BrunoC02}; all synthetic datasets are essentially
mixtures of Gaussians.

%We generated synthetic datasets using a standard mixture of Gaussians model used by STHoles \cite{BrunoC02}.All synthetic datasets consisted of $500,000$ points sampled from a mixture of Gaussians. We also report results on the Census dataset from the UCI Machine Learning Repository \cite{FrankA10}.

\begin{figure*}[ht]
  \centering
  \begin{tabular}{cccc}
    \hspace*{-20pt}\includegraphics[width=.25\textwidth]{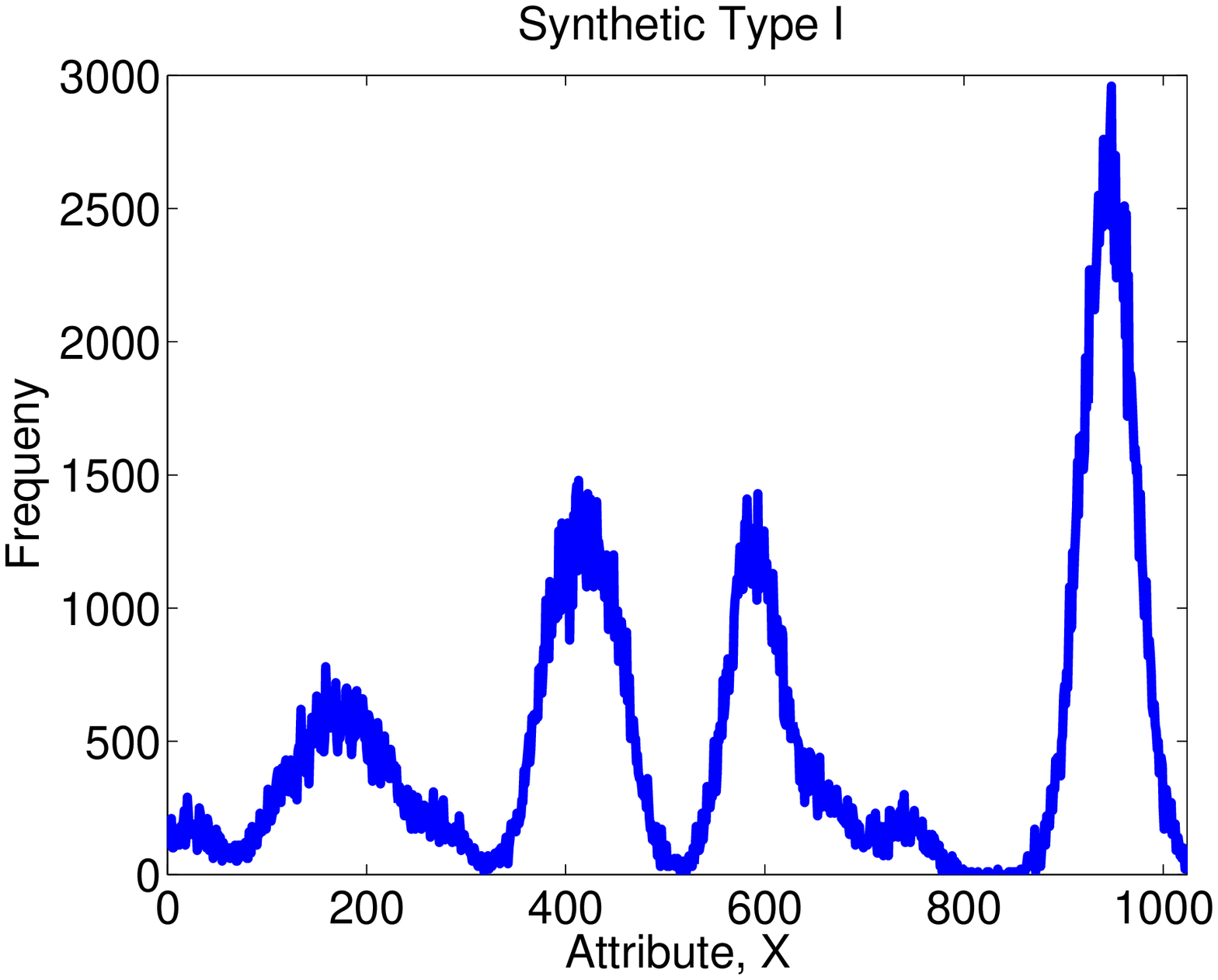}&
    \includegraphics[width=.25\textwidth]{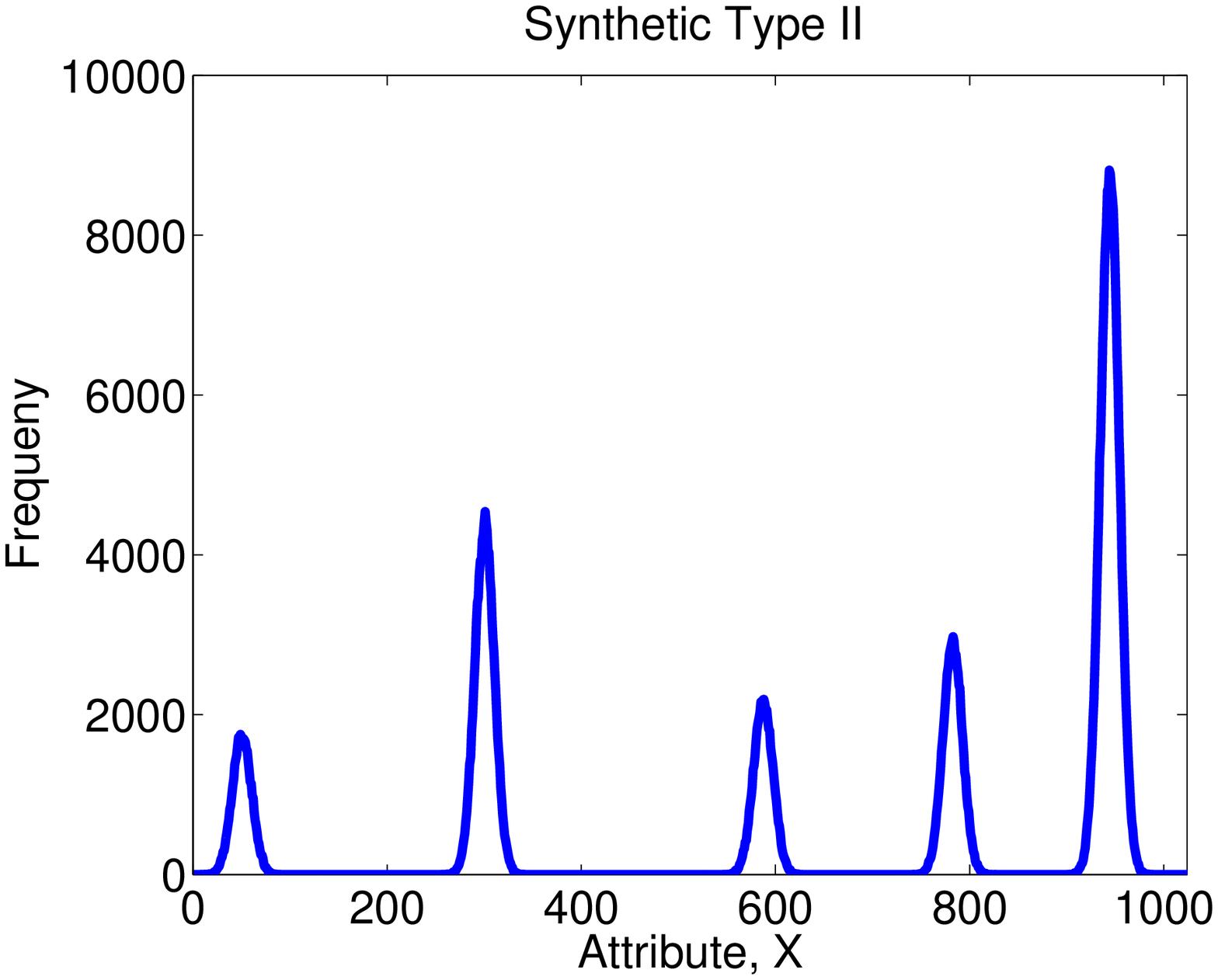}&
    \includegraphics[width=.25\textwidth]{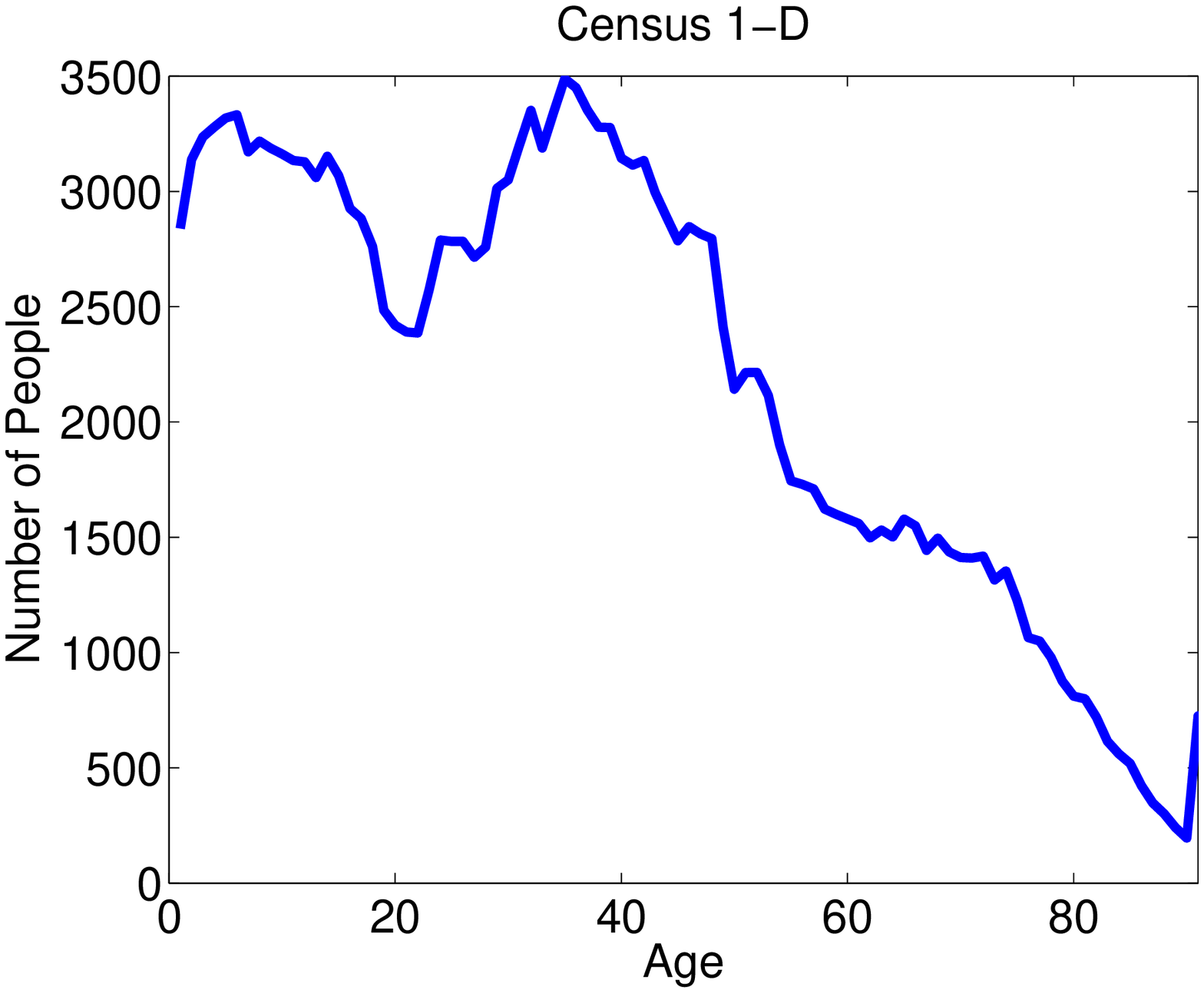}&
    \includegraphics[width=.25\textwidth]{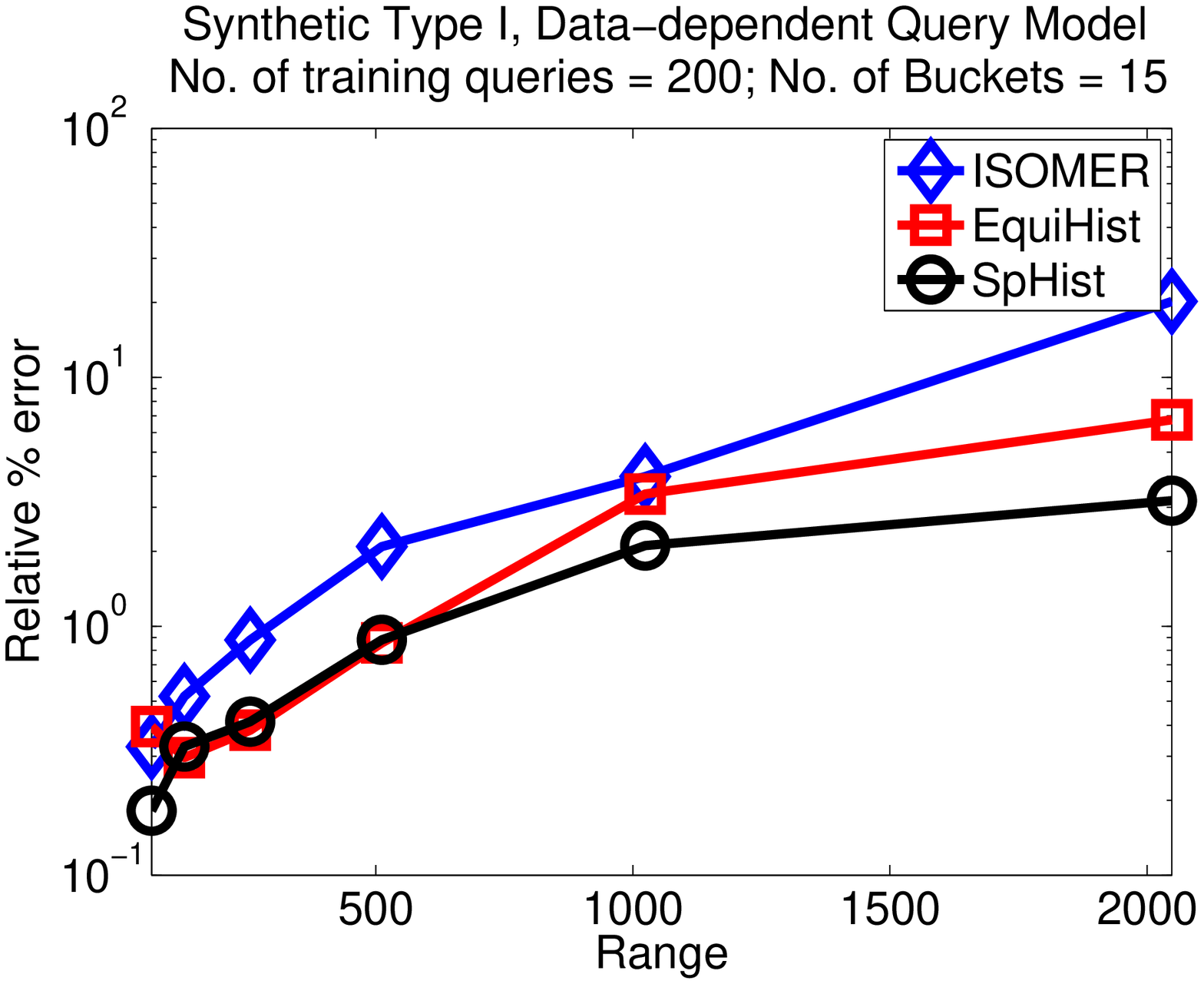}\\
    {\bf (a)}&{\bf (b)}&{\bf (c)}&{\bf (d)}\vspace*{-10pt}
  \end{tabular}
  \caption{{\bf a)} {\bf Synthetic Type I} data distribution, {\bf b)} ``spiky'' {\bf Synthetic Type II} data distribution, {\bf c)} {\bf Census $1$-D} data distribution, {\bf d)} Relative error incurred by various methods as range of the attribute in {\bf Synthetic Type I} dataset varies. Clearly, SpHist and EquiHist scales better with increasing range than ISOMER. Also, as expected due to Theorem~\ref{thm:1d}, the error increases at sub-linear rate with increasing range.}
  \label{fig:1d}\vspace*{-10pt}
\end{figure*}
We conduct experiments for {\em one-dimensional case} using two different types of synthetic data and a $1$-D projection of Census data (see Figure~\ref{fig:1d} (a),(b),(c) for the data distributions of the above three datasets):
%In the one-dimensional case, we generated two different kinds of synthetic data and use Census data
%one with wider modes covering the range, and another with small variances leading to a `spiky' distribution (most records concentrated around a small number of attribute-values). We refer to these as Synthetic Type I and Synthetic Type II, respectively. We also learn one-dimensional histograms over a 1-D projection of the Census Dataset. The data distributions in these three cases are plotted in Figure~\ref{fig:1d}.

\noindent {\bf \textbullet \ Synthetic Type I}: For this dataset, we sampled points from a mixture of seventeen Gaussians, each with variance $625$ and means selected uniformly at random from $[0,r]$. %Range $r$ of the attribute Number of Gaussians in the mixture was fixed at 17. Range of the attribute (denoted $r$ in our theory) was randomly selected between 128 to 2048. Mean of each of Gaussian was picked uniformly at random between $0$ to range $r$ and the variance was fixed at $625$.

\noindent {\bf \textbullet \ Synthetic Type II}: Here, we sampled from
a mixture of five Gaussians and means of each Gaussian is selected
uniformly at random. The variance is selected to be just $100$,
leading to ``spiky'' distribution, i.e., most records are concentrated
around a small number of attribute-values.
%The number of Gaussians in the mixture was fixed at 5 and the means were selected randomly between 0 and 1024. The variance was of each mode was set to 100.

\noindent {\bf \textbullet \ Census 1-D}: We use the {\em Age} attribute of the standard Census dataset with $199,523$ database records. Range ($r$) here is $91$. %and the queries are generated using Uniform Query model.

Similarly, for {\em multi-dimensional histogram experiments}, we generated synthetic data using multi-dimensional Gaussians and use multi-dimensional projections of Census data. That is, % We refer to these as Synthetic Multi-D. We also considered 2-D and 3-D projections of the Census data which we refer to as Census Multi-D.

\noindent {\bf \textbullet \ Synthetic Multi-D}: We generated 2 and 3
dimensional datasets for a given range by sampling from a mixture of
spherical Gaussians of corresponding number of dimensions. For
2-dimensional datasets, we used a mixture of 9 Gaussians with random
means and variance equal to 100. For 3-dimensional case we used a
mixture of 5 Gaussians with random means and variance set to 25. The
range along each attribute was fixed to 32.

\noindent {\bf \textbullet \ Census Multi-D}:  We used the $2$-dimensional dataset obtained by selecting the ``Age'' and ``Number of Weeks worked'' attributes. For the $3$-dimensional dataset we chose the attributes of ``Age'', ``Marital status'' and ``Education''.

Given the above datasets, we now describe the models to generate {\em QFRs} used for training and testing learned histograms. We used two standard models of range query generation models proposed by \cite{PagelSTW93} and later used by \cite{BrunoC02}:

\noindent {\bf \textbullet \ Data-dependent Query Model}: In this model, first query ``center'' is sampled from the underlying data distribution. Then, the query is given by a hyper-rectangle whose centroid is given by the generated ``center'' and whose volume is at most $20$\% of the total volume. 

\noindent {\bf \textbullet\ Uniform Query Model}: In this model, query ``centers'' are selected uniformly at random from the data range. Then, similar to the above query model, each query is a hyper-rectangle generated around the ``center'' and volume at most $20$\% of the total volume. 

As mentioned in \cite{BrunoC02}, the above two models are considered to be fairly realistic and mimics many real-world query workloads. 
We generated separate training and test sets (of QFRs) in all the experiments. In each of the experiments, we evaluated various methods using a test set of $5000$ QFRs.

%For all the datasets, we compare our methods with ISOMER \cite{SrivastavaHMKT06}, a state-of-the-art method for self-tuning histograms. For comparing different methods, we use average relative error as the error metric, a standard error metric for this problem. Formally, 
\noindent {\bf Implementation Details:} For experiments with one-dimensional histograms, we implemented both of our methods EquiHist and SpHist, as well as ISOMER using Matlab. We modified an C++ implementation of STHoles \cite{BrunoC02} for multi-dimensional histograms experiments. For these experiments, we implemented both ISOMER as well as our equi-width approach (EquiHist) using C++. SpHist was implemented in MATLAB. For each experiment, we report numbers averaged over 10 runs.

For solving the max-entropy problem in ISOMER, we use an iterative solver based on Bregman's method \cite{CensorZ97}. We found Bregman's method for solving max-entropy problem to be significant faster than the Iterative Scaling method used by \cite{SrivastavaHMKT06}. 

%uniformly random range queries. That is, suppose we are working in $d$ dimensions and let say range along each of the dimension is $R$. Then, the query is defined by $[\bm{\ell}, \uvec]$, where $\ell_i\sim Unif([0, R]) 1\leq i\leq d$ is the ``lower''-end of the query along $i$-th dimension and  $u_i$ is the ``upper''-end of the query. 
%For example, for 2-d query:
%\begin{verbatim}
%SELECT COUNT(*) FROM XTABLE,
%WHERE ATTRIBUTE_1>=$l1$ and ATTRIBUTE_1<
%\end{verbatim}
% $[\bm{\ell}, \uvec]$, where 
%As mentioned in \cite{BrunoC02}, the above two models are considered to be fairly realistic and models user queries reasonable well. 

%For each experiment, we report numbers averaged over 10 runs. 

\subsection{Results for One-Dimensional Histograms}
\label{sec:1D-results}

\begin{figure*}[ht]
  \centering
  \begin{tabular}{cccc}
    \hspace*{-20pt}
    \includegraphics[width=.25\textwidth]{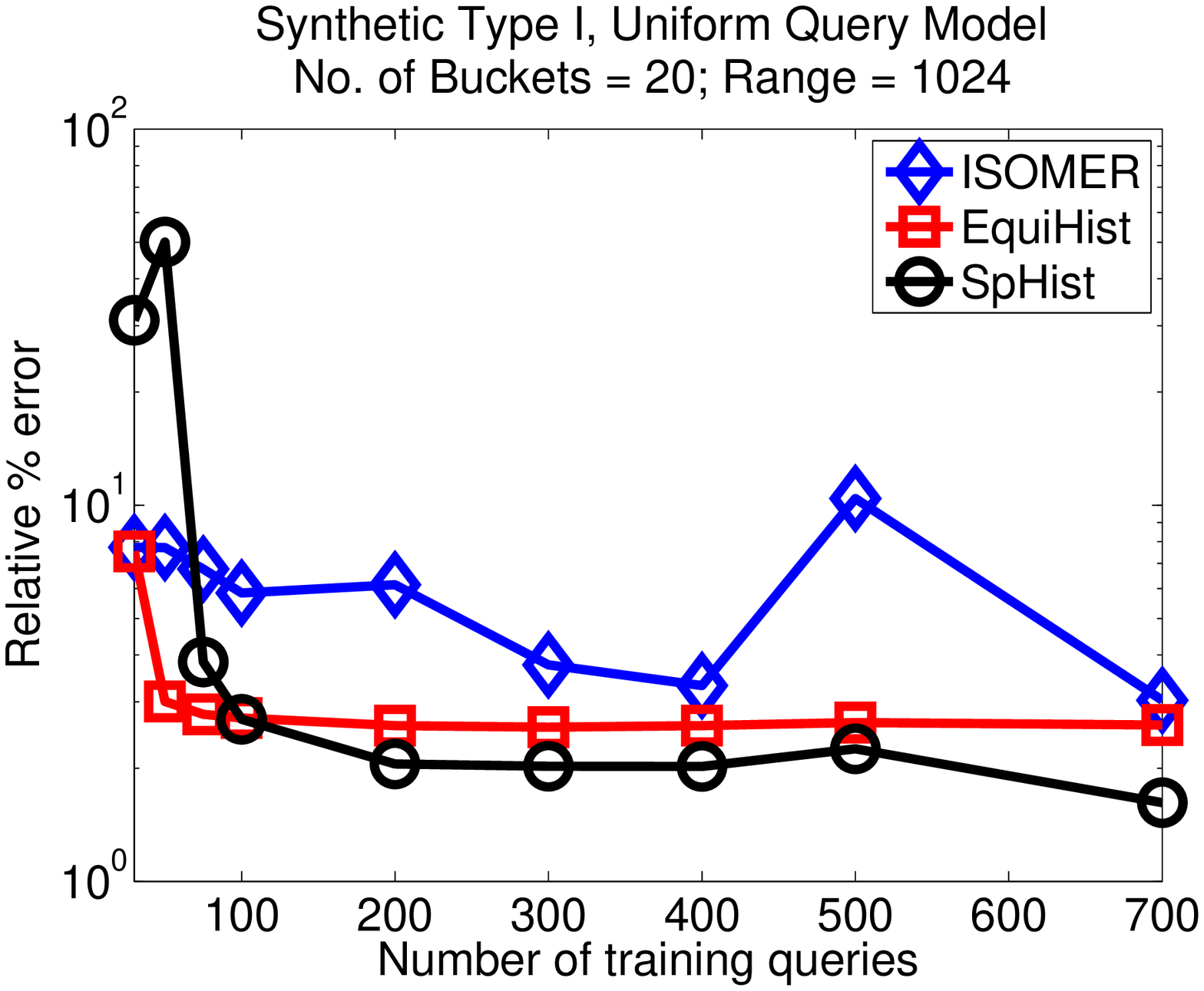}&
    \includegraphics[width=.25\textwidth]{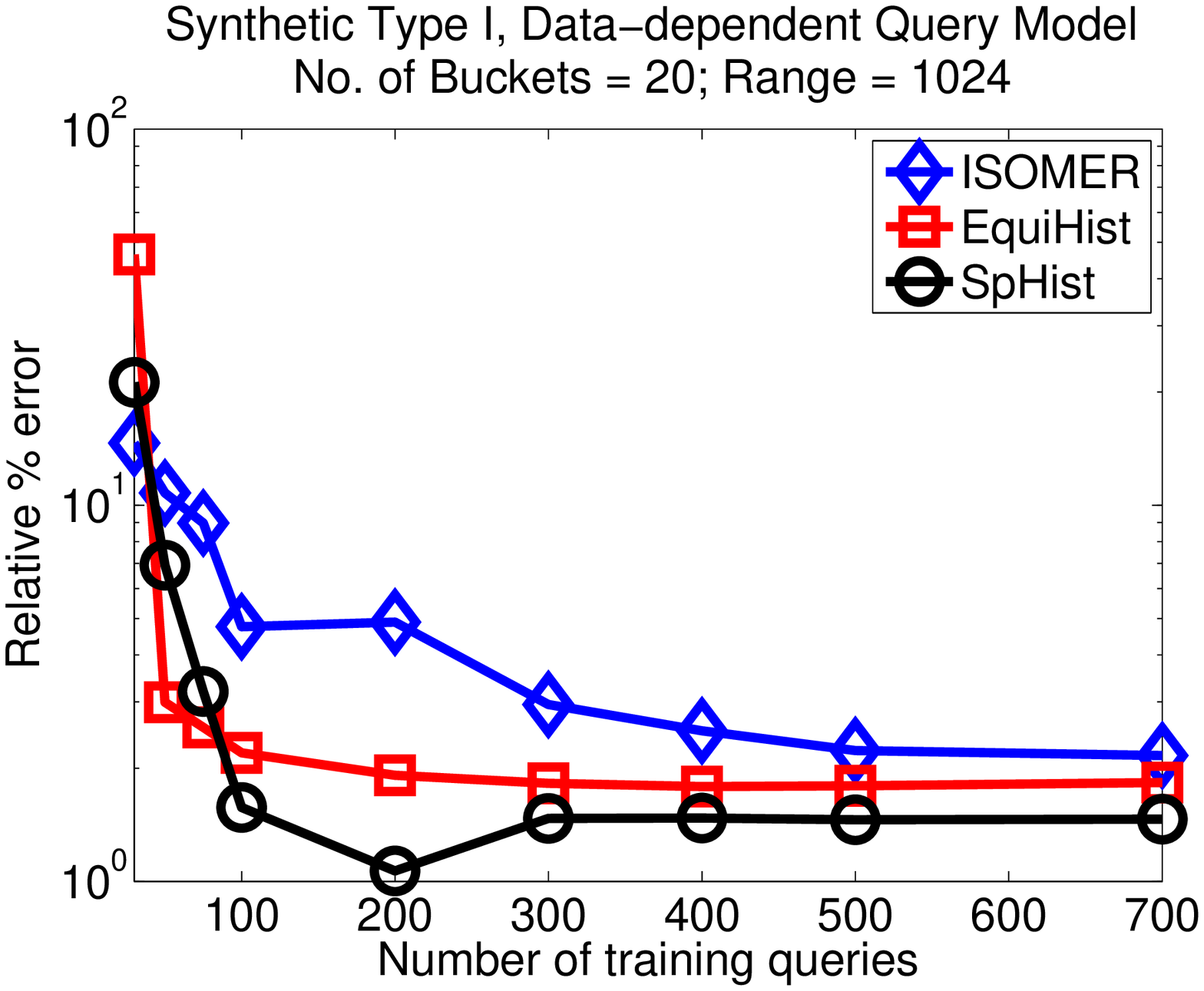}&
    \includegraphics[width=.26\textwidth]{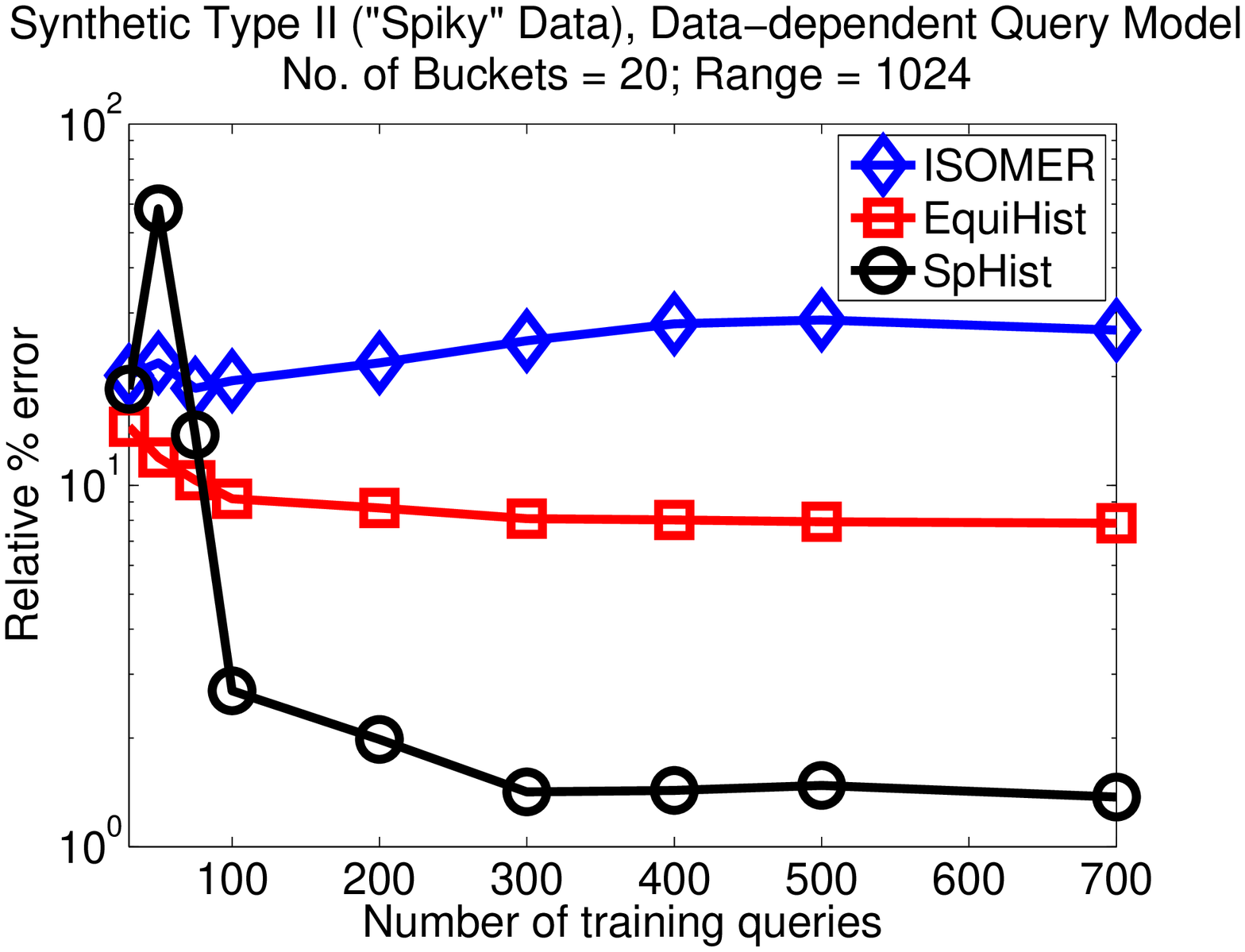}&
    \includegraphics[width=.25\textwidth]{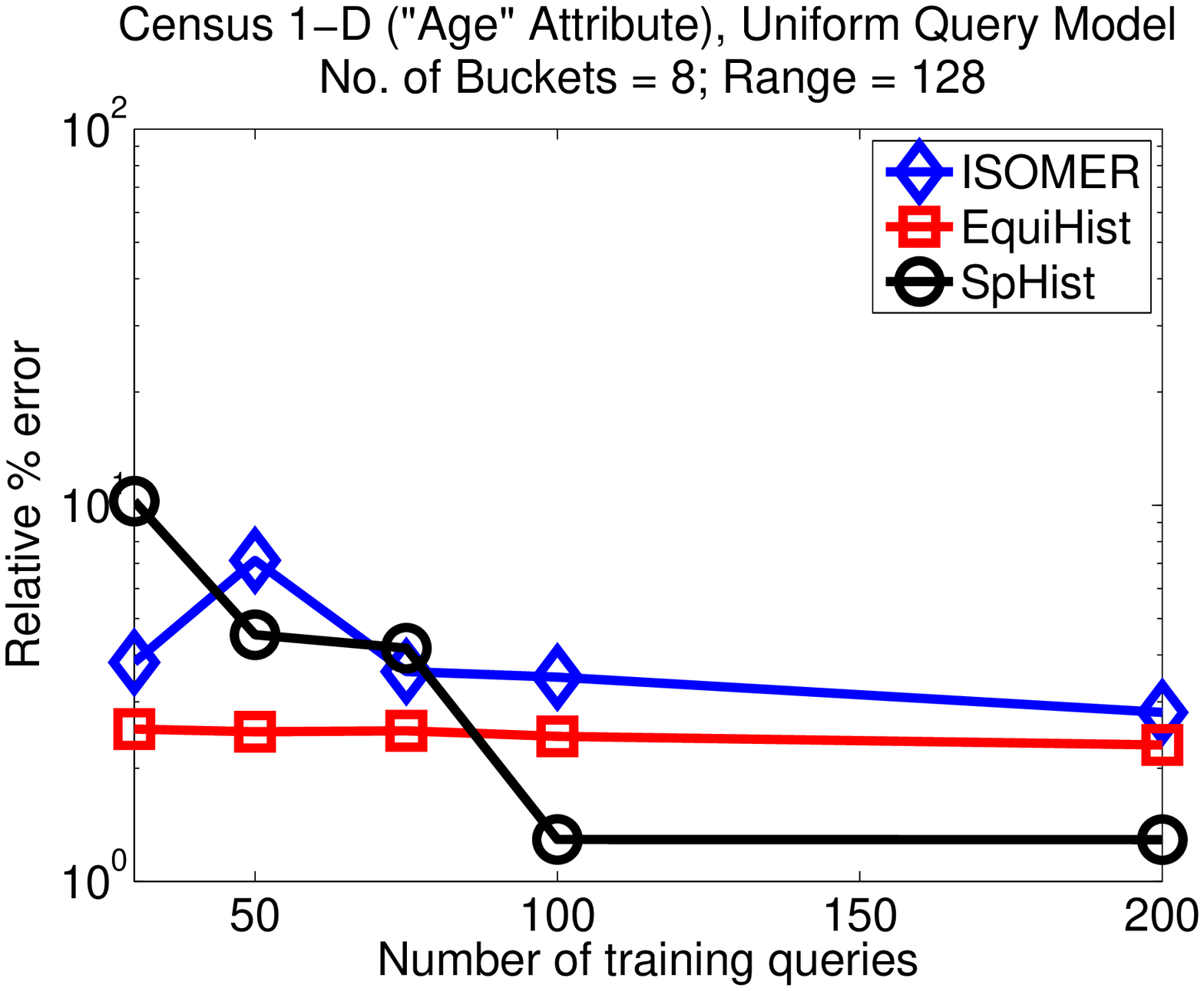}\\
    {\bf (a)}&{\bf (b)}&{\bf (c)}&{\bf (d)}\vspace*{-10pt}
  \end{tabular}
  \caption{Comparison of average relative error (on log-scale) with varying number of training queries. {\bf (a)} Test error for {\bf Synthetic Type I} dataset with queries generated from {\bf Uniform Query model}. Both EquiHist and SpHist converges at around $200$ training queries only, and incurs $0.5\%$ and $1.4\%$ less error than ISOMER, respectively, for $700$ training queries.  {\bf (b)} Test error on {\bf Synthetic Type I} dataset with queries generated from {\bf Data-dependent Query Model}. Here again, both EquiHist and SpHist converges at around $200$ training queries and are finally obtains $0.3\%$ and $1.6\%$ less error than ISOMER. {\bf (c)} Test error on ``spiky'' {\bf Synthetic Type II} dataset with queries from {\bf  Data-dependent Query model}. For $700$ training queries, ISOMER incurs $26.87\%$ error, while SpHist incurs only $1.37\%$ error. {\bf (d)} Test error on the Age attribute of {\bf Census $1$-d} data with queries generate from {\bf Uniform Query model}. For $200$ queries, EquiHist incurs approximately $0.5\%$ less error than ISOMER, while SpHist incurs $1.5\%$ less error.}
  \label{fig:var_q}\vspace*{-10pt}
\end{figure*}

We now present results for 1-D histograms and study how the performance varies under different conditions: first, as the number of training queries increases, second, as the number of buckets in the histogram being learnt increases, and finally, as the range of attribute value increases. \\[2pt]
{\bf Varying Number of Training Queries}:\\
We first compare our EquiHist and SpHist method with ISOMER for varying number of training queries. Figure~\ref{fig:var_q} (a) compares relative error incurred on test queries by the three methods on {\bf Synthetic Type I} dataset for queries generated from {\bf Uniform Query Model}. Here, we vary the number of training queries from $25$ to $700$, while the range $r$ of attribute values is fixed to be $1024$ and the number of buckets in the histogram is fixed to be $20$. Naturally, the error incurred by each of the methods decreases with increasing number of training queries. However, both of our methods are able to decrease the relative error more rapidly. For example, in around $200$ queries, error converges for both EquiHist and SpHist. In contrast, error incurred by ISOMER decreases slowly and oscillates, primary reason being in the final round ISOMER  uses only twice the number of queries (approximately) as number of buckets ($20$) and hence over-fits in some runs. Furthermore, even with $700$ queries, our SpHist method is $1.4\%$ more accurate than ISOMER, while EquiHist is $0.5\%$ more accurate. 

Next, we compare the three methods on {\bf Synthetic Type I} dataset with queries generated from {\bf Data-dependent Query Model} (See Fig.~\ref{fig:var_q}~(b)). Here again, both EquiHist and SpHist requires only $300$ training queries to converge, and are about $0.3\%$ and $1.6\%$ more accurate than ISOMER. 

\begin{figure}[ht]
  \centering
  \includegraphics[width=\columnwidth]{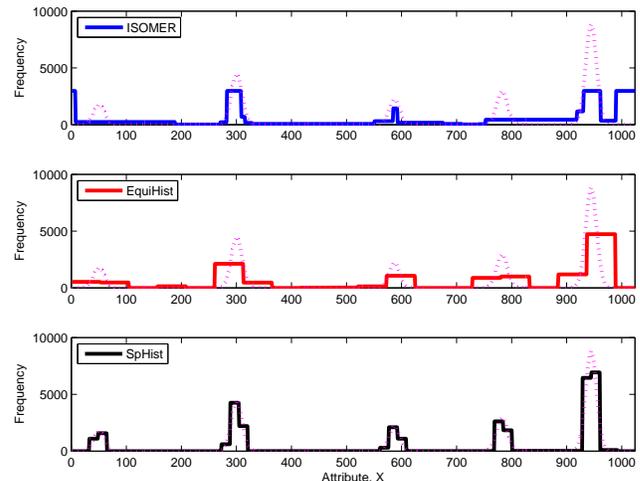}\vspace*{-10pt}
  \caption{Figure shows learned histogram by ISOMER (top plot), EquiHist (middle plot) and SpHist (bottom plot) for ``spiky'' {\bf Synthetic Type II} dataset with $700$ data-dependent queries and $20$ buckets. Clearly, bucket boundaries discovered by ISOMER do not align well with the peaks of the true frequency distribution(see Figure~\ref{fig:var_q} (c)), leading to high error. EquiHist is constrained to partition range at equal intervals, hence is mis-aligned with several peaks. In contrast, SpHist is able to accurately align bucket boundaries to the true frequency distribution, hence incurs less test error.}
\vspace*{-5pt}
  \label{fig:1d_rec_comp}%\vspace*{-10pt}
\end{figure}
In Fig.~\ref{fig:var_q}~(c) we compare performances on the spiky {\bf Synthetic Type II} dataset with queries generated from {\bf Data-dependent} {\bf Query Model}. For this experiment, all the three methods converge at about $300$ queries. However, SpHist is significantly more accurate than both ISOMER and EquiHist. Specifically, SpHist incurs only $1.37\%$ error, while EquiHist incurs $7.85\%$ error and ISOMER incurs $26.87\%$ error. EquiHist naturally is a little inaccurate as EquiHist's bin boundaries will typically be much wider than optimal histograms boundaries. Interestingly, SpHist is able to learn correct bucket boundaries with small number of training queries and hence provides a histogram very similar to the underlying distribution. Figure~\ref{fig:1d_rec_comp} shows the recovered histograms by the different methods overlayed on the true frequency distribution. We observe that SpHist is able to align bin boundaries accurately with respect to the true distribution. In comparison, ISOMER and EquiHist's buckets are not as well aligned, leading to higher test errors.

In Fig.~\ref{fig:var_q}~(d) we compare the three methods on the {\bf Census 1-D} dataset and queries generated from {\bf Uniform Query Model}. As in the previous case, SpHist incurs less error than both ISOMER and EquiHist ($1.5\%$ and $1.0\%$ less error respectively), and is able to learn from a smaller number of training queries.

\begin{figure*}[ht]
  \centering
  \begin{tabular}{cccc}
    \hspace*{-20pt}
    \includegraphics[width=.25\textwidth]{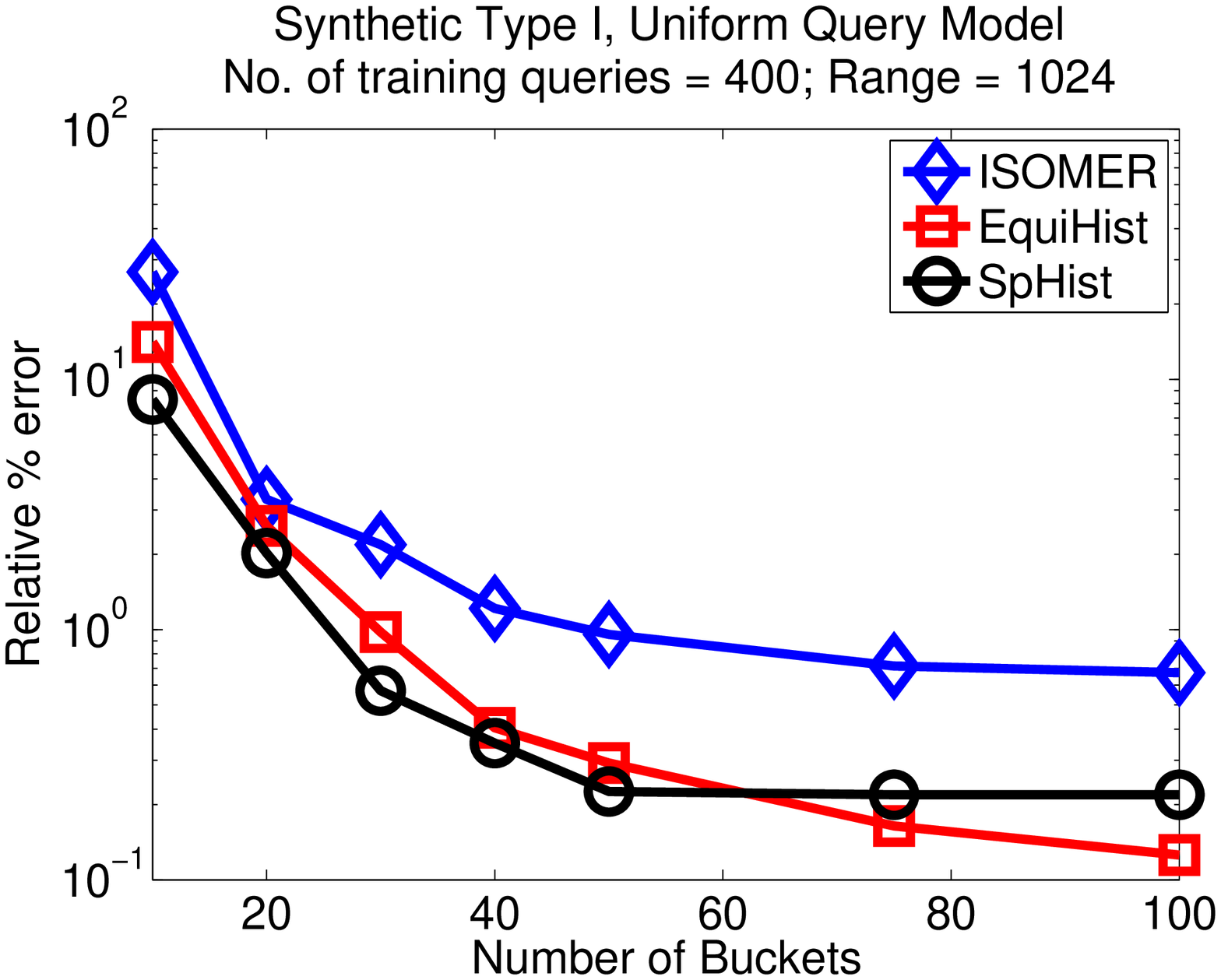}&
    \includegraphics[width=.245\textwidth]{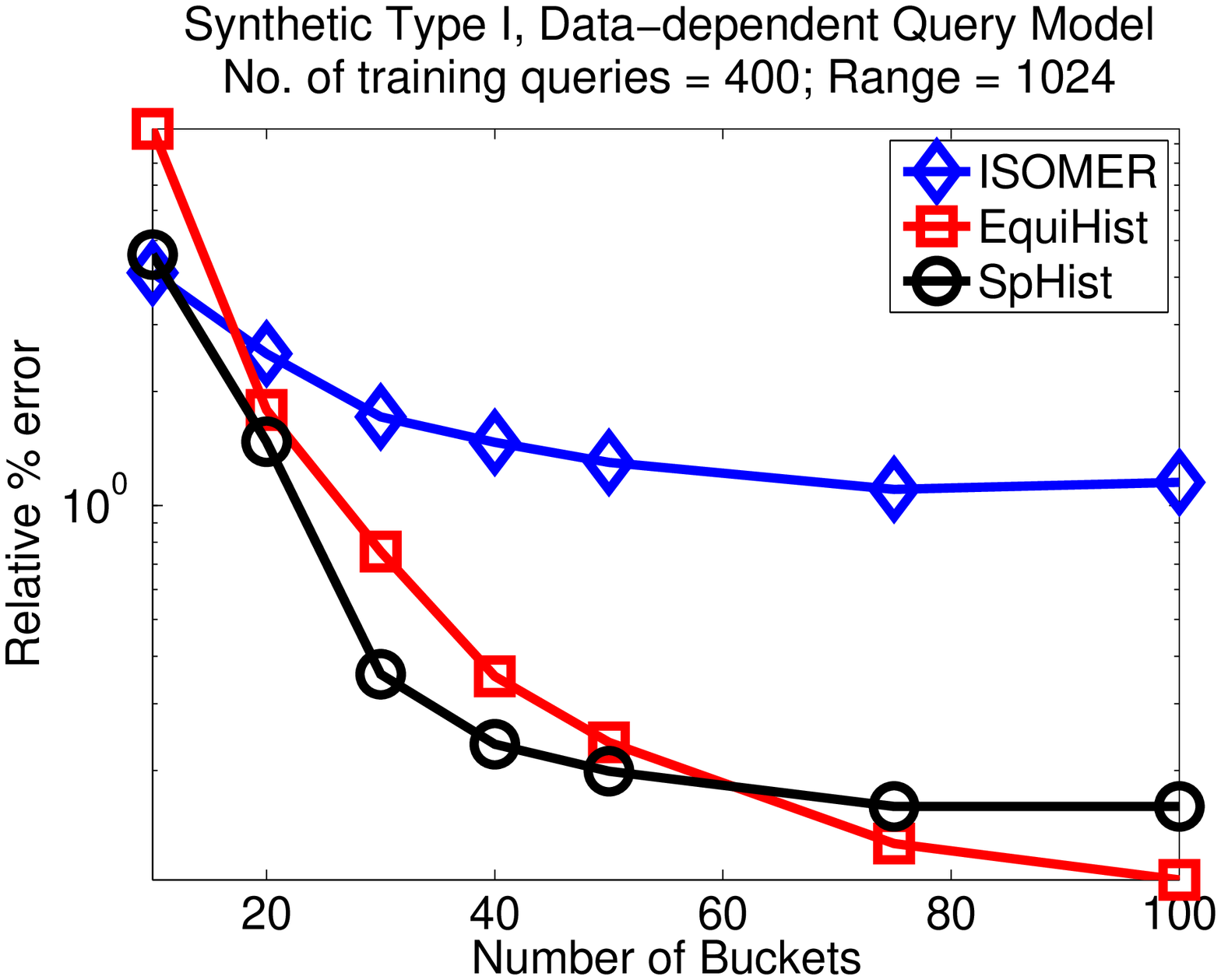}&
    \includegraphics[width=.26\textwidth]{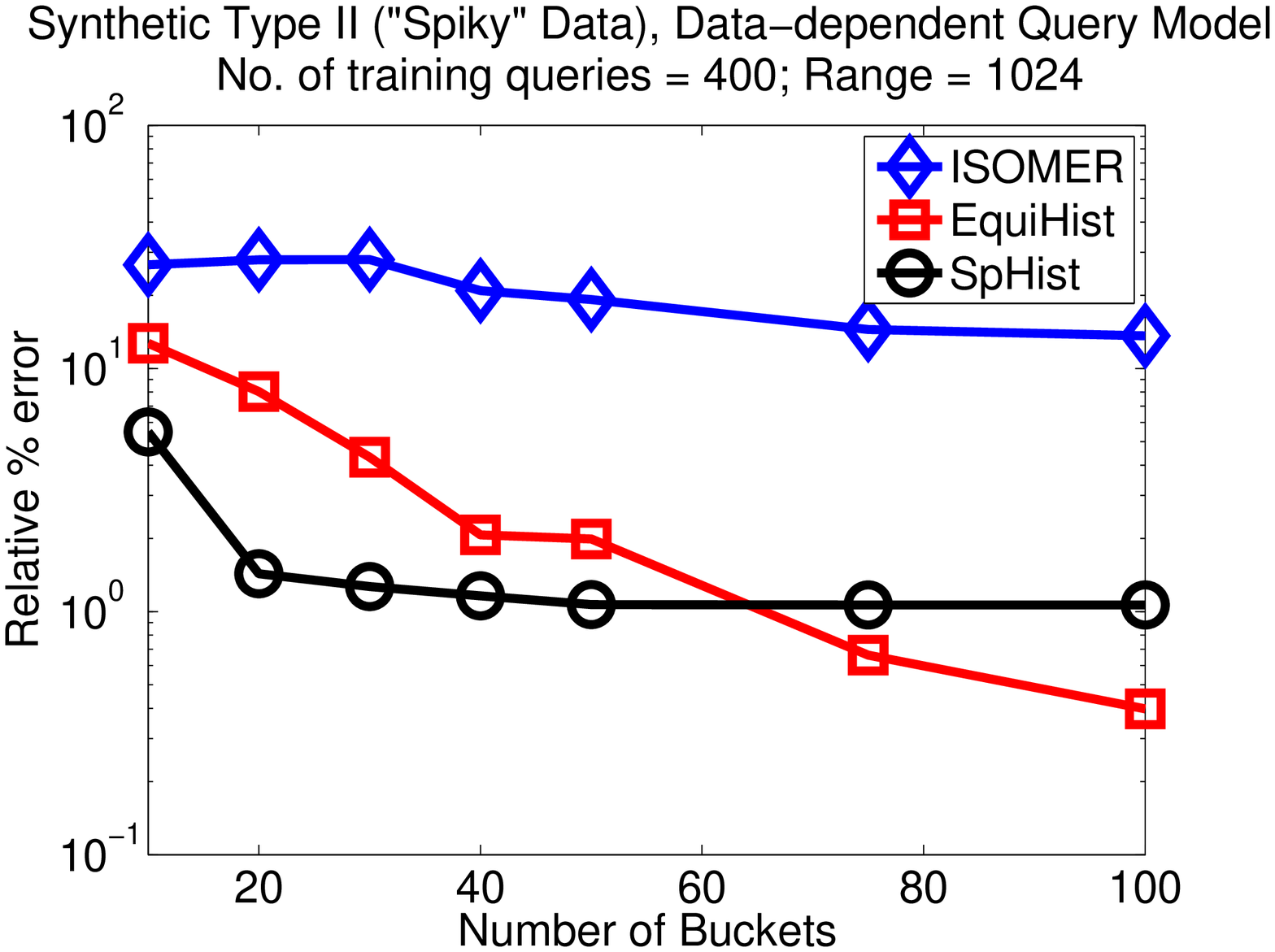}&
    \includegraphics[width=.25\textwidth]{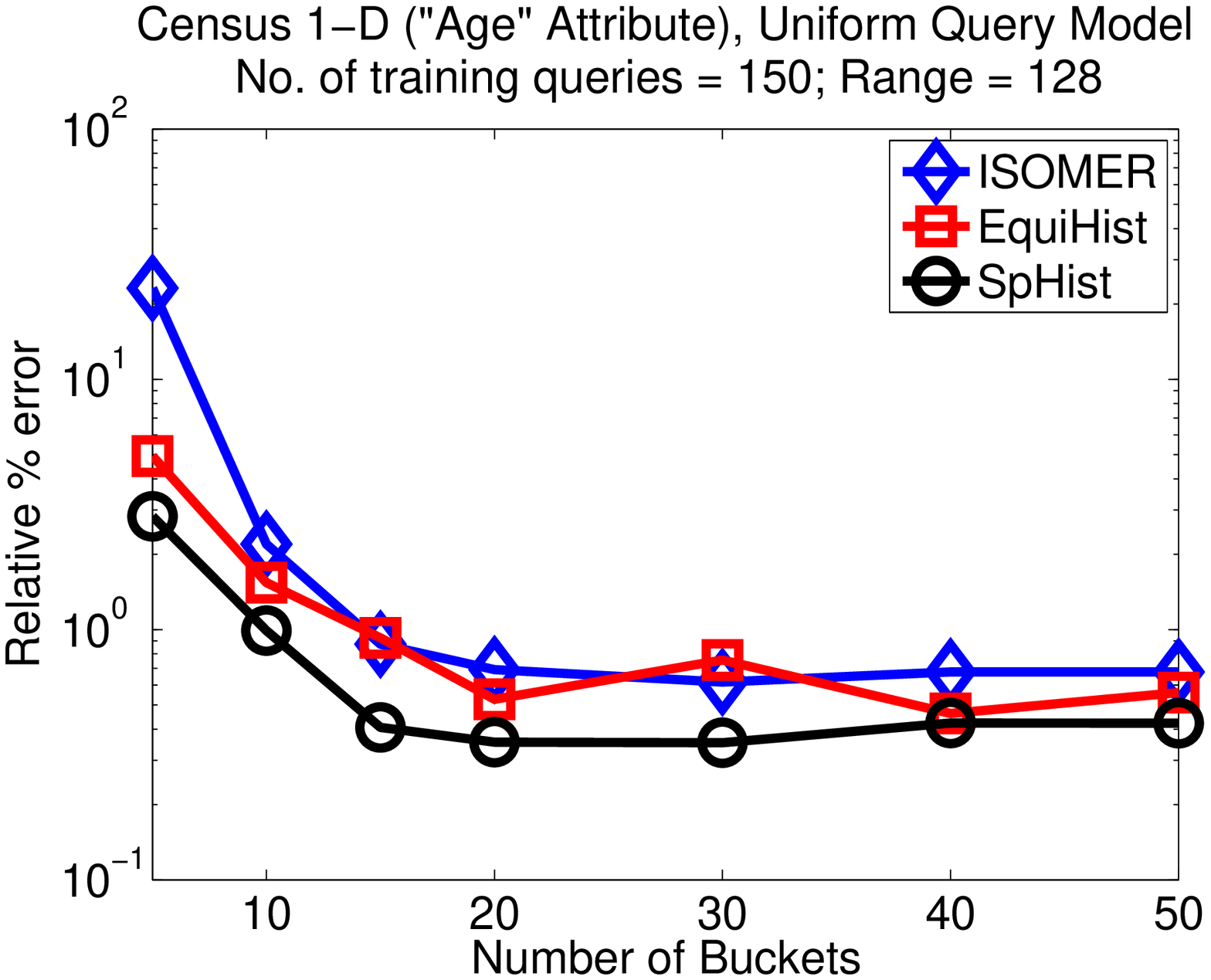}\\
    {\bf (a)}&{\bf (b)}&{\bf (c)}&{\bf (d)}\vspace*{-10pt}
  \end{tabular}
  \caption{Comparison of average relative error (on log-scale) with varying number of histogram buckets. {\bf (a)} Test error on {\bf Synthetic Type I} datasets with queries generated from {\bf Uniform Query model}. For most values, both EquiHist and SpHist are significantly more accurate than ISOMER. As expected, for small number of buckets, SpHist is more accurate than EquiHist while for large number of buckets, EquiHist incurs less error. In particular, for $10$ buckets, SpHist is $18\%$ more accurate than ISOMER and $6\%$ more accurate than EquiHist. {\bf (b)} Test error on {\bf Synthetic Type I} dataset with queries generated from {\bf Data-dependent Query Model}. Here again, SpHist is around $5\%$ more accurate than EquiHist for $10$ buckets and is around $0.3\%$ less accurate than ISOMER. {\bf (c)} Test error on {\bf Synthetic Type II} dataset with queries from {\bf Data-dependent Query model}. Here, SpHist incurs significantly less error than both ISOMER (by $21\%$) and EquiHist (by $6\%$) for $10$ buckets. {\bf (d)} Test error on {\bf Census 1-D} dataset with queries generate from {\bf Uniform Query model}. For $5$ buckets, SpHist incurs about $2\%$ less error than EquiHist and $20\%$ less error than ISOMER.}\vspace*{-10pt}
  \label{fig:var_buckets}
\end{figure*}
\noindent {\bf Varying Number of Buckets}: \\
Here, we study our methods as number of buckets in the histograms vary. First, we consider {\bf Synthetic Type I} dataset and vary number of buckets from $10$ to $100$, while range and number of training queries are fixed to be $1024$ and $400$, respectively. Figure~\ref{fig:var_buckets} (a) shows error incurred by the three methods for varying number of buckets when queries are generated using {\bf Uniform Query Model}. Clearly, for small number of buckets, SpHist achieves significantly better error rates than EquiHist and ISOMER. Specifically, for $10$ buckets, SpHist incurs $8.3\%$ error, while EquiHist incurs $14.09\%$ error and ISOMER incurs $26.84\%$ error. However, EquiHist performs better than both ISOMER and SpHist as number of buckets increase. Figure~\ref{fig:var_buckets} (b) shows a similar trend when queries are generated from {\bf Data-dependent Query Model}. Here, interestingly, for $10$ buckets, ISOMER performs significantly better than EquiHist and performs similar to SpHist. However, with larger number of buckets ISOMER converges to significantly higher error than both EquiHist and SpHist. 

Next, we consider the {\bf Synthetic Type II} dataset with {\bf Data-dependent Query Model} and vary number of buckets from $10$ to $1000$. Figure~\ref{fig:var_buckets} (c) compares test error incurred by the three methods. Here again, SpHist performs best of the three methods. In particular, for $10$ buckets, SpHist incurs $5.48\%$ error while EquiHist incurs $12.68\%$ error and ISOMER incurs $26.66\%$ error. 

Finally, we consider {\bf Census 1-D} data with queries drawn from {\bf Uniform Query Model}. We vary the number of buckets from $5$ to $50$; note that the range of Age attribute is only $91$. Similar to the above experiments, SpHist incurs significantly less error than ISOMER (by $20\%$) and EquiHist (by $2\%$). 
%\begin{figure}[ht]
%  \centering
%  \includegraphics[width=\columnwidth]{figs/SynSkew_IF_1_Buckets15_Queries200_ITER_0}
%  \caption{Relative error incurred by various methods as range of the attribute in synthetic dataset varies. Clearly, SpHist and EquiHist scales better wi%th increasing range than ISOMER. Also, as expected due to Theorem~\ref{thm:1d}, the error increases at sub-linear rate with increasing range.}
%  \label{fig:var_range}
%\end{figure}

\noindent{\bf Varying Range of Attribute Values}: \\
In the next experiment, we study performance of the different methods for varying range of attribute values. Here, we use {\bf Synthetic Type I} dataset with {\bf Data-dependent Query Model} and fix the number of buckets to $15$, the number of training queries are fixed to be $200$. Figure~\ref{fig:1d} (d) compares error incurred by SpHist and EquiHist to ISOMER on {\bf Synthetic Type I} dataset with queries from {\bf Data-dependent Query Model}. Here again, our methods are significantly better than ISOMER. Also, as predicted by our theoretical results (see Theorem~\ref{thm:1d}), EquiHist does not depend heavily on the range and is able to learn low-error histograms with small number of queries. Similar trends were observed for {\bf Uniform Query Model} and {\bf Synthetic Type II} data as well. 

We summarize our results for $1$-dimensional histogram settings as follows:

\noindent {\bf \textbullet \ \ }Both EquiHist and SpHist converge quicker than ISOMER with respect to number of queries and in general incurs less error for all training queries numbers. %\setlength\parindent{0pt}
%\item 

\noindent {\bf \textbullet \ \ }SpHist incurs significantly less error than EquiHist and ISOMER for ``spiky'' data (Synthetic Type II dataset).
%\item 

\noindent {\bf \textbullet \ \ }EquiHist and SpHist consistently outperform ISOMER with varying number of buckets.
%\item 

\noindent {\bf \textbullet \ \ }SpHist demonstrates a clear advantage over the other two methods for smaller number of buckets. For larger number of buckets, EquiHist incurs less error than SpHist.\\
%\item
\noindent {\bf \textbullet \ \ }Both EquiHist and SpHist scale well with increasing range of the attribute values. 
%\end{compactitem}
\subsection{Multi-dimensional Histograms}
\label{sec:multiD-results}
In this section, we empirically compare our EquiHist and SpHist methods with ISOMER for learning multi-dimensional histograms. For these experiments also, we use synthetic as well Census data. 
%\begin{compactitem}
%\item {\bf Synthetic Datasets}: We generate $2,3$-dimensional datasets for a given range, number of bins, and range by sampling 50000 points from a mixture of spherical Gaussians in the respective number of dimensions. 
%\item {\bf Census Datasets}: We obtain $2$-dimensional dataset using ``Age'', ``Number of Weeks worked'' attributes. We obtain $3$-dimensional dataset by using ``Age'', ``Marital status'' and ``Education''. 
%\end{compactitem}
Also, we use {\bf Data-dependent Query Model} for all the multi-dimensional histogram experiments. 
\begin{figure*}[ht]
  \centering
  \begin{tabular}{cccc}
    \hspace*{-20pt}
    \includegraphics[width=.25\textwidth]{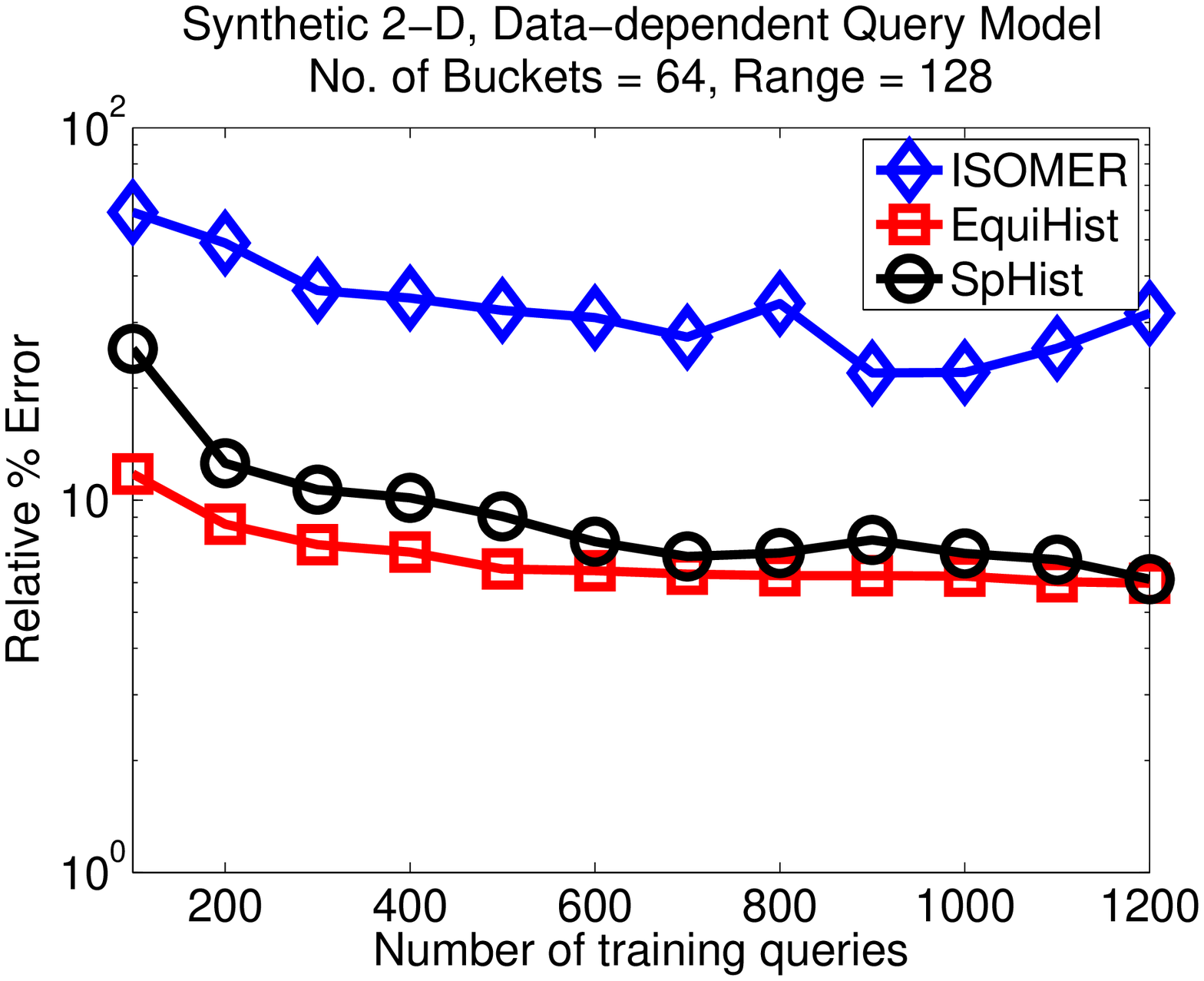}&
    \includegraphics[width=.25\textwidth]{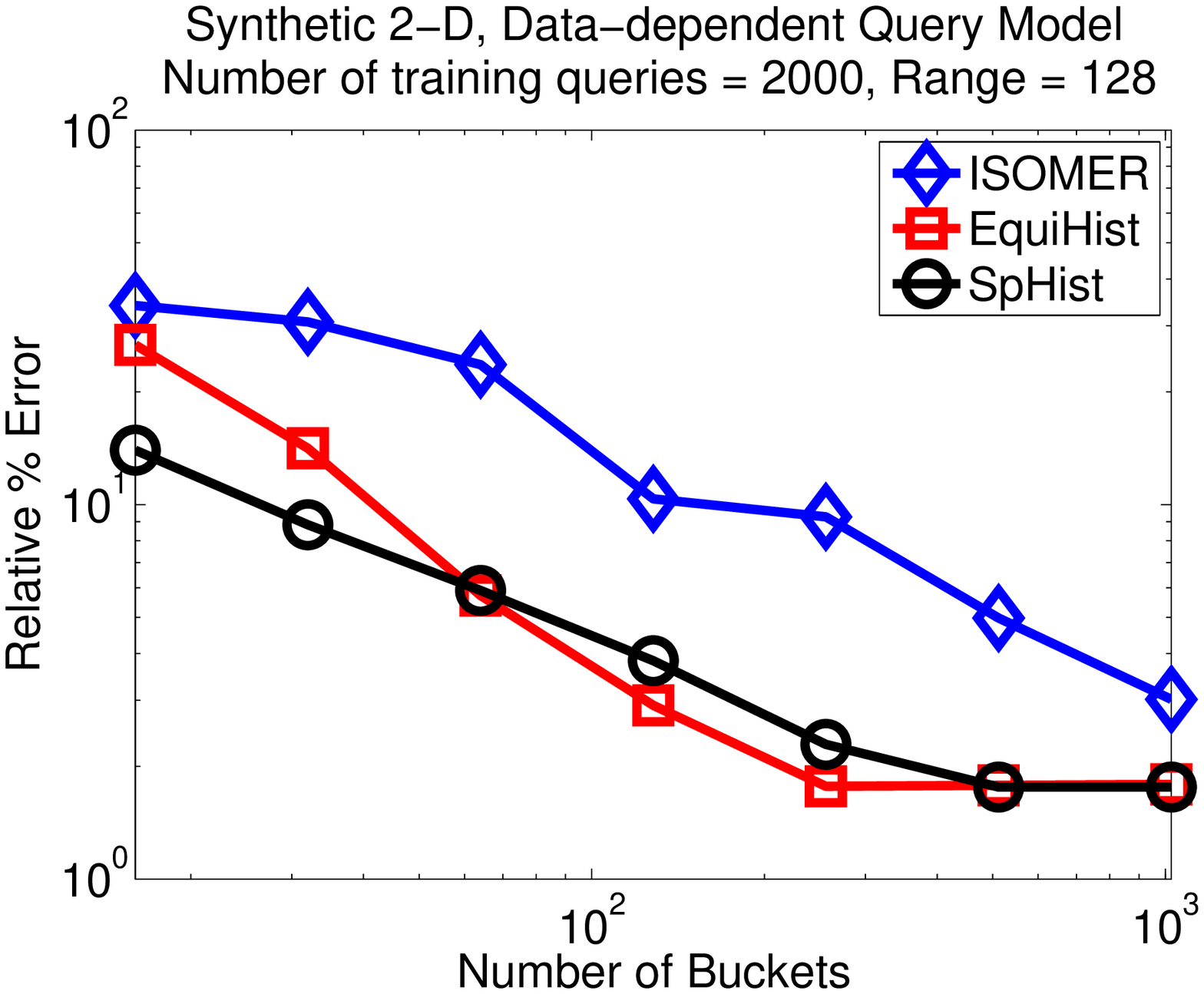}&
    \includegraphics[width=.25\textwidth]{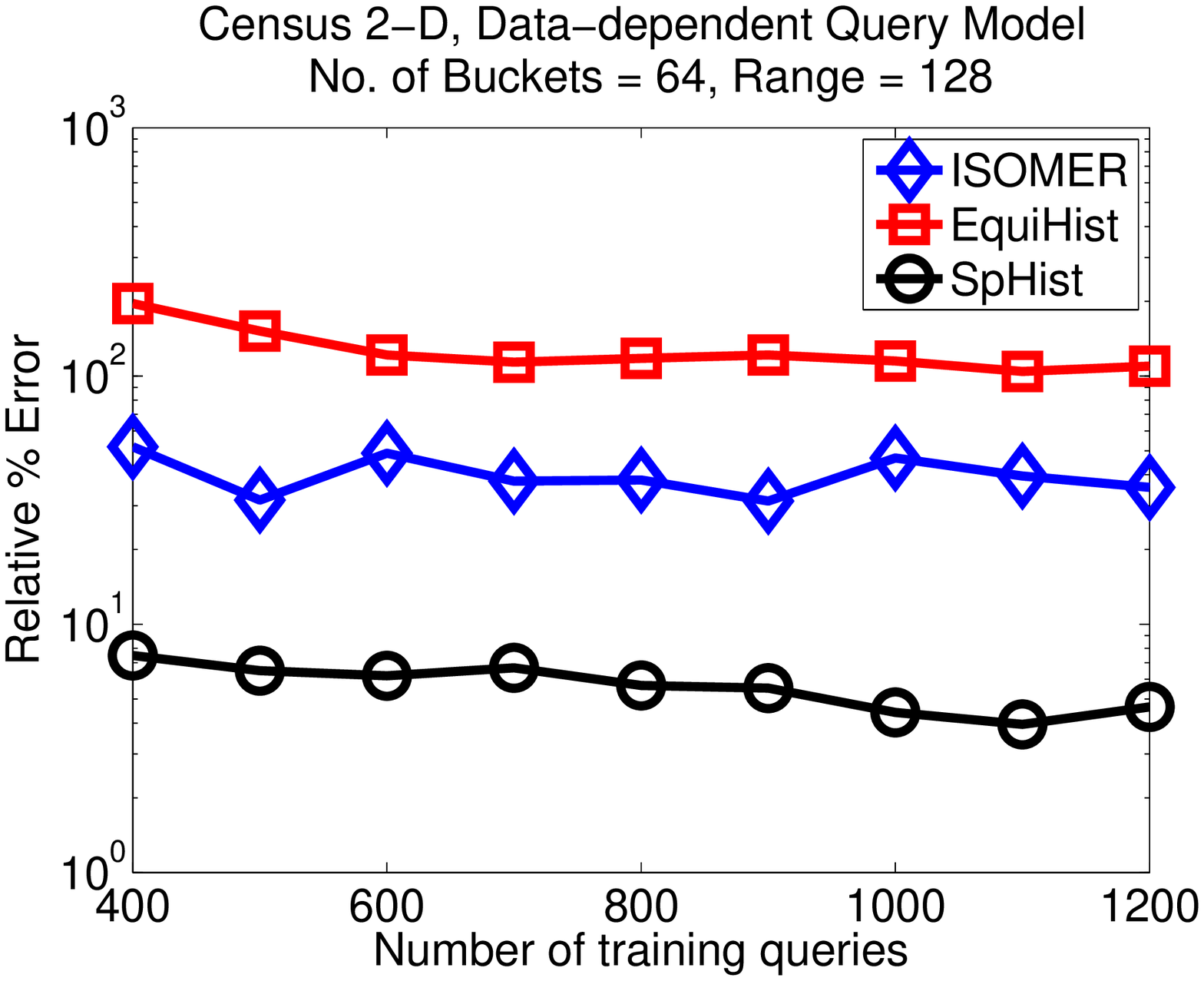}&
    \includegraphics[width=.25\textwidth]{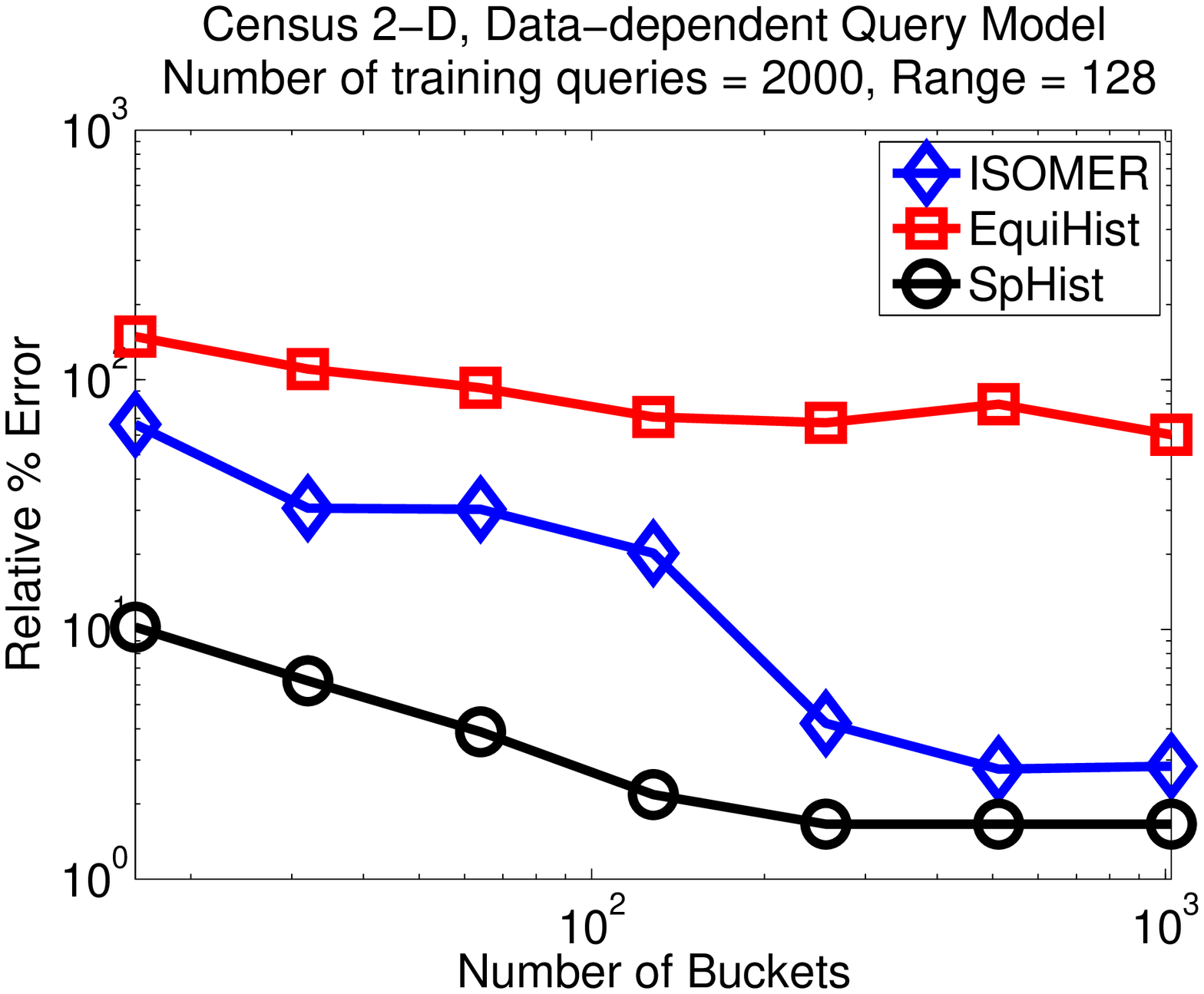}\\
    {\bf (a)}&{\bf (b)}&{\bf (c)}&{\bf (d)}\vspace*{-10pt}
  \end{tabular}
  \caption{Comparison of average relative error (on log-scale) for various methods on two-dimensional datasets with queries generated from {\bf Data-dependent Query model}. {\bf (a)} Test error on {\bf Synthetic 2-D} dataset with varying number of training queries. For $1200$ queries, both SpHist and EquiHist incurs about $26\%$ less error than ISOMER. {\bf (b)} Test error for {\bf Synthetic 2-D} dataset with varying number of buckets. For $16$ buckets, SpHist incurs $13.95\%$ error while EquiHist incurs $26.68\%$ error and ISOMER incurs $33.99\%$ error. {\bf (c)} Test error for {\bf  Census 2-D} data with varying number of training queries. For $1200$ queries, SpHist incurs $4.64\%$ error, while EquiHist incurs $109.54\%$ error and ISOMER incurs $35.55\%$ error. EquiHist incurs more error than both SpHist and ISOMER due to ``spikiness'' of the data (see Figure~\ref{fig:2d_hist} (a)). {\bf (d)} Test error for {\bf  Census 2-D} data with varying number of histogram buckets. For $16$ buckets, SpHist incurs $10.21\%$ error while ISOMER incurs $66.34\%$ error and EquiHist incurs $148.7\%$ error.  Similar to plot (b), EquiHist incurs larger error due to heavily skewed data (see Figure~\ref{fig:2d_hist} (a)).}\vspace*{-10pt}
  \label{fig:2d_var}
\end{figure*}
\begin{figure*}[ht]
  \centering
\begin{tabular}{cccc}
\hspace*{-20pt}
  \includegraphics[width=.25\textwidth]{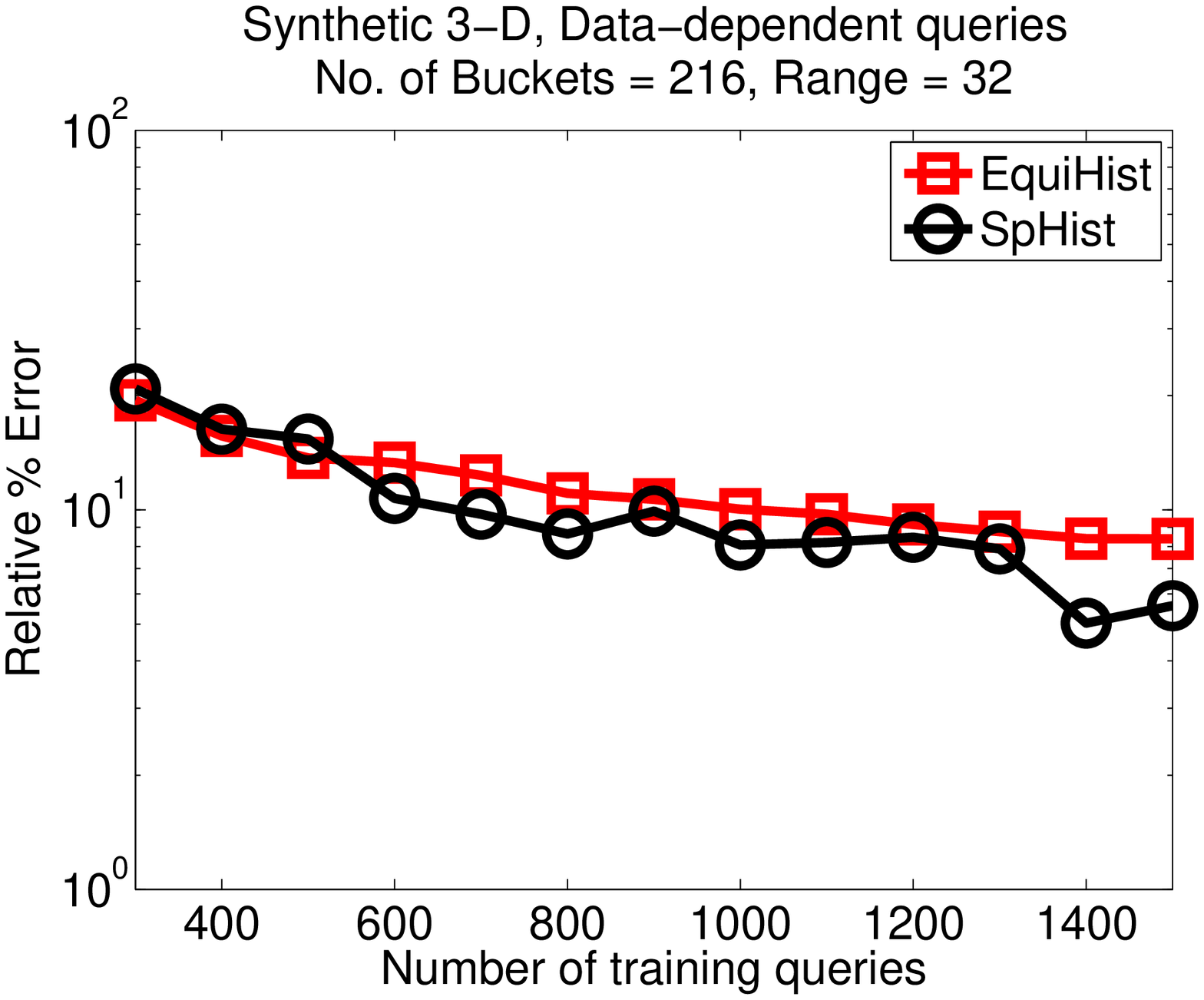}&
  \includegraphics[width=.25\textwidth]{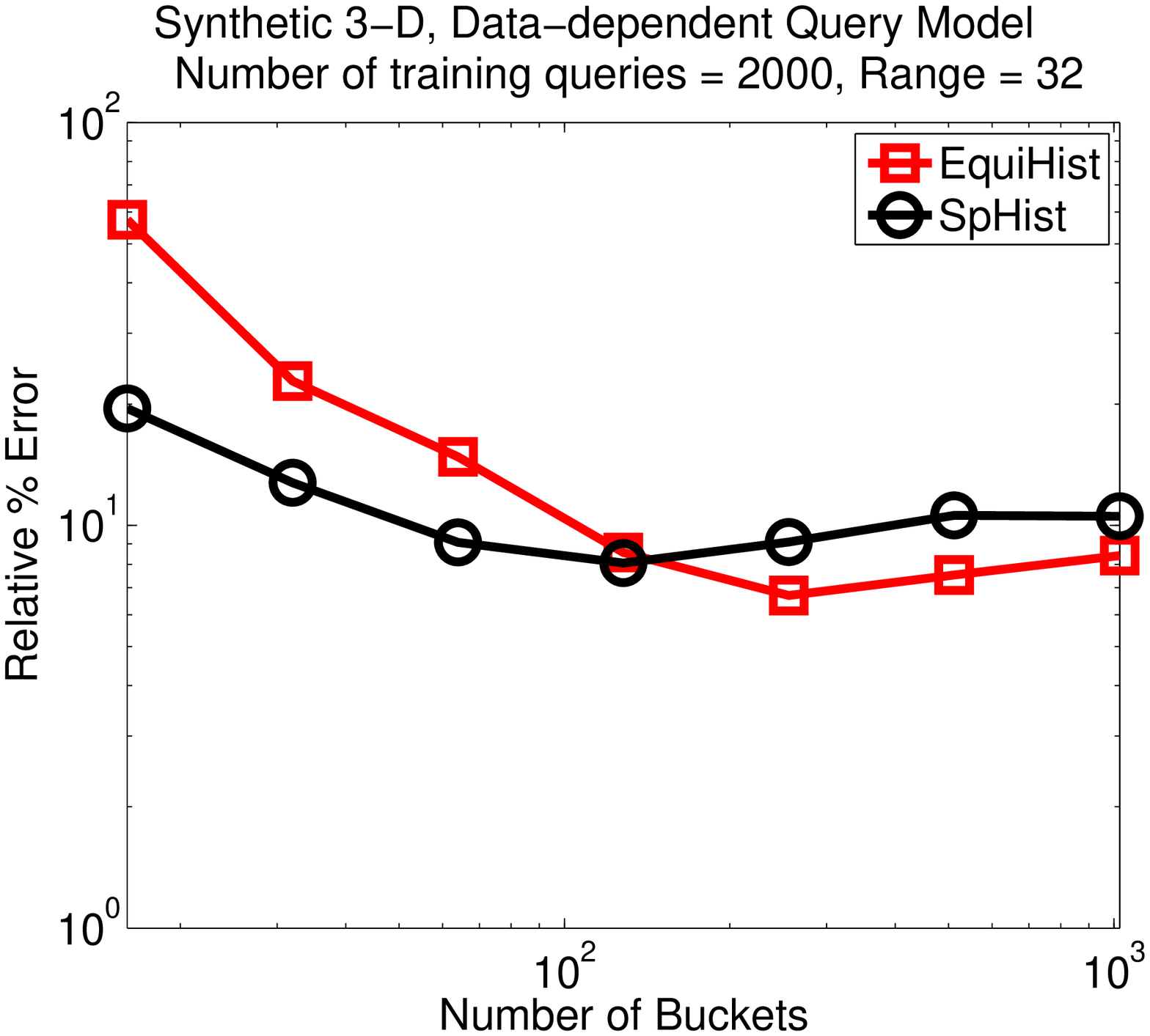}&
  \includegraphics[width=.26\textwidth]{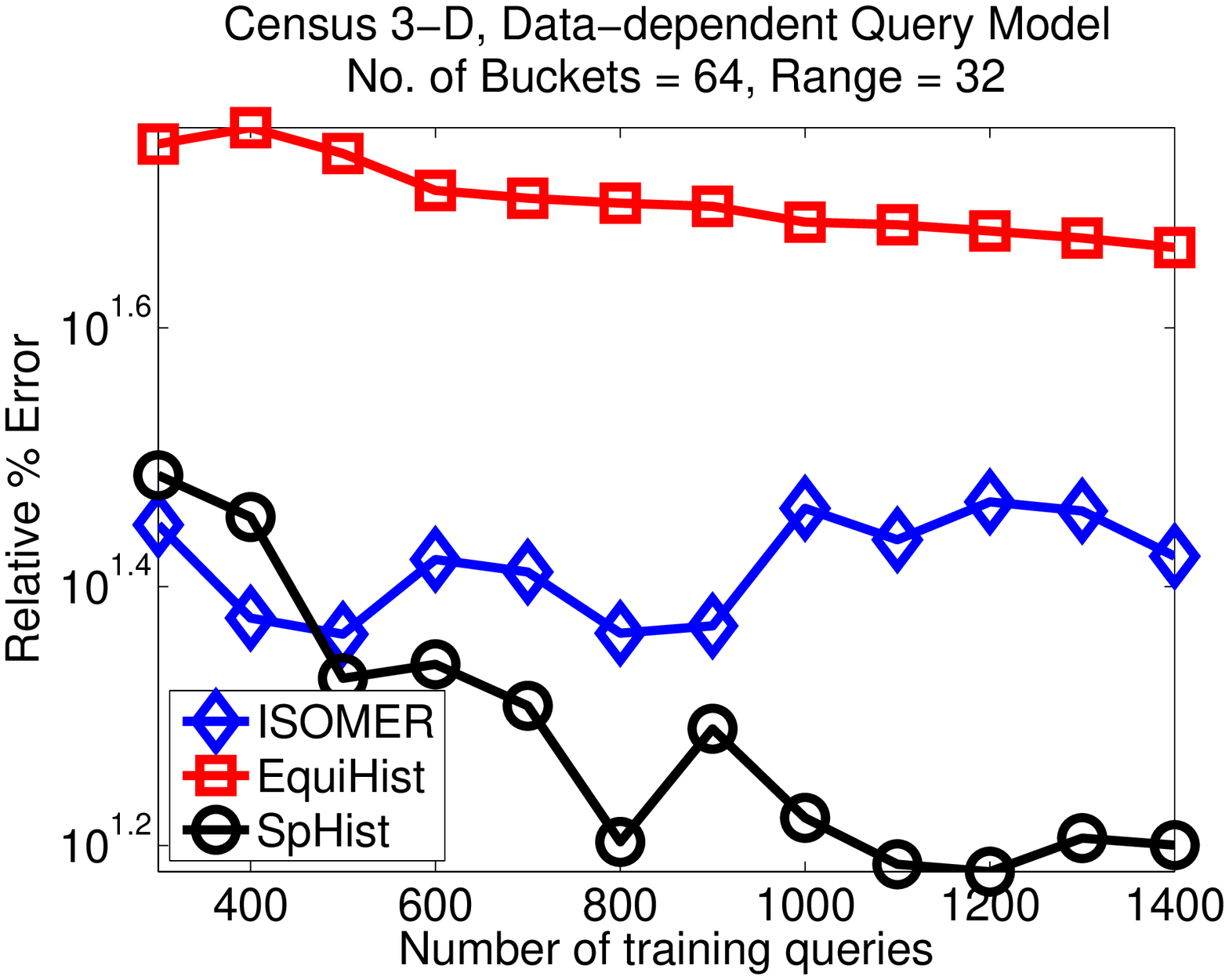} &
  \includegraphics[width=.25\textwidth]{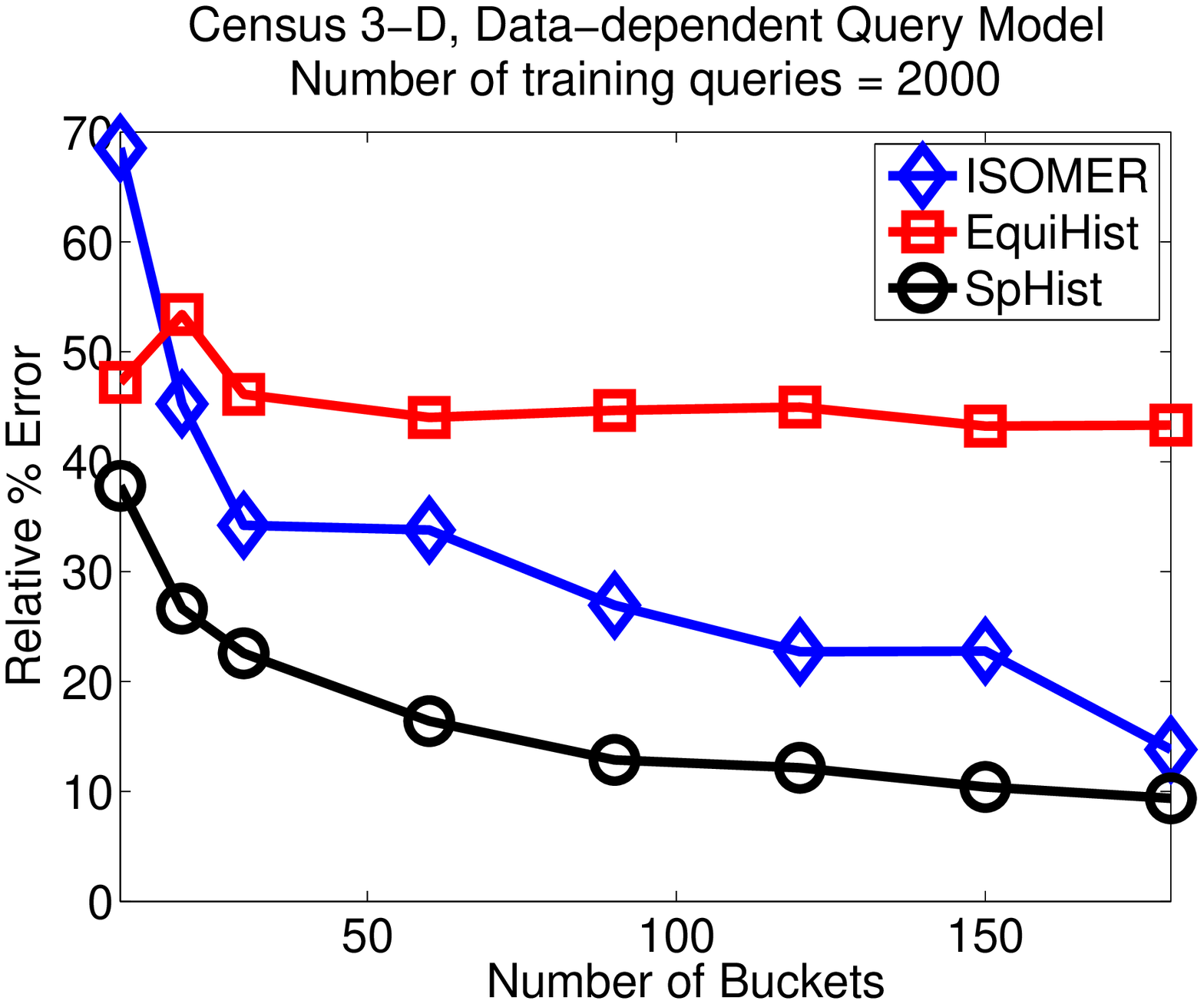} \\
{\bf (a)}&{\bf (b)}&{\bf (c)}&{\bf (d)}\vspace*{-10pt}
\end{tabular}
  \caption{Comparison of average relative error (on log-scale) for various methods on three-dimensional datasets with queries generated from {\bf Data-dependent Query model}. {\bf a)}, {\bf b)} Test error on {\bf Synthetic 3-D dataset} with varying number of training queries and varying number of buckets. Here, both EquiHist and SpHist are able to learn reasonable histograms (incurs about $10\%$ error) and follow similar trends to two-dimensional experiments (see Figure~\ref{fig:2d_var}). We do not report test error for ISOMER, as our implementation of ISOMER did not finish even after two days, {\bf c)}, {\bf d)} Test error on {\bf Census 3-D} dataset with varying number of training queries and number of buckets . Similar to {\bf Census 2-D} dataset, SpHist incurs significantly less error than both EquiHist and ISOMER. For example, in plot (c), for $1500$ queries, SpHist incurs about $11\%$ less error than ISOMER and $30\%$ less error than EquiHist.}
  \label{fig:3d_var}\vspace*{-10pt}
\end{figure*}
\begin{figure*}[ht]
  \centering
  \begin{tabular}{cccc}
    \hspace*{-20pt}
  \includegraphics[width=.25\textwidth]{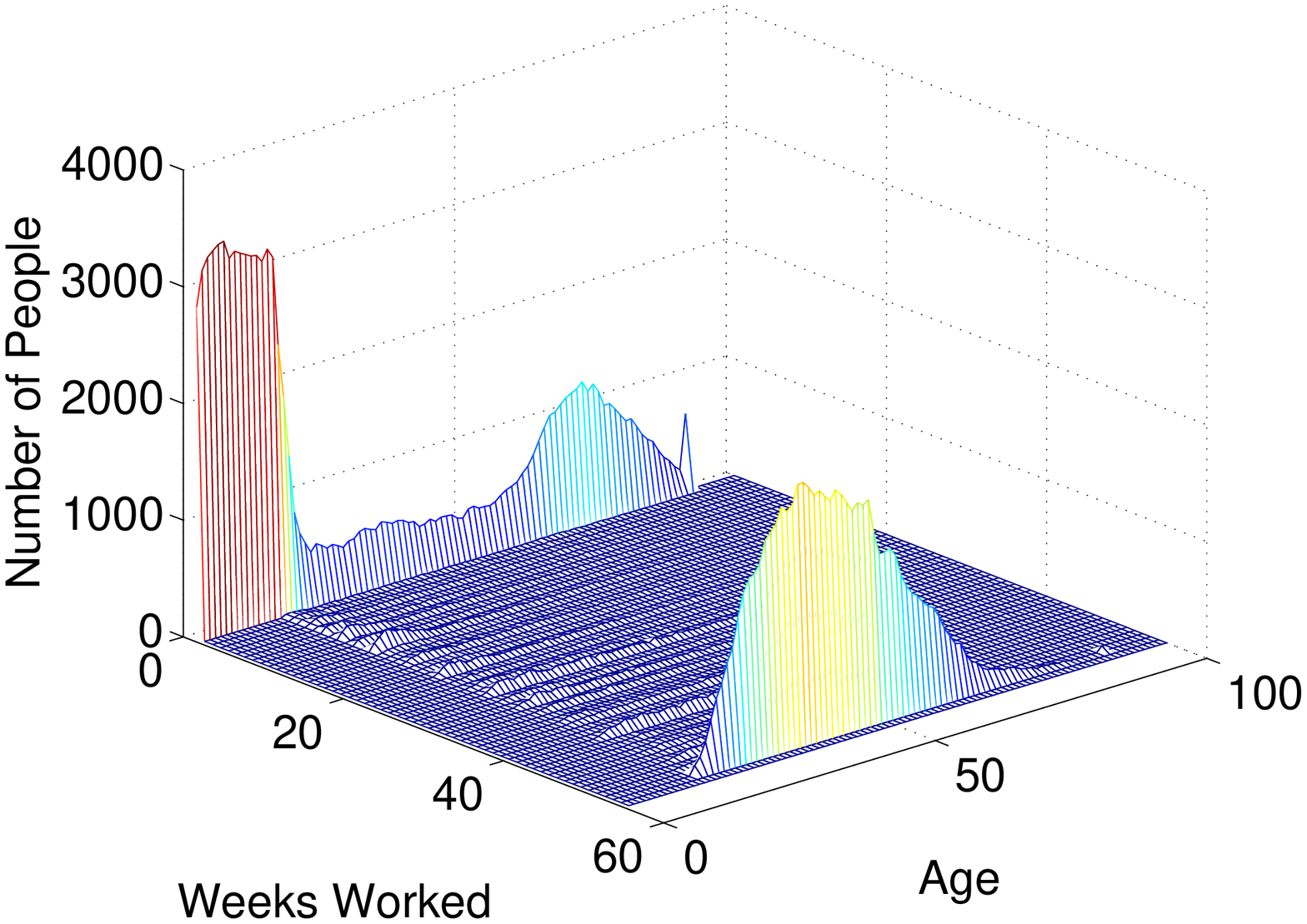}&
  \includegraphics[width=.25\textwidth]{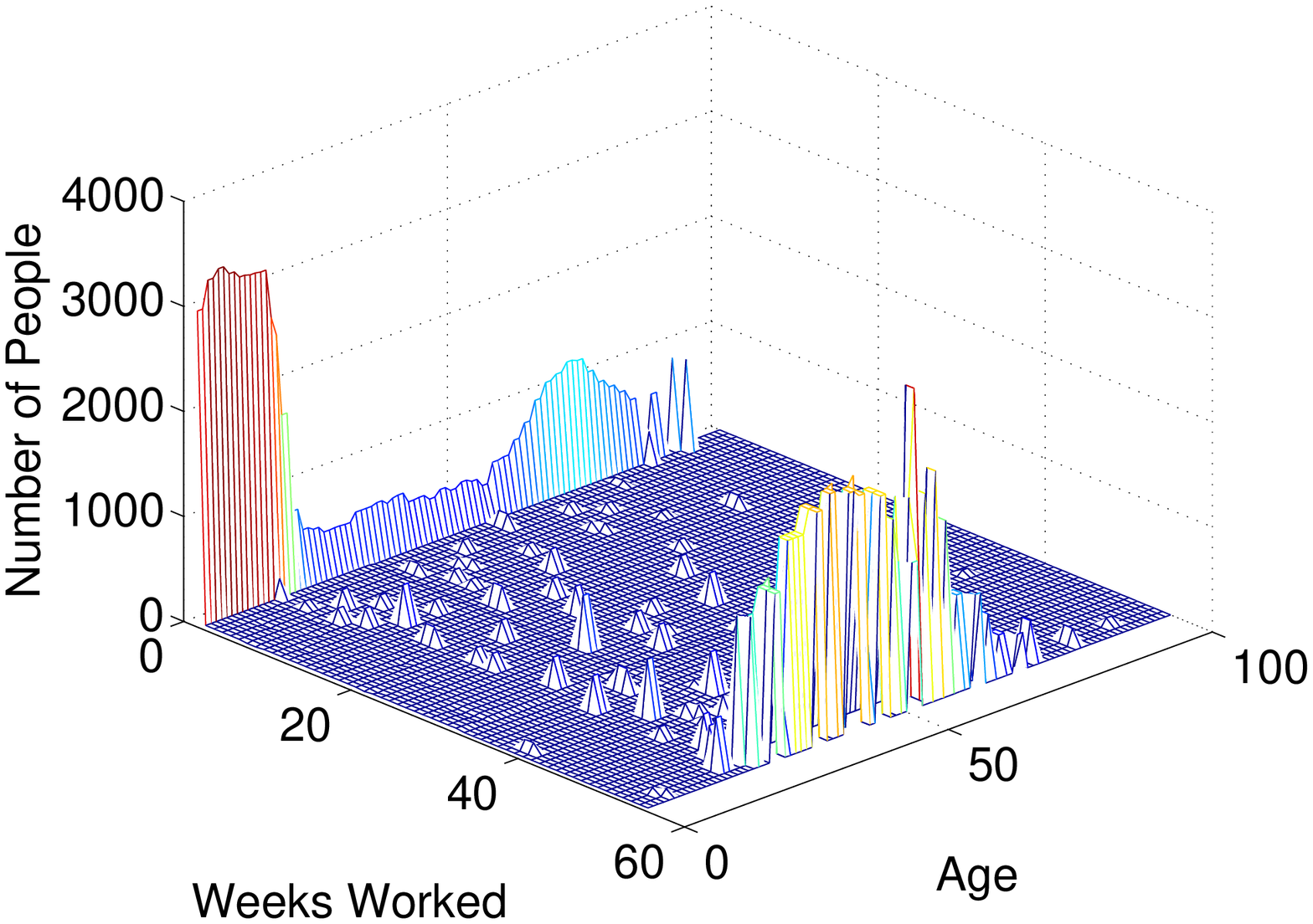}&
  \includegraphics[width=.22\textwidth]{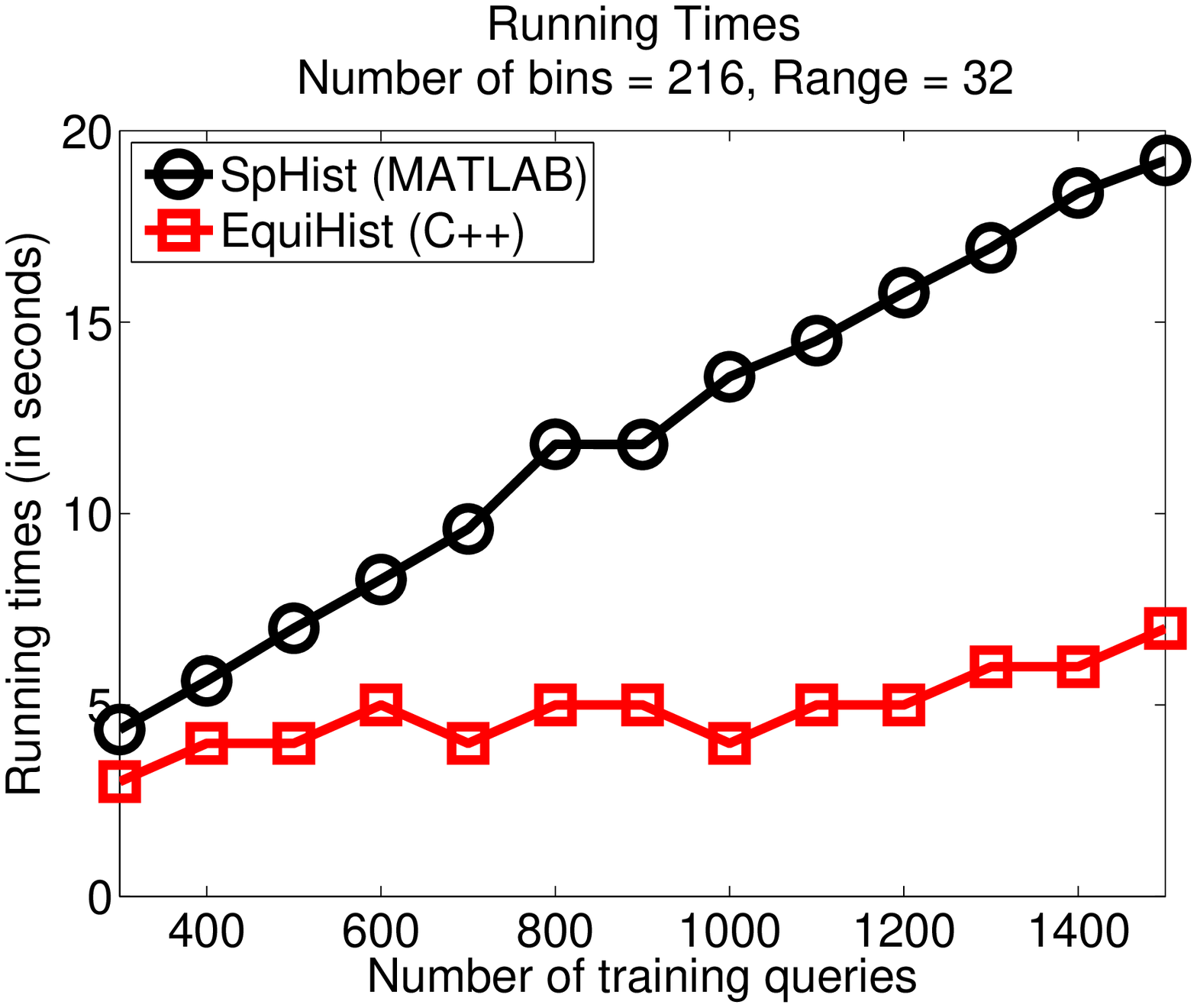}&
  \includegraphics[width=0.23\textwidth]{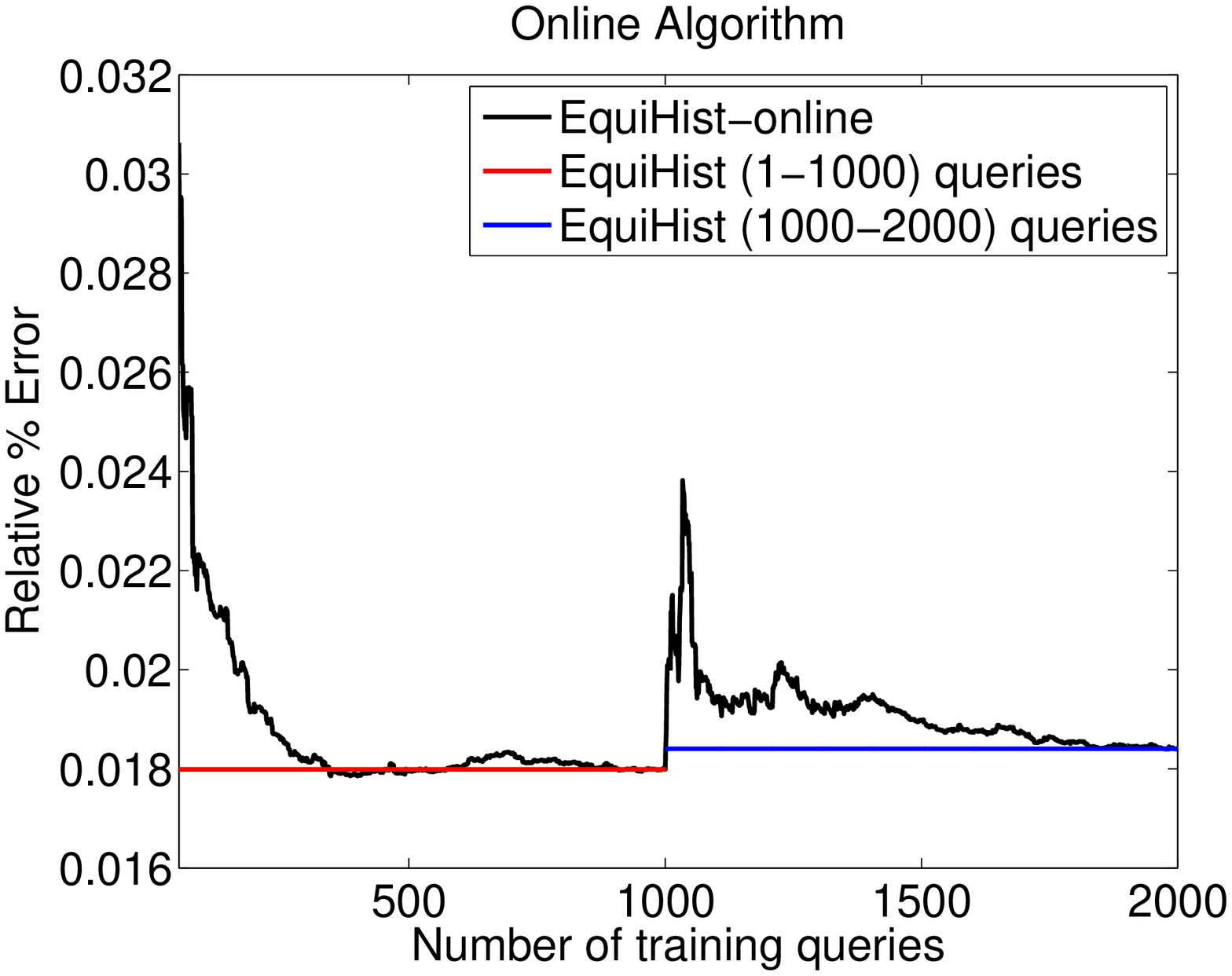}\\
 {\bf (a)}&{\bf (b)}&{\bf (c)}&{\bf (d)}\vspace*{-10pt}
\end{tabular}
  \caption{{\bf a)} Frequency distribution of ``Age'' and ``Number of Weeks Worked'' attributes of {\bf Census $2$-D} data, {\bf b)} Estimated histogram for {\bf Census $2$-D} data using SpHist method with $1000$ training queries and $64$ buckets (See plot (a) for true distribution). SpHist is able to recover the underlying distribution well using small number of buckets and training queries. In particular, the high density areas for ``Weeks Worked'' attribute values $0$ and $52$ are captured accurately, {\bf c)} Running times of SpHist and EquiHist for {\bf Synthetic $3$-D} dataset (on a dual-core 2GHz processor with 4GB RAM). Note that our methods mostly finish with in 30 seconds, while on the same experiment our implementation of ISOMER did not finish in 2 days, {\bf d)} Streaming queries and database updates experiment. Test error incurred by online version of EquiHist as well as batch version for {\bf Synthetic Type I} dataset. At each step, one training query is provided; at $1000$-th step database is updated by randomly permuting $30\%$ of the database. EquiHist (1-1000) represents batch Equihist histogram obtained by training on first $1000$ queries, while EquiHist(1001-2000) trains on the batch of $1001-2000$-th queries.}\vspace*{-10pt}
  \label{fig:2d_hist} 
\end{figure*}
%\begin{figure}[ht]
%  \centering
%  \includegraphics[width=\columnwidth]{figs/rechist5kQ_512bnew}
%  \caption{Estimated histogram for 2-d Census data (``Age'' and ``Number of Weeks Worked'' attributes) using SpHist method with $3000$ training queries and $2048$ buckets (See Figure~\ref{fig:2d_true} for true frequency distribution). Clearly, SpHist is able to recover the underlying distribution well using small number of buckets and training queries. In particular, the high density areas for ``Weeks Worked'' attribute values $0$ and $52$ are captured accurately.}
%  \label{fig:2d_rec_sp}
%\end{figure}

\noindent {\bf Experiments with 2-dimensional Datasets}: 
We first compare our methods to ISOMER for varying number of training queries on {\bf Synthetic 2-D} dataset. Figure~\ref{fig:2d_var} (a) shows the test error obtained by all the three methods for different number of training queries, where queries are generated using {\bf Data-dependent Query Model} and the number of histogram buckets is fixed to be $64$. Clearly, our methods outperform ISOMER and are able to reduce error rapidly with increasing number of training queries. For example, for $1000$ training queries, both SpHist and EquiHist incurs about $6.0\%$ error while ISOMER incurs $31\%$ error. %Also, we could not report error incurred by ISOMER for more than $1000$ queries as our implementation of ISOMER did not terminate even after running for one day. Primary reason being, even in 2-dimensions, number of variables in ISOMER's max-entropy problem becomes large, e.g., for our experiments with $1000$ queries ISOMER had $11,000$ variables in the first round. Furthermore, ISOMER needs to iteratively re-optimize after throwing out a fixed number of queries, thus further increasing run-times. By contrast, both of our methods are very efficient, and need to run the optimization only once. We observed that in all our two-dimensional experiments, our methods terminated in less than $5$ minutes (All methods were run on same machine).

Next, we compare the three methods on {\bf Synthetic 2-D} dataset, while varying number of buckets, with number of training queries is fixed at $2000$ (see Figure~\ref{fig:2d_var} (b)). Here again, both EquiHist and SpHist outperform ISOMER for small number of buckets. For $128$ buckets, ISOMER incurs $10.36\%$ error while EquiHist and SphHist incur around $2.91\%$ and $3.83\%$ error respectively. %EquiHist overfits to the training data when the number of buckets approaches number of training queries. However, SpHist is still able to recover reasonably accurate histogram ($2.48\%$ error for $2048$ queries). 

%In the next set of experiments, we compare the performance of the three algorithms on real-word 2D-data. The {\bf Synthetic 2-D} experiments already show that EquiHist and SpHist achieve significantly lower average relative error when the 2-D data distribution was smooth (mixture of 2-D Gaussians). Similar results were obtained on 2-D real Census data when we picked ``Age'' and ``Salary'' as the 2 attributes, where again the 2-D distribution was relatively smooth over most of the range of attribute values. However, a more interesting 2-D dataset for our study was when we selected ``Age'' and ``Number of Weeks Worked'' as the two attributes from Census data (what we call as {\bf Census 2-D}), since most data points were concentrated around the two extremes of 0 or 53 weeks (see Fig.~\ref{fig:2d_true}). This dataset was able to demonstrate when SpHist substantially outperforms both EquiHist and ISOMER. We report these results below. 

In the next set of experiments, we compare the performance of the three algorithms on real-world {\bf Census 2-D} dataset (see Figure~\ref{fig:2d_var} (c)). Recall that {\bf Census 2-D} dataset projects Census data on ``Age'' and ``Number of Weeks Worked'' attributes. Now, for most database records,  ``Number of Weeks Worked'' are concentrated around either $0$ or $53$ weeks. That is, the data is extremely ``spiky''. However, EquiHist still tries to approximate the entire space using equi-width buckets leading to several ``empty'' buckets. Consequently, EquiHist incurs large error ($109\%$ for $1200$ queries), while ISOMER also incurs $35.55\%$ error. However, SpHist is still able to approximate the underlying distribution well and incurs only $4.64\%$ error. Figure~\ref{fig:2d_hist} (b) shows the histogram estimated by SpHist with $1000$ queries and $64$ buckets. Clearly, SpHist is able to capture high-density regions well; there are small peaks in low-density areas which contribute to the error that SpHist incurs.

%significantly more accurate than both ISOMER and EquiHist. For example, for $1200$ training queries, SpHist incurs $3.98\%$ error while EquiHist incurs $18.94\%$ error and ISOMER incurs $68.54\%$ error. While most data points are concentrated along only a few ``Weeks Worked'' attribute values, EquiHist tries to place fixed-sized (large) buckets and hence incurs large error. By contrast, SpHist is able to fix accurate bin boundaries and hence leads to accurate histograms. For example, Figure~\ref{fig:2d_rec_sp} shows the histogram estimated by SpHist with $1000$ queries and $64$ buckets. Clearly, SpHist is able to capture high-density regions well; there are small peaks in low-density areas which contribute to the error that SpHist incurs. 
%Figure~\ref{fig:2d_var} (c), compares the three methods on the {\bf Census 2-D} dataset. Clearly, 

Finally, we report test error incurred by the three methods on {\bf Census-2D} dataset in Figure~\ref{fig:2d_var} (d), as we vary the number of buckets from $5$ to $2048$, while number of training queries is fixed at $2000$. For this data as well, SpHist outperforms both EquiHist and ISOMER significantly, especially for small number of buckets. Specifically, for $16$ buckets, SpHist incurs only $10.21\%$ relative error while  EquiHist and ISOMER incur about $148.7\%$ and $66.34\%$ relative error, respectively. \\[2pt]
{\bf 3-dimensional Datasets}: We now consider $3$-dimensional datasets to study scalability of our methods with increasing dimensions. 

We first conduct experiments on the {\bf Synthetic 3-D} dataset, which is drawn from a mixture of spherical 3-D Gaussians.  Figure~\ref{fig:3d_var} (a) shows relative error incurred by our methods when queries are generated from {\bf Data-dependent Query Model} and the number of training queries vary, while number of buckets is fixed at $216$. For $2000$ queries, SpHist incurs $5.59\%$ error while EquiHist incurs $8.39\%$ error. Note that we are not able to report error incurred by ISOMER as our implementation of ISOMER did not terminate even after running for two days. %; number of variables in the first max-entropy problem for ISOMER was approximately $27,000$, hence each iteration (each max-entropy problem) of ISOMER ran for a very long time. In contrast, both of our methods produced results in under $1$ minute. 
Primary reason being, even in 3-dimensions, number of variables in ISOMER's max-entropy problem becomes large, e.g., for our experiments with $1000$ queries ISOMER had $27,000$ variables in the first round. Furthermore, ISOMER needs to iteratively re-optimize after throwing out a fixed number of queries, thus further increasing run-times. In contrast, both of our methods are very efficient, and need to run the optimization only once. We observed that in all our experiments, our methods terminated in less than $1$ minutes (see Figure~\ref{fig:2d_hist}(c)).

Next, we vary number of buckets while number of training queries is fixed to be $2000$. Figure~\ref{fig:3d_var} (b) shows relative error incurred by our methods. Clearly, for small number of buckets SpHist is significantly better than EquiHist. For example, for $16$ buckets SpHist incurs $19.51\%$ error while EquiHist incurs $57.36\%$ error. %For $1024$ buckets, number of buckets is larger than number of training queries, leading to overfitting by EquiHist; SpHist incurs $14.76\%$ error in this case. 

Finally, we repeat these experiments on our {\bf Census 3-D} dataset (which, considers the ``Age'', ``Marital Status'' and ``Education'' as the 3 attributes), i.e., we vary number of training queries (Figure~\ref{fig:3d_var} (c)) as well as number of buckets (Figure~\ref{fig:3d_var}(d)).  Overall, we observe the same trend as 2-dimensional case with SpHist being most accurate and EquiHist being most inaccurate, e.g., for $200$ buckets SpHist incurs $7\%$ error while ISOMER incurs $20\%$ error and EquiHist incurs $38\%$ error.  %First, we fix the number of buckets to be $64$ while varying the number of training queries. Figure~\ref{fig:3d_var} (c) plots relative error incurred by the three methods and overall reflects same trends as 2-dimensional Census dataset. In particular, SpHist is significantly more accurate than ISOMER, while EquiHist incurs the highest error. Next, we fix the number of trianing queries to be $3000$ while varying number of buckets. Here again, trends are similar to the 2-dimensional case with SpHist obtaining smallest error (for $200$ buckets SpHist incurs $7\%$ error while ISOMER incurs $20\%$ error and EquiHist incurs $38\%$ error) and EquiHist being most inaccurate. 
% We also, present similar plot for $4$-dimensional synthetic and Census dataset; however, ISOMER has trouble scaling to these datasets as the number of ``buckets'' increases rapidly for ISOMER and even after $30$-minutes ISOMER didn't produce any result. 
%Next, Figure~\ref{fig:md_b_synth_unif} shows the error obtained by both ISOMER as well as EquiHist for varying number of buckets. Here again, we see similar trend. Similarly, Figure~\ref{fig:md_r_synth_unif} shows EquiHist scales better than ISOMER as range of the dataset increases. 
\subsection{Dynamic QFRs and data updates} 
\label{sec:results_streaming}
We now study the performance of online version of EquiHist
(Section~\ref{sec:online}) for dynamic QFRs and in the presence of
data updates. (The performance of online SpHist is similar when the
set $\Scal$ of non-zero wavelet coefficients is kept unchanged.) The
goal is to show that our online updates are effective, converge to the
optimal batch solution quickly, and are robust to database updates.

%In Subsection~\ref{sec:online}, we discussed a streaming query model where at each step a training query is given and we provided fast updates for our histograms using the latest query. In this experiment, the goal is to show that our online updates are effective and converge to the optimal batch solution quickly. Furthermore, we also study performance of our online learning algorithm when database is updated. 

For this experiment, we consider {\bf Synthetic Type I} dataset with queries generated from the {\bf Uniform Query Model}. We compare our online version of EquiHist against the batch-version of EquiHist. After each update of the online version, we measure relative error on $5000$ test queries. We also, measure relative test error incurred by the {\em batch EquiHist method} (that can observe all the training queries beforehand). Figure~\ref{fig:2d_hist} (d) compares relative error incurred by online EquiHist to batch method. For $1$ to $1000$ steps, the database remains the same. Clearly, online learning algorithm quickly (in around $250$ steps) converges to relative error similar to batch method (red line) with $1000$ training queries. 

Now at $1000$-th step, we update our database by randomly perturbing $30\%$ of the database. This leads to larger error for online EquiHist  for a few steps after $1000$-th step (see Figure~\ref{fig:2d_hist} (d)), however it quickly converges to the optimal batch solution (blue line) for the updated database (using queries $1001$ to $2000$). 
%\begin{figure}[h]
%  \begin{center}
%    \includegraphics[width=0.4\textwidth]{figs/online_experiment}
%  \end{center}
%  \caption{Streaming queries and database updates experiment. Plot shows relative error incurred by online version of EquiHist as well as batch version. A%t each step, one training query is provided; at $1000$-th step database is updated by randomly permuting $20\%$ of the database. EquiHist (1-1000) represe%nts batch Equihist solution obtained by simultaneously training on first $1000$ queries, while EquiHist(1001-2000) denotes the batch EquiHist solution usi%ng $1001-2000$ queries.}
%  \label{fig:online}
%\end{figure}
\subsection{Comparison to ISOMER}
Here, we summarize our experimental results and discuss some of the advantages that our method enjoys over ISOMER. 

\noindent {\bf \textbullet \ } ISOMER reduces the number of histogram buckets by removing queries. Hence for small number of buckets, ISOMER removes almost all the queries and ends up over-fitting to a few queries, leading to high relative error. In contrast, our methods use all the queries and hence are able to generalize significantly better. 

%\noindent {\bf \textbullet \ } ISOMER reduces histogram buckets by sequentially removing queries whose corresponding dual variables are the smallest. Now, dual problem of ISOMER is not a strongly convex problem, hence optima of dual of ISOMER is not unique and hence this step becomes non-deterministic and ad-hoc. That is, implementation bias might lead to very different results for this step. Furthermore, the step greedily reduces the number of buckets and often leads to sub-optimal solutions. In contrast, our methods restrict the number of buckets in a principled manner, hence leading to theoretical guarantees as well as better empirical results.  

\noindent {\bf \textbullet \ } ISOMER has to re-run max-entropy solver after each query-reduction step, hence time required by ISOMER is significantly larger than our methods. 

\noindent {\bf \textbullet \ } For multi-dimensional case, ISOMER's data structure forms large number of buckets even though the final number of buckets required is small. For this reason, ISOMER doesn't scale well to high-dimensions; in contrast, our method apriori fixes the number of buckets and hence scales fairly well with high-dimensions. 

\noindent {\bf \textbullet \ } Database updates lead to inconsistent QFRs, which which needs to be thrown away by ISOMER in a heuristic manner. In comparison, our methods easily extend to database updates. 
%%% Local Variables: 
%%% mode: latex
%%% TeX-master: "histestimation-main"
%%% End: 

\section{Related Work}
\label{sec:related}

Simplicity and efficiency of histograms have made them the choice
data-structure to summarize information for cardinality estimation, a
critical component of query optimization. Consequently, histograms are
used in most commercial database systems. Most of the prior work on
histograms has focused on constructing and maintaining histograms from
data, and we refer the reader to \cite{Ioannidis03} for a survey.

% With increase in databases sizes and dynamic nature of modern
% databases, self-tuning histograms have become popular tool for
% selectivity estimation. Self-tuning histograms are formed just using
% query workload and query answers, and typically do not scan the entire
% database. This has several advantages over traditional histograms
% constructed using a scan of complete database: 1) more accurate for
% dynamic datasets where fixed histograms might become obsolete and
% hence inaccurate quickly, 2) tuned to query workload, hence are more
% accurate and efficient for future queries, 3) efficient for
% multi-dimensional databases, as typically queries for
% multi-dimensional databases are ``sparse'', i.e., cover few attribute
% values, 4) better suited for remote databases or sensitive databases,
% for which direct access to databases is not possible.  See
% \cite{ChaudhuriN07} for a more detailed discussion on advantages of
% self-tuning histograms.

Self-tuning histograms were first introduced by \cite{AboulnagaC99} to
exploit workload information for more accurate cardinality estimation.
The method proposed in \cite{AboulnagaC99} is typically referred to as
STGrid and it merges and splits buckets according to bin
densities. However, it does not have any known provable
bounds on the expected relative error and is restricted by the grid
structure for high-dimensional cases. \cite{BrunoCG01} introduced
STHoles data-structure which is significantly more powerful than the
simple grid structure used by STGrid.  However, this method requires
fine-grained query feedback corresponding to each bucket in the
current histogram and therefore imposes a nontrivial overhead while
collecting feedback. Furthermore, if the number of queries is small then
STHoles is not ``consistent'', i.e., different ways of constructing
data-structure can lead to different query cardinality estimations .
This consistency problem was addressed by \cite{SrivastavaHMKT06} who
designed a maximum entropy based method called ISOMER. ISOMER uses a
data-structure similar to STHoles, but learns frequency values in each
bucket using a maximum-entropy solution. Note that, ISOMER is
considered to be the state-of-the-art method for self-tuning
histograms \cite{ChaudhuriN07} and hence we compare against the same
both theoretically as well as empirically. Recent work
\cite{KaushikS09, ReS10} has extended to maximum entropy approach to
handle feedback involving distinct values. \cite{KaushikS09} also
present a histogram construction that is similar to EquiHist, but this
algorithm only handles 1-dimensional histograms. Self-tuning
histograms are part of a larger effort that seeks to leverage
execution feedback for query optimization~\cite{MarklLR03}.

Wavelets are a popular signal-processing tool for compressing
signals. Haar wavelets are one of the most popular and simple wavelets
that are especially effective for piecewise constant signals.  As
histograms are piecewise constant signals, Haar wavelets are
extensively used in the context of databases. Specifically,
\cite{MatiasVW98} introduced a wavelet based histogram that can be
used for selectivity estimation. Similarly, \cite{ChakrabartiGRS01}
also defined a method for selectivity estimation using
wavelets. \cite{GarofalakisG02} introduced a probabilistic method to
decide the wavelet coefficients to be used and provides error
guarantees for the same.  However, most of these methods compute
wavelets coefficients using a complete scan of the database and are
not self-tuning. In contrast, we introduce a method that uses
sparse-vector recovery techniques to learn appropriate wavelet
coefficients for estimating self-tuning histograms.

%%% Local Variables: 
%%% mode: latex
%%% TeX-master: "histestimation-main"
%%% End: 

\section{Conclusions}
\label{sec:conclusion}
In this paper, we introduced a learning theoretic framework for the problem of self-tuning histograms. We cast the problem in an empirical loss minimization framework, and propose two different approaches in this framework. Our first approach (EquiHist) efficiently learns well-known equi-width histograms. We also show that the equi-width approach, despite limitations, can still solve our histogram estimation problem up to an additive approximation factor while  requiring only a finite number of training queries. To the best of our knowledge, this is the first theoretical guarantee for equi-width histograms in the context of self-tuning histograms.  %This reduced the problem to a simple convex optimization which can be solved efficiently. Furthermore, surprisingly, we show that equi-binning approach despite limitations is still able to solve our histogram estimation problem upto an additive approximation factor while only requiring a finite number of training queries. 

However, in high-dimensions where data is sparse or for ``spiky'' datasets, equi-width approach suffers 
as it wastes many buckets on empty region. Our second approach
(SpHist) handles this problem where, by using Haar wavelet transform, we
cast the problem as that of learning a sparse vector. Next, we adapt
the popular Orthogonal Matching Pursuit (OMP) method \cite{TroppG07} for solving the
transformed problem. 

%However, theoretical analysis of the method for our problem is not straightforward and is left as work for future research.

Both of our techniques can be easily extended to multi-dimensional
settings, dynamic QFRs and database updates. To
demonstrate effectiveness of our methods in all these scenarios, we
provide a variety of empirical results. Overall, our empirical results
show that SpHist is consistently better than EquiHist as
well as ISOMER, especially for multi-dimensional datasets as well as
for small number of buckets---both critical parameters for real-world
databases. For example, SpHist is able to recover back the true distribution reasonably well for {\bf Census 2-D} dataset (see Figure~\ref{fig:2d_hist}(b)). 

% for is significantly better than both of our approaches (EquiHist and SpHist) outperform ISOMER \cite{SrivastavaHMKT06} significantly for both synthetic and real-world Census data in $1$-dimensions. For multi-dimensional case, as expected EquiHist suffers on real-world datasets, but our SpHist technique significantly outperforms both ISOMER and EquiHist on all the datasets and scenarios considered. We also empirically demonstrate effectiveness of our online extension of EquiHist method, in particular we show that even with streaming queries our online method quickly converges to batch method's solution and furthermore, if database is updated then our methods can adapt to the new database reasonably quickly.

For future work, we intend to work on theoretical analysis of SpHist. Another interesting direction is further study of the multi-dimensional buckets output by SpHist, so as to further improve its efficiency. Finally, we intend to apply our techniques to real-world query workloads. 
%%% Local Variables: 
%%% mode: latex
%%% TeX-master: "histestimation-main"
%%% End: 

\bibliographystyle{abbrv}
\bibliography{histestim}  
\newpage
\clearpage 
\appendix
\section{Proofs of Equi-width Approach}
\subsection{Proof of Theorem~\ref{thm:1d}}
\label{sec:proof1d}
\begin{proof}
Now, 
\begin{align}
  F(\hh)-F(\hvec^*)&= F(\hh)-F(\tilde{\hvec})+F(\tilde{\hvec})-F(\hvec^*), \\
&=E_1+E_2,
\label{eq:tb}
\end{align}
where $\tilde{\hvec}\in \Ccal'$ is given by $\tilde{\hvec}=B\tilde{\wvec}$, $\tilde{\wvec}$ being the optimal solution to \eqref{eq:g}, $E_1= F(\hh)-F(\tilde{\hvec})$ is the excess generalization error incurred by $\hh$ compared to the optimal solution $\thh$ and $E_2=F(\tilde{\hvec})-F(\hvec^*)$ is the difference between optimal error achievable by histograms in $\Ccal'$ to the histograms in $\Ccal$. Intuitively, $E_2$ measures how expressive set $\Ccal'$ is w.r.t. $\Ccal$. Below we bound both the error individually and finally combine the two errors to obtain error bound on $F(\hh)-F(\hvec^*)$. \\
{\bf Bound on $E_1$}: Let $\thh=B\tw$. Now, since $G_\Dcal(\cdot)$ is the regularized expected risk under a convex risk function ($f(\qvec^TB\wvec;s_q)$) we can use standard results from stochastic convex optimization to bound the generalization error. In particular, using Theorem~\ref{thm:sco} by \cite{Shalev-ShwartzSSS09}, with probability $1-\delta$:
\begin{multline}
  \label{eq:e1_1}
  \mathbb{E}_{\qvec\sim \Dcal}[f(\qvec^TB\hw; s_\qvec)]-\lambda H(\frac{r}{Mb}\hw)\leq \mathbb{E}_{\qvec\sim \Dcal}[f(\qvec^TB\tw; s_\qvec)]\\-\lambda H(\frac{r}{Mb}\tw) + O\left(\frac{\Omega^2L_f^2M^2b\log\frac{1}{\delta}}{r\Delta\lambda N}\right),
\end{multline}
where:
\begin{compactitem}
  \item $\Omega=\max_{\qvec\sim \Dcal}\|B^T\qvec\|$. Assuming each query covers only $|Q|$ attribute values, $\Omega=|Q|$. 
  \item $L_f$ is the Lipschitz constant of function $f(u;s_q)$ w.r.t. $u$. Note that as $\Kcal$ is a compact set, $L_f\leq \max_{u\in \Omega\|\Kcal\|_2}\|\grad f(u)\|_2$. 
  \item $\lambda>0$ is a constant. Also, note that entropy function $H(\frac{r}{Mb}\wvec)$ is $\frac{r\Delta}{M^2b}$-strongly convex. 
\end{compactitem}

Now, $H(\frac{r}{Mb}\hw)\leq \log b$. Hence, using \eqref{eq:e1_1},
\begin{align}
F(\hw)-F(\tw)&\leq \lambda \log b+O\left(\frac{|Q|^2L_f^2M^2b\log \frac{1}{\delta}}{r\Delta\lambda N}\right),\nonumber\\
&\leq O\left(\sqrt{\frac{b\log b\log \frac{1}{\delta}}{r\Delta N}}M|Q|L_f\right),
\label{eq:e12}
\end{align}
where the second inequality follows by selecting $$\lambda=\sqrt{\frac{b \log \frac{1}{\delta}}{r\log b\Delta N}}M|Q|L_f.$$

That is, 
\begin{multline}
\label{eq:be1}
E_1=F(\hh)-F(\thh)=F(\hw)-F(\tw)\\\leq O(\sqrt{\frac{b\log b\log \frac{1}{\delta}}{r\Delta N}}M|Q|L_f).
\end{multline}
{\bf Bound on $E_2$}: To bound $E_2$, we first observe that $$F(\thh)\leq F(\hvec)+\lambda \log b, \forall \hvec\in \Ccal'.$$
Hence, 
\begin{equation}\label{eq:be21}
E_2\leq F(\hvec)-F(\hvec^*)+E_1, \forall \hvec\in \Ccal'.
\end{equation}
Now, given any $\hvec^*$ we construct a new vector $\bh \in \Ccal'$ for which we can bound the error $F(\bh)-F(\hvec^*)$. 

Now, $\bh$ has $b$ buckets and each the value in each bucket ( i.e., $\bw_i, 1\leq b$) is the average of the histogram heights of $\hvec^*$ in that particular bucket. Formally, 
\begin{equation}
  \label{eq:hvect}
  \bw_i=\frac{1}{r/b}\sum_{j=\frac{r}{b}(i-1)}^{\frac{r}{b}\cdot i}\hvec^*_{j}.
\end{equation}
See Figure~\ref{fig:convert} for an illustration of our conversion scheme from $\hvec^*$ to $\bh$. 
\begin{figure}[ht]
  \centering
  \includegraphics[width=\columnwidth]{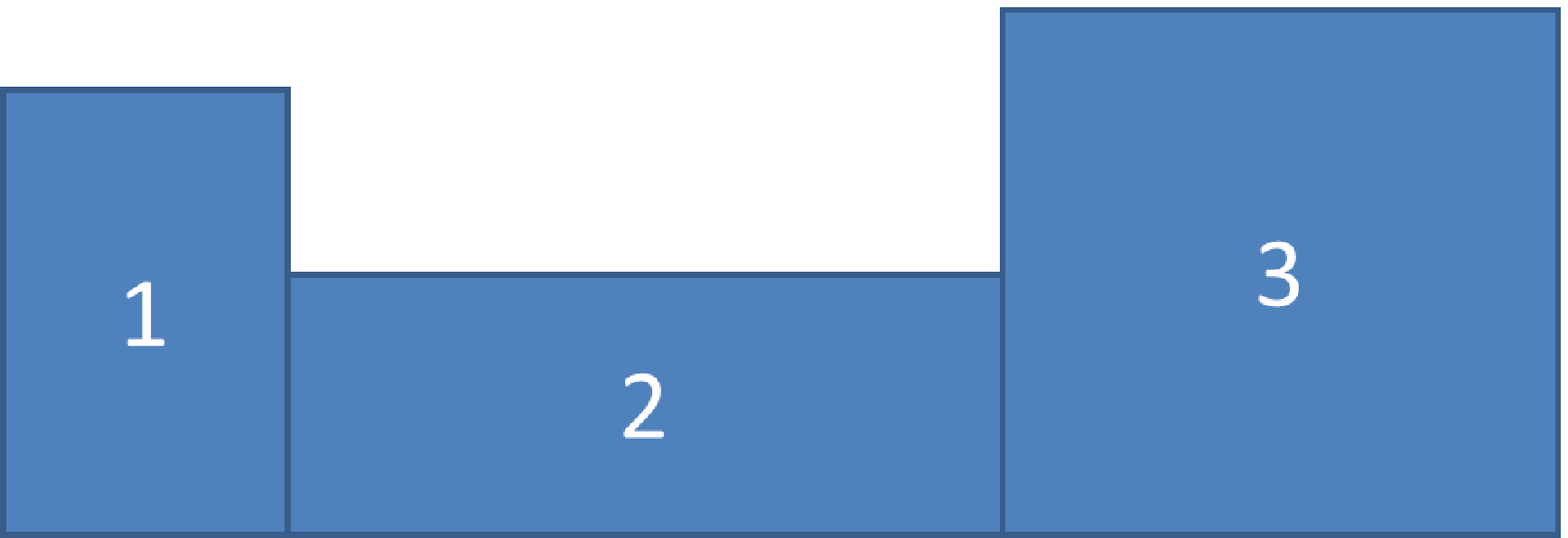}
  \includegraphics[width=\columnwidth]{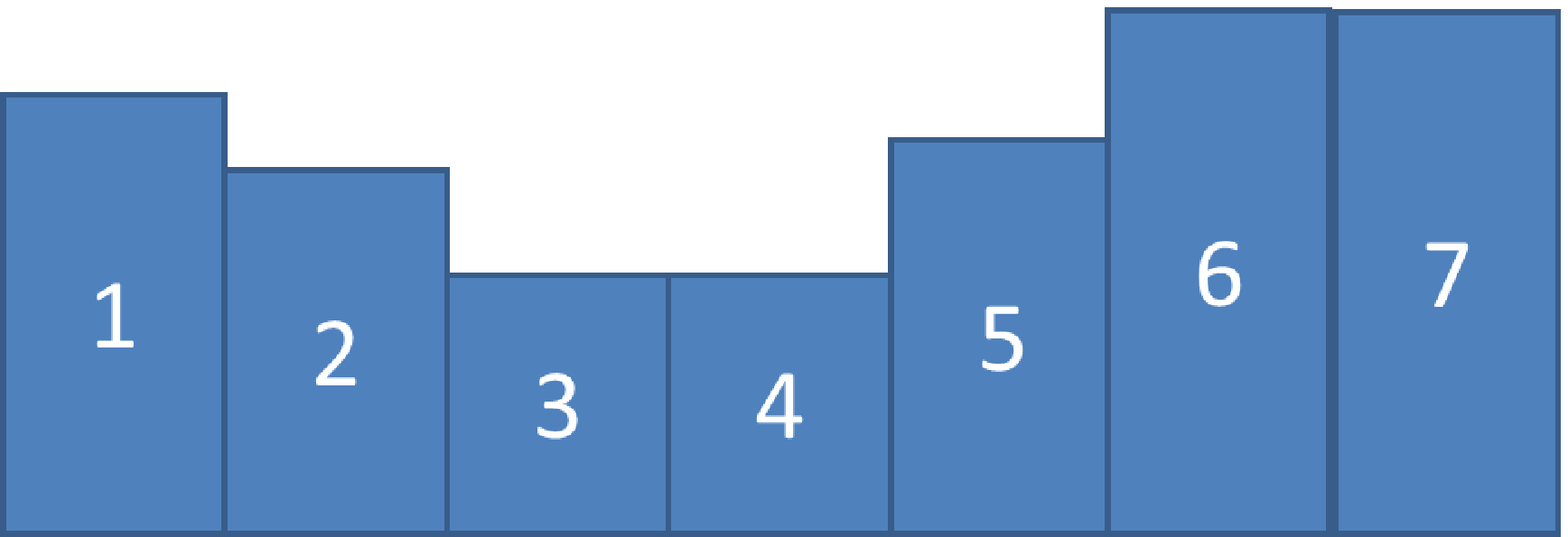}
  \caption{Conversion of $\hvec^*$ to $\bh$. Top figure shows $\hvec^*$ while the bottom one shows $\bh$. Note that buckets 1,3,4,6,7 of $\bh$ lie with-in buckets of $\hvec^*$, hence we assign same heights to them as the corresponding buckets in $\hvec^*$. Buckets 2, 5 of $\bh$ are at the intersection of of buckets 1 and 2, 2 and 3 of $\hvec^*$, hence a convex combination of heights of those buckets are assigned to Buckets 2, 5 of $\bh$.}
  \label{fig:convert}
\end{figure}

Now, note that, $$\sum_{j=1}^r\bh_j=\sum_{j=1}^r\hvec^*=M.$$
Furthermore, assuming $\frac{r}{b}\leq \Delta$ (i.e., width of buckets in $\bh$ is smaller than the smallest bucket of $\hvec^*$), average is over at most two different heights of $\hvec^*$ in each bucket of $\bh$. Since there are only $k$ buckets in $\hvec^*$, only $k-1$ buckets in $\bh$ are at the intersection of buckets in $\hvec^*$. Let $I$ be the set of all the buckets in $\bh$ that are at intersection of buckets in $\hvec^*$. Now consider $\|\hvec^*-\bh\|_2^2$. Clearly, if a bucket in $\bh$ do not lie in $\hvec^*$ then it do not contribute to $\|\hvec^*-\bh\|_2^2$, as value of attributes in that bucket is same as the value of attributes in the corresponding bucket of $\hvec^*$. Thus,
\begin{equation}
  \label{eq:ce1}
  \|\hvec^*-\bh\|_2^2 = \sum_{J\in I}\sum_{i\in J}(\hvec^*_i-\bh_i)^2\leq  \sum_{J\in I}\sum_{i\in J}(\hvec^*_i)^2\leq \frac{r}{b}\sum_{J\in I}h_i^*\|\hvec^*\|_\infty,. 
\end{equation}
where last inequality follows using Cauchy-Schwarz inequality. Now, $\|\hvec^*\|_\infty\leq \frac{M}{\Delta}$, as each bucket is of at least $\Delta$ width and there are at most $M$ records to fill in one bucket. Similarly, 
$$\Delta\sum_{J\in I}h_i^*  \leq \|\hvec^*\|_1 =M.$$
Hence, using \eqref{eq:ce1}, with the above observations
\begin{equation}
  \label{eq:ce2}
  \|\hvec^*-\bh\|_2\leq \frac{M}{\Delta}\sqrt{\frac{r}{b}}. 
\end{equation}
Now, we bound $F(\hvec^*)-F(\bh)$:
\begin{align}
  F(\hvec^*)-F(\bh)&=\mathbb{E}_{\qvec\sim \Dcal}[f(\qvec^T\hvec^*;s_q)-f(\qvec^T\bh;s_q)],\nonumber\\
&\leq\mathbb{E}_{\qvec\sim \Dcal}[L_f|\qvec^T\bh-\qvec^T\hvec^*|],\nonumber\\
&\leq\mathbb{E}_{\qvec\sim \Dcal}[L_f\|\qvec\|_2\|\hvec^*-\bh\|_2],\nonumber\\
&\leq \sqrt{|Q|}L_f  \frac{M}{\Delta}\sqrt{\frac{r}{b}},
\label{eq:ce3}
\end{align}
where last inequality follows from the fact that $\qvec$ is over $|Q|$ attributes only and using \eqref{eq:ce2}. 

Now using \eqref{eq:ce3} and \eqref{eq:be21}: 
\begin{equation}
  \label{eq:be2}
  E_2\leq \sqrt{|Q|}L_f  \frac{M}{\Delta}\sqrt{\frac{r}{b}} + E_1. 
\end{equation}
Hence, by combining \eqref{eq:tb}, \eqref{eq:be1}, \eqref{eq:be2}:
\begin{align}
  F(\hh)-F(\hvec^*) &\leq \sqrt{|Q|}L_f  \frac{M}{\Delta}\sqrt{\frac{r}{b}} \nonumber\\&\qquad+ O\left(\sqrt{\frac{b\log b\log \frac{1}{\delta}}{r\Delta N}}M|Q|L_f\right), \nonumber\\
&\leq M\left(L_f\left(\frac{|Q|}{\Delta}\right)^{3/4}\left(\frac{\log \frac{1}{\delta}}{N}\right)^{1/4}\right),\nonumber\\
&\leq M \epsilon,
  \label{eq:tb1}
\end{align}
where last inequality follows by selecting N using:
\begin{equation}
  \label{eq:bn}
  N\geq C_1\left(\frac{|Q|}{\Delta}\right)^3 \frac{L_f^4\log \frac{1}{\delta}}{\epsilon^4},
\end{equation}
where $C_1>0$ is a constant.
Now, second inequality in \eqref{eq:tb1} follows by selecting $b$ using:
\begin{multline}
b=C_2r\left(\left(\frac{1}{|Q|\Delta}\right)^{1/2}\left(\frac{N}{\log \frac{1}{\delta}}\right)^{1/2}\right)\\\geq C_2r\frac{|Q|L_f^2}{\Delta^2\epsilon^2}\geq C_2k\frac{|Q|L_f^2}{\Delta\epsilon^2},
\label{eq:bb}
\end{multline}
where last inequality follows using $k\geq \frac{r}{\Delta}$ as $\Delta$ is the minimum bucket width and $C_2>0$ is a constant.  

Hence proved.
%\begin{align}
  %F(\hw)-F(\tw)&\leq\lambda %\log b+O\left(\frac{\|\Ccal'\|^2M^2b\log\frac{1}{\delta}}{r\Delta\lambda N}\right),\nonumber\\
%&\leq %\sqrt{\frac{b\log b\log \frac{1}{\delta}}{r\Delta N}}M. 
 % \label{eq:e1_2}
%\end{align}
\end{proof}
\subsection{Proof of Theorem~\ref{thm:2d}}
\label{sec:proof2d}
\begin{proof}
Similar to proof of Theorem~\ref{thm:1d}:
\begin{align}
  F(\hhh)-F(H^*)&= F(\hhh)-F(\tilde{H})+F(\tilde{H})-F(H^*), \\
&=E_1+E_2,
\label{eq:tb_2d}
\end{align}
where $\tilde{H}\in \Ccal'$ is given by $\tilde{H}=B\tilde{W}B^T$. 
 
Recall that, 
\begin{equation}
\label{eq:g_2d}
G_\Dcal(W)=E_\Dcal\left[f(\langle Q,BWB^T\rangle;s_Q)\right]-\lambda H(\frac{r\cdot r}{Mb}W),  
\end{equation}
and $\tilde{W}$ is the optimal solution to \eqref{eq:g_2d}. Also, $E_1= F(\hhh)-F(\tilde{H})$ is the excess generalization error incurred by $\hhh$ compared to the optimal solution $\thhh$ and $E_2=F(\tilde{H})-F(H^*)$ is the difference between optimal error achievable by histograms in $\Ccal'$ to the histograms in $\Ccal$. Below we bound both the errors individually:\\
{\bf Bound on $E_1$}: Recall that $\thhh=B\tilde{W}B^T$. Also, note that \\$\langle Q, BWB^T\rangle=\langle B^TQB, W \rangle$ is the inner product function over matrices. As $G_\Dcal(\cdot)$ is the regularized empirical risk under a convex risk function $f$, using Theorem~\ref{thm:sco}, with probability $1-\delta$:
\begin{multline}
  \label{eq:e1_2d}
  \mathbb{E}_{Q\sim \Dcal}[f(\langle Q,B\hww B^T\rangle;s_Q)]-\lambda H(\frac{r^2}{Mb}\hww)\\\leq \mathbb{E}_{Q\sim \Dcal}[f(\langle Q,B\tww B^T\rangle;s_Q)]-\lambda H(\frac{r^2}{Mb}\tww)\\ + O\left(\frac{\Omega^2L_f^2M^2b\log\frac{1}{\delta}}{r\Delta\lambda N}\right),
\end{multline}
where:
\begin{compactitem}
  \item $L_f$ is the Lipschitz constant of function $f(u;s_q)$ w.r.t. $u$, over a finite set. Hence, $L_f\leq \max_{|u|\leq \Omega\|\Kcal\|_2}\|\grad f(u)\|_2$
  \item $\Omega=\max_{\qvec\sim \Dcal}\|B^TQB\|_F$. Assuming each query covers only $|Q|$ attribute values, $\Omega=|Q|$. 
  \item $\lambda$ is a constant. Also, note that entropy function $H(\frac{r^2}{Mb}W)$ is $\frac{r^2\Delta}{M^2b}$-strongly convex. 
\end{compactitem}

Now, $H(\frac{r^2}{Mb}\hww)\leq \log b$. Hence, using \eqref{eq:e1_2d},
\begin{align}
F(\hw)-F(\tw)&\leq \lambda \log b+O\left(\frac{L_f^2|Q|^2M^2b\log \frac{1}{\delta}}{r^2\Delta\lambda N}\right),\nonumber\\
&\leq O(\sqrt{\frac{b\log b\log \frac{1}{\delta}}{r^2\Delta N}}L_fM|Q|),
\label{eq:e12_2d}
\end{align}
where the second inequality follows by selecting $$\lambda=\sqrt{\frac{b \log \frac{1}{\delta}}{r^2\log b\Delta N}}M|Q|L_f.$$

That is, 
\begin{multline}
\label{eq:be1_2d}
E_1=F(\hhh)-F(\thhh)=F(\hww)-F(\tww)\\\leq O\left(\sqrt{\frac{b\log b\log \frac{1}{\delta}}{r^2\Delta N}}L_fM|Q|\right).
\end{multline}
{\bf Bound on $E_2$}: To bound $E_2$, we first observe that $$F(\thhh)\leq F(H)+\lambda \log b, \forall\ H\in \Ccal'.$$
Hence, 
\begin{equation}\label{eq:be21_2d}
E_2\leq F(H)-F(H^*)+E_1, \forall H\in \Ccal'.
\end{equation}
Now, given any $H^*$ we construct a new vector $\bhh \in \Ccal'$ for which we can bound the error $F(\bhh)-F(H^*)$. 

Note that along any of the axis, the number of 1-dimensional buckets are at most $k$. Hence, using error analysis from 1-dimensional case (see Equation~\ref{eq:ce2}), 
\begin{equation}
  \label{eq:ce2_2d}
  \|H^*-\bhh\|_F\leq \frac{M}{\Delta}\frac{r}{\sqrt{b_1}}. 
\end{equation}
Now, we bound $F(\hvec^*)-F(\bh)$:
\begin{align}
  F(H^*)-F(\bhh)&=\mathbb{E}_{Q\sim \Dcal}[f(\langle Q,H^*\rangle;s_Q)-f(\langle Q,\bhh\rangle;s_Q)],\nonumber\\
&\leq\mathbb{E}_{Q\sim \Dcal}[L_f|\langle Q,\bhh\rangle-\langle Q,H^*\rangle|],\nonumber\\
&\leq\mathbb{E}_{Q\sim \Dcal}[L_f\|Q\|_F\|H^*-\bhh\|_F],\nonumber\\
&\leq L_f\sqrt{|Q|}  \frac{M}{\Delta}\frac{r}{\sqrt{b_1}},
\label{eq:ce3_2d}
\end{align}
where the second inequality follows using Lipschitz property of $f$, and last inequality follows from the fact that $\qvec$ is over $|Q|$ attributes only and using \eqref{eq:ce2_2d}. 

Now using \eqref{eq:ce3_2d} and \eqref{eq:be21_2d}: 
\begin{equation}
  \label{eq:be2_2d}
  E_2\leq L_f\sqrt{|Q|}  \frac{M}{\Delta}\frac{r}{\sqrt{b_1}} + E_1. 
\end{equation}
Hence, by combining \eqref{eq:tb_2d}, \eqref{eq:be1_2d}, \eqref{eq:be2_2d}:
\begin{align}
  F(\hhh)-F(H^*) &\leq L_f\sqrt{|Q|}  \frac{M}{\Delta}\frac{r}{\sqrt{b_1}} \nonumber\\&\qquad+ O\left(\sqrt{\frac{b\log b\log \frac{1}{\delta}}{r\Delta N}}M|Q|L_f\right), \nonumber\\
&\leq M\left(L_f\left(\frac{|Q|}{\Delta}\right)^{3/4}\left(\frac{\log \frac{1}{\delta}}{N}\right)^{1/4}\right),\nonumber\\
&\leq M \epsilon,
  \label{eq:tb1_2d}
\end{align}
where last inequality follows by selecting N using:
\begin{equation}
  \label{eq:bn_2d}
  N\geq \left(\frac{|Q|}{\Delta}\right)^3 \frac{L_f^4\log \frac{1}{\delta}}{\epsilon^4}. 
\end{equation}
Now, second inequality in \eqref{eq:tb1_2d} follows by selecting $b$ using:
\begin{equation}
b=r^4\left(\left(\frac{1}{|Q|\Delta}\right)\left(\frac{N}{\log \frac{1}{\delta}}\right)\right)\geq r^4\frac{|Q|^2}{\Delta^4\epsilon^4}\geq k^2\frac{|Q|^2L_f^2}{\epsilon^4},
\label{eq:bb_2d}
\end{equation}
where last inequality follows using $k\geq \frac{r^2}{\Delta^2}$ as $\Delta$ is the minimum bucket width. 
Hence proved.
\end{proof}

%%% Local Variables: 
%%% mode: latex
%%% TeX-master: "histestimation-main"
%%% End: 

%\balancecolumns
%\appendix
%Appendix A
%\section{Headings in Appendices}

\end{document}